\documentclass[10pt]{scrartcl}
\pdfoutput=1

\usepackage[T1]{fontenc}

\usepackage{amsmath,amsfonts}
\usepackage{booktabs}

\usepackage[margin=0.8cm, font=small, labelfont=bf]{caption}
\usepackage[list=true]{subcaption}
\DeclareCaptionSubType*[arabic]{figure}
\DeclareCaptionSubType*[arabic]{table}
\usepackage{etoolbox}

\let\theparentequation\theequation
\patchcmd{\theparentequation}{equation}{parentequation}{}{}

\renewenvironment{subequations}[1][]{
  \refstepcounter{equation}%
  \setcounter{parentequation}{\value{equation}}
  \setcounter{equation}{0}
  \def\theequation{\theparentequation\alph{equation}}%
  \let\parentlabel\label
  \ifx\\#1\\\relax\else\label{#1}\fi
  \ignorespaces
}{%
  \setcounter{equation}{\value{parentequation}}
  \ignorespacesafterend
}

\newcommand*{\nextParentEquation}[1][]{
  \refstepcounter{parentequation}
  \setcounter{equation}{0}
  \ifx\\#1\\\relax\else\parentlabel{#1}\fi
}

\usepackage{jheppub}

\let\theparentequation\theequation
\patchcmd{\theparentequation}{equation}{parentequation}{}{}

\def\unity{{\bf 1}}
\def\zero{{\bf 0}}
\newcommand{\IE}{\textit{i.\,e.}}
\newcommand{\I}{{i\mkern1mu}}
\newcommand{\E}{\mathrm{e}}
\newcommand{\Real}[1]{\Re\hspace{-1pt}\mathfrak{e}{\left[#1\right]}}
\newcommand{\Realtilde}[1]{\tilde{\Re\hspace{2pt}}\hspace{-3pt}\mathfrak{e}{\left[#1\right]}}
\newcommand{\Imag}[1]{\Im\hspace{-1pt}\mathfrak{m}{\left[#1\right]}}
\newcommand{\MZ}{\mathbf{M}^{\mathbf{Z}}}
\newcommand{\TZ}{T^Z}
\newcommand{\st}{\tilde{t}}
\newcommand{\sbottom}{\tilde{b}}

\newcommand{\Anull}[1]{A_{0}{\left(#1\right)}}
\newcommand{\Bnull}[1]{B_{0}{\left(#1\right)}}
\newcommand{\Cnull}[1]{C_{0}{\left(#1\right)}}
\newcommand{\Beins}[1]{B_{1}{\left(#1\right)}}
\newcommand{\Bnnull}[1]{B_{00}{\left(#1\right)}}
\newcommand{\DBnull}[1]{B_{0}^{\prime}{\left(#1\right)}}

\newcommand{\Log}[1]{\operatorname{L}{\left(#1\right)}}
\newcommand{\dilog}[1]{\operatorname{Li_{2}}{\left(#1\right)}}

\newcommand*{\Tadpole}{T}

\newcommand\ontop[2]{\genfrac{}{}{0pt}{}{#1}{#2}}

\usepackage{setspace}

\newcommand{\mathsym}[1]{{}}
\newcommand{\unicode}[1]{{}}

\BeforeTOCHead[toc]{{\pdfbookmark[1]{\contentsname}{toc}}}

\begin{document}


\thispagestyle{empty}

\def\thefootnote{\fnsymbol{footnote}}

\begin{flushright}
MPP--2014--338
\end{flushright}

\vspace{2cm}

\begin{center}

{\large\sc {\bf Higgs boson masses and mixings in the complex MSSM}}

\vspace{0.4cm}

{\large\sc {\bf with two-loop top-Yukawa-coupling corrections}}

\vspace{1cm}

Wolfgang Hollik\footnote{email: hollik@mpp.mpg.de}
and
Sebastian Pa{\ss}ehr\footnote{email: passehr@mpp.mpg.de}

\vspace*{.7cm}

{\sl
Max-Planck-Institut f\"ur Physik \\
(Werner-Heisenberg-Institut) \\
F\"ohringer Ring 6, 
D--80805 M\"unchen, Germany
}

\end{center}

\vspace*{2cm}

\begin{abstract}{}
Results for the leading two-loop corrections of
$\mathcal{O}{\left(\alpha_t^2\right)}$ from the Yukawa sector to the
Higgs-boson mass spectrum of the MSSM with complex parameters are
presented, with details of the analytical calculation performed in the
Feynman-diagrammatic approach using a mixed
$\left.\text{on-shell}\middle/\,\overline{\text{DR}}\right.$ scheme that
can be directly matched onto the higher-order terms in the code {\tt
  FeynHiggs}.  Numerical results are shown for the masses and mixing
effects in the neutral Higgs-boson sector and their variation with the
phases of the complex parameters.  Furthermore, the analytical
expressions of the two-loop self-energies and the required
renormalization constants are recorded.  The new results can
consistently be implemented in {\tt FeynHiggs}.
\end{abstract}

\def\thefootnote{\arabic{footnote}}
\setcounter{page}{0}
\setcounter{footnote}{0}

\newpage
\tableofcontents\newpage
\section{Introduction}

The recently discovered new boson~\cite{Aad:2012tfa,Chatrchyan:2012ufa} 
with a mass around~$125.6$~GeV by the experiments ATLAS and CMS
at the LHC has given rise to substantial investigations to reveal the nature
of this particle as a Higgs boson responsible for
electroweak symmetry breaking.
Within the present experimental uncertainties
the measured properties of this new boson 
are consistent with the corresponding expectations for
the Standard Model Higgs boson~\cite{Landsberg:2013ina};
on the other hand, 
a large variety of other interpretations is possible where
the Higgs particle belongs to an extended model
connected to physics beyond the Standard Model. 
Within the theoretically well motivated 
minimal supersymmetric Standard Model~(MSSM),
the observed particle could be interpreted as a light state within
a richer  spectrum of scalar particles. 
The Higgs sector of the MSSM
consists of two complex scalar doublets leading
to five physical Higgs bosons and three (would-be)  Goldstone bosons.
At the tree-level, the physical states are given by  
the neutral $CP$-even $h,H$ and  $CP$-odd $A$ bosons, 
together with the charged $H^{\pm}$ bosons, and can be
parametrized in terms of the $A$-boson mass $m_A$ and
the ratio of the two vacuum expectation values, 
$\tan\beta = \left.v_2\middle/v_1\right.$.
In the MSSM with complex parameters, the cMSSM,
$CP$-violation is induced in the Higgs sector
via loop contributions involving complex parameters from other SUSY sectors
leading to mixing between $h,\,H$ and $A$ in the 
mass eigenstates~\cite{Pilaftsis:1998pe}.
 
Masses and mixings in the neutral sector are strongly affected by
loop contributions. A  lot of work has been invested 
into higher-order calculations of the mass spectrum from the SUSY parameters,
in the case of the real
MSSM~\cite{Heinemeyer:1998jw,Heinemeyer:1998np,Heinemeyer:1999be,Heinemeyer:2004xw,Borowka:2014wla,mhiggsFD3l,Zhang:1998bm,Espinosa:2000df,Brignole:2001jy,Casas:1994us,Degrassi:2002fi,Heinemeyer:2004gx,Allanach:2004rh,Martin:2001vx}
as well as for the cMSSM~\cite{Demir:1999hj,Pilaftsis:1999qt,Carena:2000yi,Heinemeyer:2007aq,Frank:2006yh}. 
The largest loop contributions originate from the Yukawa sector with the
large top Yukawa coupling~$h_t$, or~\mbox{$\alpha_t=\left.h_t^2\middle/(4\pi)\right.$}.
The class of leading two-loop Yukawa-type corrections  of $\mathcal{O}{\left(\alpha_{t}^{2}\right)}$ 
has been calculated for the case of real parameters ~\cite{Espinosa:2000df,Brignole:2001jy}, 
applying the method of the effective potential.
Together with the full one-loop result~\cite{Frank:2006yh} and the leading 
$\mathcal{O}{\left(\alpha_{t}\alpha_{s}\right)}$ terms~\cite{Heinemeyer:2007aq}, 
both accomplished in the Feynman-diagrammatic approach including complex parameters, 
it has been implemented in the public program 
{\tt FeynHiggs}~\cite{Heinemeyer:1998np,Degrassi:2002fi,Frank:2006yh,Heinemeyer:1998yj,Hahn:2010te}.
A calculation of the $\mathcal{O}{\left(\alpha_{t}^{2}\right)}$ terms
for the complex version of the MSSM, however, was not available so far; 
it is the content of this article.

In a recent paper~\cite{Hollik:2014wea}
we have shown first results for the impact of the $\mathcal{O}{\left(\alpha_{t}^{2}\right)}$ 
contributions within the cMSSM on the mass of the lightest neutral
Higgs boson. Here we give details of the calculation and list the
analytic results entering the evaluation of the one- and two-point
functions of the Higgs sector and the required  counterterms.     
The calculation is done in the Feynman-diagrammatic approach,
extending the on-shell renormalization scheme of Ref.~\cite{Frank:2006yh}
to the two-loop level. This ensures that the obtained analytical
results for the renormalized two-loop self-energies can consistently be incorporated 
in {\tt FeynHiggs}. 
In the numerical analysis, we show results for the masses and
$CP$-mixing of the three neutral Higgs bosons of the cMSSM and 
their dependence on the complex phases of the relevant parameters.

The paper is organized as follows: section~\ref{sec:HiggsSect} provides the
theoretical framework of the calculation and renormalization for getting
the dressed propagators of the neutral Higgs sector up to the two-loop level.
The necessary one-loop subrenormalization is described in section~\ref{sec:subren},
 and numerical results are shown in section~\ref{sec:numeric}.
The Appendix contains all couplings and counterterm vertices needed for
the calculation of the contributing Feynman diagrams, as well as a 
complete list of the counterterms and of the analytical expressions  
for the tadpoles and the self-energies of the Higgs sector.

\section{The Higgs sector of the complex MSSM\label{sec:HiggsSect}}
\subsection{Tree-level relations for masses and mixing}

The two scalar $SU(2)$-doublets are conventionally expressed in terms
of their components in the following way,
\begin{alignat}{2}
  \label{eq:Higgsfields}
  \mathcal{H}_{1} &= \begin{pmatrix} v_{1} + \frac{1}{\sqrt{2}}(\phi_{1} - \I \chi_{1})\\ -\phi^{-}_{1}\end{pmatrix},&\quad
  \mathcal{H}_{2} &= \E^{\I \xi}\begin{pmatrix} \phi^{+}_{2}\\ v_{2} + \frac{1}{\sqrt{2}}(\phi_{2} + \I \chi_{2})\end{pmatrix} .
\end{alignat}
Making use of the 
notation \mbox{$\phi^{-}_{1} = \left(\phi^{+}_{1}\right)^{\dagger}, \,
\phi^{-}_{2} = \left(\phi^{+}_{2}\right)^{\dagger}$},
the Higgs potential can be written as a polynomial in the field components,
\begin{align}
  \begin{split}
    V_{H} &= -T_{\phi_{1}} \phi_{1} - T_{\phi_{2}} \phi_{2} - T_{\chi_{1}} \chi_{1} - T_{\chi_{2}} \chi_{2}\\
         &\quad + \frac{1}{2}\begin{pmatrix} \phi_{1}, & \phi_{2}, & \chi_{1}, & \chi_{2} \end{pmatrix}
            \begin{pmatrix}\mathbf{M}_{\phi} & \mathbf{M}_{\phi\chi}\\ \mathbf{M}_{\phi\chi}^{\dagger} & \mathbf{M}_{\chi} \end{pmatrix}
            \begin{pmatrix} \phi_{1}\\ \phi_{2}\\ \chi_{1}\\ \chi_{2}\end{pmatrix}
            + \begin{pmatrix} \phi^{-}_{1}, & \phi^{-}_{2}\end{pmatrix} \mathbf{M}_{\phi^{\pm}} \begin{pmatrix} \phi^{+}_{1}\\ \phi^{+}_{2}\end{pmatrix} + \dots  \ ,
  \end{split}
\end{align}
where the third and fourth powers in the fields have been dropped.
Explicit expressions for the tadpole coefficients $T_i$ and for the mass matrices $\mathbf{M}$ 
can be found in Ref.~\cite{Frank:2006yh}. They are parametrized by the
phase $\xi$,  the real SUSY-breaking quantities 
\mbox{$m_{1,2}^{2} = \tilde{m}_{1,2}^{2} + \lvert\mu\rvert^{2}$},
and the complex SUSY-breaking quantity $m_{12}^{2}$. 
The latter can be redefined as real~\cite{Dimopoulos:1995kn} 
with the help of a Peccei--Quinn transformation~\cite{Peccei:1977hh} 
leaving only the phase $\xi$ as a source of $CP$-violation at the tree-level. 
The requirement of minimizing $V_{H}$ at the vacuum
expectation values $v_{1}$ and $v_{2}$  induces vanishing tadpoles at the tree level, 
which in turn leads to $\xi = 0$. As a consequence, also $\mathbf{M}_{\phi\chi}$ 
is equal to zero and~$\phi_{1,2}$ are decoupled from $\chi_{1,2}$ at the tree-level. 
The remaining $(2\times 2)$-matrices  $\mathbf{M}_{\phi}$, $\mathbf{M}_{\chi}$, $\mathbf{M}_{\phi^{\pm}}$ 
can be transformed into the mass eigenstate basis with the help of orthogonal
matrices~\mbox{$D(x) = \big(\begin{smallmatrix} -s_{x} & c_{x}\\ c_{x} & s_{x}\end{smallmatrix}\big)$}, 
using the abbreviations \mbox{$s_{x} \equiv \sin{x}$}, \mbox{$c_{x} \equiv \cos{x}$},
\begin{alignat}{6}
  \label{eq:higgsmixing}
  \begin{pmatrix} h\\ H \end{pmatrix} &= D(\alpha) \begin{pmatrix} \phi_{1}\\ \phi_{2}\end{pmatrix},&\quad
  \begin{pmatrix} A\\ G \end{pmatrix} &= D(\beta_{n}) \begin{pmatrix} \chi_{1}\\ \chi_{2}\end{pmatrix},&\quad
  \begin{pmatrix} H^{\pm}\\ G^{\pm} \end{pmatrix} &= D(\beta_{c}) \begin{pmatrix} \phi^{\pm}_{1}\\ \phi^{\pm}_{2}\end{pmatrix}.
\end{alignat}
The Higgs potential in this basis can be expressed as follows,
\begin{align}
\label{eq:HiggsPotential}
  \begin{split}
    V_{H} &= -T_h \, h- T_H \, H - T_A \, A - T_G\,  G\\
         &\quad + \frac{1}{2}\begin{pmatrix} h, & H, & A, & G \end{pmatrix}
            \mathbf{M}_{hHAG}
            \begin{pmatrix} h \\ H\\ A\\ G\end{pmatrix}
            + \begin{pmatrix} H^{-}, & G^{-}\end{pmatrix} 
                \mathbf{M}_{H^\pm G^\pm} 
                \begin{pmatrix} H^{+}\\ G^{+}\end{pmatrix} + \dots\ 
  \end{split}
\end{align}
with the tadpole coefficients $T_{h,H,A,G}$
and the mass matrices
\begin{align}
\label{eq:mmatrices}
\mathbf{M}_{hHAG} &= \begin{pmatrix}
        m^2_{h} & m^2_{hH} & m^2_{hA} & m^2_{hG} \\
        m^2_{hH} & m^2_{H} & m^2_{HA} & m^2_{HG} \\
        m^2_{hA} & m^2_{HA} & m^2_{A} & m^2_{AG} \\
        m^2_{hG} & m^2_{HG} & m^2_{AG} & m^2_{G} \end{pmatrix} , \qquad
\mathbf{M}_{H^\pm G\pm} \,= \begin{pmatrix}
        m^2_{H^\pm}  &  m^2_{H^-G^+} \\
        m^2_{G^-H^+} &  m^2_{G^\pm} \end{pmatrix}  ;
\end{align} 
explicit expressions for the entries are given in Ref.~\cite{Frank:2006yh}.

\noindent
At lowest order, the tadpoles and the non-diagonal entries of the mass
matrices vanish, 
\begin{align}
\label{eq:mmatricesdiag}
\mathbf{M}_{ h H A G}^{(0)}  & =\, \mathrm{diag} \left( m_h^2,\, m_H^2,\, m_A^2,\, m_G^2 \right) , \quad
\mathbf{M}_{H^\pm G^\pm}^{(0)} \, =\,  \mathrm{diag} \left( m_{H^\pm}^2,\, m_{G^\pm}^2 \right) ,
\end{align}
for $\beta = \beta_{n} = \beta_{c}$, with  $\beta$  
given in terms of the vacuum expectations values,
\begin{align}
\label{eq:deftanbeta}
\tan\beta &\equiv t_\beta = \frac{v_2}{v_1} \ , 
\end{align}
and for the second mixing angle $\alpha$ 
(with $-\frac{\pi}{2} < \alpha < 0$)  determined by
\begin{align}
  \label{eq:alpha}
  \tan (2\alpha) &= \frac{m_{A}^2 + m_Z^2}{m_A^2 - m_Z^2}\, \tan(2\beta)\, .
\end{align}
The Goldstone bosons are massless, $m_{G^\pm} = m_G = 0$, and 
the masses $m_{H^\pm}, m_A, m_h, m_H$  fullfil the relations 
\begin{subequations}
\label{eq:treelevelmasses}
\begin{align}
  m_{H^{\pm}}^{2} &= m_{A}^{2} + M_{W}^{2} \, , \\
  m_{h,\,H}^2 &= \frac{1}{2}\left(m_{A}^{2} +
    M_{Z}^{2}\mp\sqrt{\left(m_{A}^{2} + M_{Z}^{2}\right)^{2} - 4 m_{A}^{2} M_{Z}^{2}\, c_{2\beta}^{2}}\right) ,
\end{align}
\end{subequations}
including the vector-boson masses $M_W$ and $M_Z$.

\subsection{Masses and mixing beyond lowest order}
At lowest order, the irreducible two-point vertex functions of the
neutral Higgs sector 
\begin{align}
\label{eq:irredgammazero}
\Gamma^{(0)}_{hHAG}(p^2) & = \, i \, \Big[ p^2 \unity - \mathbf{M}^{(0)}_{hHAG} \Big] 
\end{align}
are diagonal, and the entries of the mass matrices in Eq.~\eqref{eq:mmatricesdiag}
provide the poles of the diagonal lowest-order propagators
\begin{align}
\Delta^{(0)}_{hHAG} (p^2) &=\, - \Big[ \Gamma^{(0)}_{hHAG}(p^2) \Big]^{-1} \, .
\end{align}
At higher order, the irreducible two-point functions are dressed by adding
 the renormalized self-energies,
\begin{align}
\label{eq:homassmatrix}
 p^2 \unity - \mathbf{M}^{(0)}_{hHAG} & \quad \to \quad
 p^2 \unity - \mathbf{M}^{(0)}_{hHAG}  +\mathbf{\hat{\Sigma}}_{hHAG}(p^2) \; \equiv \;
p^2 \unity - \mathbf{M}_{hHAG}(p^2) \, ,
\end{align}
yielding the renormalized two-point vertex functions $\hat{\Gamma}_{hHAG}(p^2)$,
which contain in general mixing of all fields with equal quantum numbers. 
The dressed propagators are obtained accordingly
by inverting the matrix $\hat{\Gamma}_{hHAG}(p^2)$.

In our case, we evaluate the momentum-dependent neutral 
``mass matrix'' in Eq.~\eqref{eq:homassmatrix} at the two-loop level,
%
\begin{align}
  \label{eq:masscorr}
    \mathbf{M}_{ h H A G}^{(2)} (p^2) &= \mathbf{M}_{ h H A G}^{(0)} - \mathbf{\hat{\Sigma}}_{h H A G}^{(1)}(p^2) - \mathbf{\hat{\Sigma}}_{h H A G}^{(2)}(0) \ .
\end{align}
Therein, $\mathbf{\hat{\Sigma}}_{h H A G}^{(k)}$ 
denotes the matrix of the renormalized diagonal and non-diagonal 
self-energies for the $h, H, A, G$ fields at loop order $k$.
For the complex MSSM, the one-loop self-energies are completely known~\cite{Frank:2006yh},
and the leading two-loop $\mathcal{O}{\left(\alpha_t \alpha_s\right)}$ contributions
have been obtained in the approximation of zero
external momentum ~\cite{Heinemeyer:2007aq}. Within the 
same approximation, treating the two-loop self-energies at $p^2=0$,
we derive the leading Yukawa contributions of~$\mathcal{O}{\left(\alpha_t^2\right)}$.

In order to obtain the physical Higgs-boson masses from the dressed propagators
in the considered approximation,
it is sufficient to derive explicitly the entries of the
$(3\times 3)$-submatrix of Eq.~\eqref{eq:masscorr} corresponding to the
$(hHA)$-components. Mixing with the Goldstone boson yields subleading
two-loop contributions; also Goldstone--$Z$ mixing occurs in principle,
which is related to the other Goldstone mixings by 
Slavnov--Taylor identities~\cite{Baro:2008bg,Williams:2011bu} and 
of subleading type as well~\cite{Hollik:2002mv}. 
However, mixing with Goldstone bosons has to be taken into account 
inside the loop diagrams and for a consistent renormalization. 

The masses of the three neutral Higgs bosons
including the new $\mathcal{O}{\left(\alpha_{t}^{2}\right)}$ contributions are given by the real parts of the
poles of the $hHA$-propagator matrix, obtained as the zeroes of the determinant of the 
renormalized  two-point vertex function, 
\begin{alignat}{4}
  \label{eq:higgspoles}
   \operatorname{det}\hat{\Gamma}_{hHA}{\left(p^2\right)} &= 0, &\quad
   \hat{\Gamma}_{hHA}{\left(p^2\right)} &= \I \left[p^2 {\unity} - \mathbf{M}_{hHA}^{(2)}{\left(p^2\right)}\right],
\end{alignat}
involving the corresponding $(3\times 3)$-submatrix of Eq.~\eqref{eq:masscorr}.
The impact of the self-energies on the mixing and couplings of the
various Higgs bosons can be obtained with he same formalism as described
in Ref.~\cite{Frank:2006yh}.

\subsection{Renormalized self-energies at the two-loop level}
For obtaining the renormalized self-energies in Eq.~\eqref{eq:masscorr},
counterterms have to be introduced up to second order in the loop
expansion, for the  tadpoles 
\begin{align}
\label{eq:tadpolct}
  T_i  &\rightarrow  T_i + \delta^{(1)}T_i + \delta^{(2)} T_i \, ,  \quad  i=h,\,H,\,A,\,G\,  ,
\end{align}
and for the mass matrices in
Eq.~\eqref{eq:HiggsPotential}, 
{\allowdisplaybreaks
\begin{subequations}
\begin{align}  
\mathbf{M}_{ h H A G}\label{eq:counterterms}
  &\rightarrow
  \mathbf{M}_{ h H A G}^{(0)} + \delta^{(1)} \mathbf{M}_{ h H A G} + \delta^{(2)} \mathbf{M}_{ h H A G}\ ,  \\
  \label{eq:dm2Lneutral}
  \delta^{(k)}\mathbf{M}_{hHAG} &= \begin{pmatrix}\delta^{(k)}m_{h}^{2} & \delta^{(k)}m_{hH}^{2} & \delta^{(k)}m_{hA}^{2} & \delta^{(k)}m_{hG}^{2}\\
    \delta^{(k)}m_{Hh}^{2} & \delta^{(k)}m_{H}^{2} & \delta^{(k)}m_{HA}^{2} & \delta^{(k)}m_{HG}^{2}\\
    \delta^{(k)}m_{Ah}^{2} & \delta^{(k)}m_{AH}^{2} & \delta^{(k)}m_{A}^{2} & \delta^{(k)}m_{AG}^{2}\\
    \delta^{(k)}m_{Gh}^{2} & \delta^{(k)}m_{GH}^{2} &
    \delta^{(k)}m_{GA}^{2} & \delta^{(k)}m_{G}^{2}\end{pmatrix},\\
   \nextParentEquation\parentlabel{eq:countertermscharged}
  \mathbf{M}_{H^\pm G^\pm}  
  &\rightarrow
  \mathbf{M}_{H^\pm G^\pm}^{(0)}  +\, \delta^{(1)} \mathbf{M}_{H^\pm G^\pm}   
  +\, \delta^{(2)} \mathbf{M}_{H^\pm G^\pm}   \, ,    \\
  \label{eq:dm2Lcharged}
  \delta^{(k)}\mathbf{M}_{H^\pm G^\pm} &= \begin{pmatrix}\delta^{(k)}m_{H^{\pm}}^{2} & \delta^{(k)}m_{H^{\pm}G^{\pm}}^{2}\\ \delta^{(k)}m_{G^{\pm}H^{\pm}}^{2} & \delta^{(k)}m_{G^{\pm}}^{2}\end{pmatrix}.
\end{align}
\end{subequations}
}%
For getting the proper counterterms for the mass matrices in Eq.~\eqref{eq:mmatrices}
one has to distinguish between the rotation angles $\beta_n, \beta_c$
from Eqs.~\eqref{eq:higgsmixing} and $\beta$ in Eq.~\eqref{eq:deftanbeta} when
generating the expressions for  the matrix elements in Eq.~\eqref{eq:mmatrices}.
Whereas $\alpha$, $\beta_n$ and $\beta_c$ are not renormalized, $\beta$ gets
counterterms $\beta \to \beta +\delta\beta$ according to the
renormalization of $\tan\beta$,
\begin{align}
  \label{eq:tanbetarenormalization}
  t_\beta &\rightarrow t_\beta + \delta^{(1)}t_\beta + \delta^{(2)}t_\beta\ .
\end{align}
In the resulting expressions for the counterterm matrices,
the identification $\beta_c =\beta_n = \beta$ is done afterwards at each order.
Details at the one-loop level can be found in Ref.~\cite{Frank:2006yh}.

In addition to the parameter renormalization described above,
field-renormalization constants
$  Z_{\mathcal{H}_{i}}  = 1 + \delta^{(1)}Z_{\mathcal{H}_{i}} +
\delta^{(2)}Z_{\mathcal{H}_{i}} $
are introduced up to two-loop order for each of the scalar doublets of Eqs.~\eqref{eq:Higgsfields} 
through the transformation
\begin{subequations}
\begin{align}
  \mathcal{H}_{i} \rightarrow \sqrt{Z_{\mathcal{H}_{i}}}\mathcal{H}_{i} &= \left[ 1 + \frac{1}{2}\delta^{(1)}Z_{\mathcal{H}_{i}} + \frac{1}{2}\Delta^{(2)}Z_{\mathcal{H}_{i}}
  \right]   \mathcal{H}_{i} \, ,\\
  \Delta^{(2)}Z_{\mathcal{H}_{i}} &= \delta^{(2)}Z_{\mathcal{H}_{i}} - \frac{1}{4}\left(\delta^{(1)}Z_{\mathcal{H}_{i}}\right)^{2}.
\end{align}
\end{subequations}
The field-renormalization constants in the mass-eigenstate basis of Eqs.~\eqref{eq:higgsmixing} are obtained by
\begin{subequations}
\label{eq:neutralhiggsfieldren}
\begin{align}
  \begin{pmatrix}h \\ H \end{pmatrix} &\rightarrow
  D(\alpha)\begin{pmatrix}\sqrt{Z_{\mathcal{H}_{1}}} & 0\\ 0 & \sqrt{Z_{\mathcal{H}_{2}}}\end{pmatrix} D(\alpha)^{-1} \begin{pmatrix} h \\ H \end{pmatrix}
   \; \equiv \; \mathbf{Z}_{hH} \begin{pmatrix}h\\ H \end{pmatrix},\\[0.2cm]
  \begin{pmatrix}A\\ G \end{pmatrix} &\rightarrow D(\beta_{n})\begin{pmatrix}\sqrt{Z_{\mathcal{H}_{1}}} & 0\\ 0 & \sqrt{Z_{\mathcal{H}_{2}}}\end{pmatrix} D(\beta_{n})^{-1} \begin{pmatrix}A\\ G \end{pmatrix}
   \; \equiv \; \mathbf{Z}_{AG}  \begin{pmatrix}A\\ G\end{pmatrix},\\[0.2cm]
  \begin{pmatrix}H^{\pm}\\ G^{\pm}\end{pmatrix} &\rightarrow D(\beta_{c})\begin{pmatrix}\sqrt{Z_{\mathcal{H}_{1}}} & 0\\ 0 & \sqrt{Z_{\mathcal{H}_{2}}}\end{pmatrix} D(\beta_{c})^{-1} \begin{pmatrix}H^{\pm}\\ G^{\pm}\end{pmatrix}
   \; \equiv \;  \mathbf{Z}_{H^{\pm}G^{\pm}}  \begin{pmatrix}H^{\pm}\\  G^{\pm} \end{pmatrix} .
\end{align}
\end{subequations}
One-loop expressions for the entries in the $\mathbf{Z}$-matrices in Eqs.~\eqref{eq:neutralhiggsfieldren}
are given in in Ref.~\cite{Frank:2006yh}); their extension up to two-loop order  
is listed in the following,
{\allowdisplaybreaks
\begin{subequations}
\label{eq:fieldcorr}
\begin{align}
\mathbf{Z}_{hH} &= \mathbf{1} + \delta^{(1)}\mathbf{Z}_{hH} + \delta^{(2)}\mathbf{Z}_{hH}\ ,\ \ 
  \delta^{(i)}\mathbf{Z}_{hH} = \frac{1}{2}\begin{pmatrix}\delta^{(i)}Z_{hh} & \delta^{(i)}Z_{hH}\\ \delta^{(i)}Z_{Hh} & \delta^{(i)}Z_{HH}\end{pmatrix},\\
  \delta^{(2)}Z_{hh} &= \left(s_{\alpha}^{2}\Delta^{(2)}Z_{\mathcal{H}_{1}} + c_{\alpha}^{2}\Delta^{(2)}Z_{\mathcal{H}_{2}}\right),\\
  \delta^{(2)}Z_{HH} &= \left(c_{\alpha}^{2}\Delta^{(2)}Z_{\mathcal{H}_{1}} + s_{\alpha}^{2}\Delta^{(2)}Z_{\mathcal{H}_{2}}\right),\\
  \delta^{(2)}Z_{hH} &= \delta^{(2)}Z_{Hh} = c_{\alpha}s_{\alpha}\left(\Delta^{(2)}Z_{\mathcal{H}_{2}} - \Delta^{(2)}Z_{\mathcal{H}_{1}}\right),\\
  \quad\notag\\
  \mathbf{Z}_{AG} &= \mathbf{1} + \delta^{(1)}\mathbf{Z}_{AG} + \delta^{(2)}\mathbf{Z}_{AG}\ ,\ \ 
  \delta^{(i)}\mathbf{Z}_{AG} = \frac{1}{2}\begin{pmatrix}\delta^{(i)}Z_{AA} & \delta^{(i)}Z_{AG}\\ \delta^{(i)}Z_{GA} & \delta^{(i)}Z_{GG}\end{pmatrix},\\
  \delta^{(2)}Z_{AA} &= \left(s_{\beta_{n}}^{2}\Delta^{(2)}Z_{\mathcal{H}_{1}} + c_{\beta_{n}}^{2}\Delta^{(2)}Z_{\mathcal{H}_{2}}\right),\\
  \delta^{(2)}Z_{GG} &= \left(c_{\beta_{n}}^{2}\Delta^{(2)}Z_{\mathcal{H}_{1}} + s_{\beta_{n}}^{2}\Delta^{(2)}Z_{\mathcal{H}_{2}}\right),\\
  \delta^{(2)}Z_{AG} &= \delta^{(2)}Z_{GA} = c_{\beta_{n}}s_{\beta_{n}}\left(\Delta^{(2)}Z_{\mathcal{H}_{2}} - \Delta^{(2)}Z_{\mathcal{H}_{1}}\right),\\
  \quad\notag\\
  \mathbf{Z}_{H^{\pm}G^{\pm}} &= \mathbf{1} + \delta^{(1)}\mathbf{Z}_{H^{\pm}G^{\pm}} + \delta^{(2)}\mathbf{Z}_{H^{\pm}G^{\pm}}\ ,\ \ 
  \delta^{(i)}\mathbf{Z}_{H^{\pm}G^{\pm}} = \frac{1}{2}\begin{pmatrix}\delta^{(i)}Z_{H^{\pm}H^{\pm}} & \delta^{(i)}Z_{H^{\pm}G^{\pm}}\\ \delta^{(i)}Z_{G^{\pm}H^{\pm}} & \delta^{(i)}Z_{G^{\pm}G^{\pm}}\end{pmatrix},\\
  \delta^{(2)}Z_{H^{\pm}H^{\pm}} &= \left(s_{\beta_{c}}^{2}\Delta^{(2)}Z_{\mathcal{H}_{1}} + c_{\beta_{c}}^{2}\Delta^{(2)}Z_{\mathcal{H}_{2}}\right),\\
  \delta^{(2)}Z_{G^{\pm}G^{\pm}} &= \left(c_{\beta_{c}}^{2}\Delta^{(2)}Z_{\mathcal{H}_{1}} + s_{\beta_{c}}^{2}\Delta^{(2)}Z_{\mathcal{H}_{2}}\right),\\
  \delta^{(2)}Z_{H^{\pm}G^{\pm}} &= \delta^{(2)}Z_{G^{\pm}H^{\pm}} = c_{\beta_{c}}s_{\beta_{c}}\left(\Delta^{(2)}Z_{\mathcal{H}_{2}} - \Delta^{(2)}Z_{\mathcal{H}_{1}}\right).
\end{align}
\end{subequations}
}%

\begin{figure}[t]
  \centering
  \includegraphics[width=0.85\textwidth]{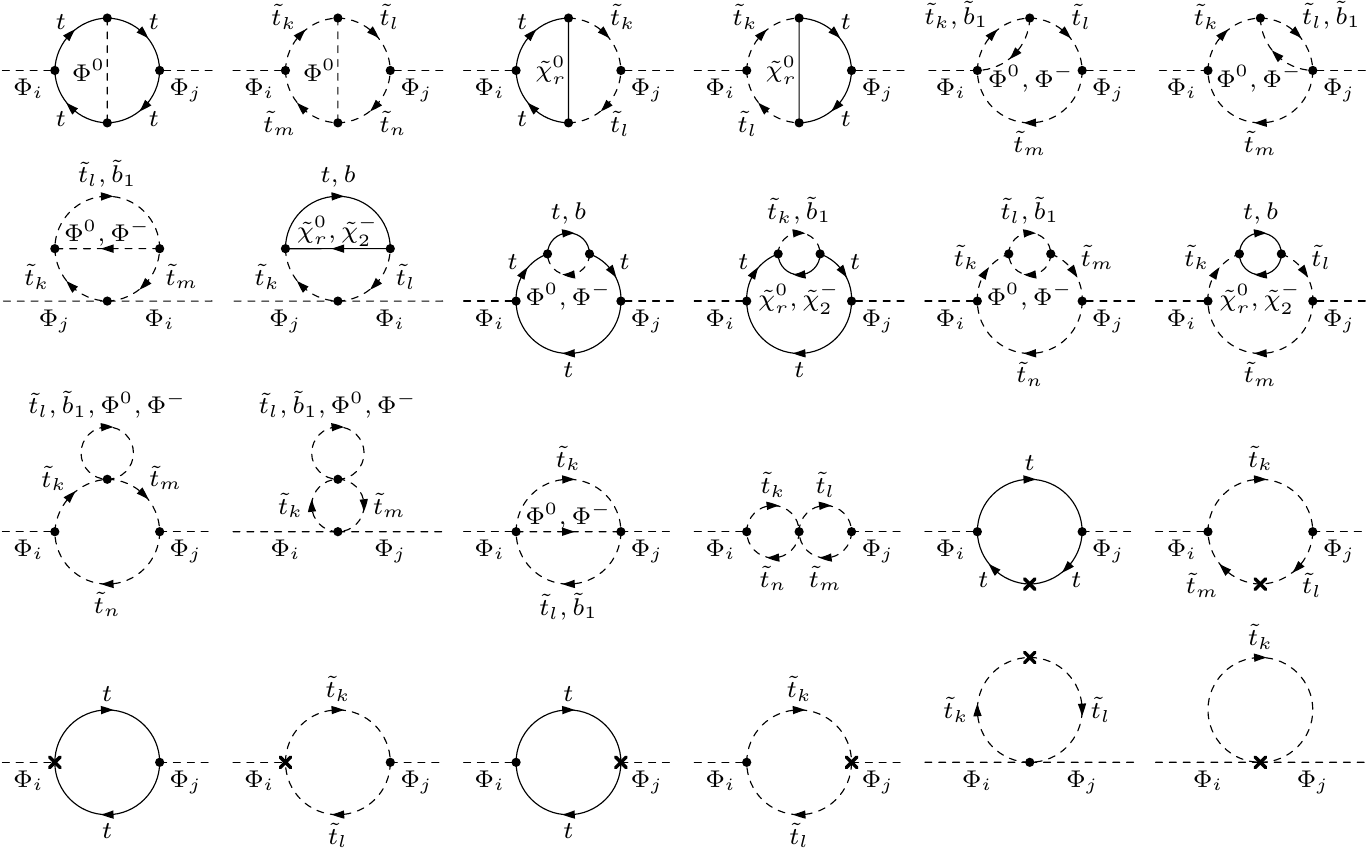}
  \caption{\label{fig:selfenergiesneutral} List of two-loop self-energy diagrams for the neutral Higgs bosons.
   One-loop counterterm insertions are denoted by a cross.
 \mbox{$\Phi_{i} = h,\,H,\,A$}; \mbox{$\;\Phi^0 = h,\,H,\,A,\,G$};  \mbox{$\;\Phi^- =H^-,\,G^-$}.  }

\end{figure}

\medskip%
\noindent
The next step is the determination of the renormalized self-energies $\mathbf{\hat{\Sigma}}$. 
The one-loop self-energies $\mathbf{\hat{\Sigma}}_{h H A  G} ^{(1)}(p^2)$ are 
contained in Ref.~\cite{Frank:2006yh}.
The renormalized two-loop self-energies can be written in compact matrix notation
as follows,
\begin{align}
\label{eq:renselfenergies}
\mathbf{\hat{\Sigma}}_{h H A G} ^{(2)} (p^2)  
   & =
\mathbf{\Sigma}_{h H A G} ^{(2)} (p^2) - \delta^{(2)} \MZ_{hHAG} \, ,
\end{align}
where $\mathbf{\Sigma}_{h H A G} ^{(2)}$ denotes the unrenormalized
self-energies corresponding to the sum of the genuine two-loop diagrams
and one-loop diagrams with subrenormalization. The
symbol $\delta^{(2)} \MZ_{hHAG}$ comprises all the two-loop counterterms
for  $\mathbf{\Sigma}_{h H A G} ^{(2)}$, resulting from parameter and field renormalization,
\begin{align}
\label{eq:higgs2Lct}
  \begin{split}
    \delta^{(2)}\MZ_{hHAG} &= 
      \begin{pmatrix} \delta^{(2)}\mathbf{Z}_{hH}^T & \mathbf{0} \\
                                \mathbf{0}  & \delta^{(2)}\mathbf{Z}_{AG}^T \end{pmatrix} 
     \left(\mathbf{M}_{hHAG}^{(0)} - p^2 \unity\right) 
      + \left(\mathbf{M}_{hHAG}^{(0)} - p^2 \unity\right)  
     \begin{pmatrix} \delta^{(2)}\mathbf{Z}_{hH} & \mathbf{0} \\ \mathbf{0}  & \delta^{(2)}\mathbf{Z}_{AG} \end{pmatrix}\\
      &\quad + \begin{pmatrix} \delta^{(1)}\mathbf{Z}_{hH}^T &
        \mathbf{0} \\ \mathbf{0}  & \delta^{(1)}\mathbf{Z}_{AG}^T \end{pmatrix} 
       \delta^{(1)}\mathbf{M}_{hHAG} + \delta^{(1)}\mathbf{M}_{hHAG} 
     \begin{pmatrix} \delta^{(1)}\mathbf{Z}_{hH} & \mathbf{0} \\ \mathbf{0}  & \delta^{(1)}\mathbf{Z}_{AG} \end{pmatrix}\\
      &\quad + \begin{pmatrix} \delta^{(1)}\mathbf{Z}_{hH}^T &
        \mathbf{0} \\ \mathbf{0}  &
        \delta^{(1)}\mathbf{Z}_{AG}^T \end{pmatrix} 
       \mathbf{M}_{hHAG}^{(0)} \begin{pmatrix}
         \delta^{(1)}\mathbf{Z}_{hH} & \mathbf{0} \\ 
         \mathbf{0}  & \delta^{(1)}\mathbf{Z}_{AG} \end{pmatrix} 
       \,+\, \delta^{(2)}\mathbf{M}_{hHAG} \ .
  \end{split}
\end{align}
Besides the field-renormalization constants from~Eqs.~\eqref{eq:fieldcorr}, 
and products of one-loop quantities, 
the two-loop mass counterterms are required, which are derived from
the Higgs potential. A complete list is given in App.~\ref{app:genuine2L};
this list is valid for the general two-loop case and not restricted to the
Yukawa approximation. 

\medskip
For the concrete calculation of the quantities entering Eq.~\eqref{eq:higgspoles} 
we evaluate the two-loop self-energies at \mbox{$p^2=0$}
in the top-Yukawa approximation, which neglects contributions from the gauge sector (gaugeless limit)
as well as the Yukawa coupling of the bottom quark setting the $b$-quark mass to zero.
In this approximation, only those Feynman diagrams that are depicted
in Fig.~\ref{fig:selfenergiesneutral} are calculated. 
The couplings utilized for their evaluation are listed in
App.~\ref{sec:couplings}. 
The diagrammatic calculation of the self-energies has been performed with
{\tt FeynArts}~\cite{Hahn:2000kx} for the generation of the Feynman diagrams 
and {\tt TwoCalc}~\cite{Weiglein:1995qs} for 
the two-loop  tensor reduction and trace evaluation.
The renormalization constants 
have been obtained with the help of {\tt FormCalc}~\cite{Hahn:1998yk}. 
The analytical result for the contribution from the genuine two-loop diagrams
can be found in App.~\ref{sec:self2L}, and the result from the diagrams
with subrenormalization in App.~\ref{sec:selfCT}.

In our approximation, the required two-loop mass counterterms 
are simplifications of those in App.~\ref{app:genuine2L} and read as follows, 
 \begin{subequations}\label{eq:atatMassCT_2L}
\begin{align}
  \begin{split}
    \delta^{(2)}m_{h}^{2} &= m_{H^{\pm}}^{2}\,c_{\beta}^{4} \left(\delta^{(1)}t_{\beta}\right)^{2} - \frac{e}{2\,M_{W}\,s_{\text{w}}}\,c_{\beta}^{2}\,\delta^{(1)}t_{\beta}\,\delta^{(1)}T_{H}\\
                       &\quad -\frac{e}{2\,M_{W}\,s_{\text{w}}} \left[\delta^{(2)}T_{h} + \delta^{(1)}T_{h}\,\delta^{(1)}Z_{\text{w}}\right]\ ,
  \end{split}\\
  \delta^{(2)}m_{H}^{2} &= \delta^{(2)}m_{H^{\pm}}^{2}\ ,\\
  \delta^{(2)}m_{A}^{2} &= \delta^{(2)}m_{H^{\pm}}^{2}\ ,\label{eq:dmA_2L}\\
  \begin{split}
    \delta^{(2)}m_{hH}^{2} &= m_{H^{\pm}}^{2}\,c_{\beta}^{2}\,\delta^{(2)}t_{\beta} + c_{\beta}^{2}\,\delta^{(1)}m_{H^{\pm}}^{2}\,\delta^{(1)}t_{\beta} - m_{H^{\pm}}^{2}\,c_{\beta}^{3}\,s_{\beta} \left(\delta^{(1)}t_{\beta}\right)^{2}\\
                         &\quad -\frac{e}{2\,M_{W}\,s_{\text{w}}} \left[\delta^{(2)}T_{H} + \delta^{(1)}T_{H}\,\delta^{(1)}Z_{\text{w}}\right]\ ,
  \end{split}\\
  \delta^{(2)}m_{hA}^{2} &= -\frac{e}{2\,M_{W}\,s_{\text{w}}} \left[\delta^{(2)}T_{A} + \delta^{(1)}T_{A}\,\delta^{(1)}Z_{\text{w}}\right]\ ,\\
  \delta^{(2)}m_{HA}^{2} &= 0\ .
\end{align}
\end{subequations}
The quantities~\mbox{$\delta^{(1)}e,\, \delta^{(1)}M_{W}$} and $\delta^{(1)}s_{\text{w}}$
 always occur in terms of the combination
\begin{align}
 \delta^{(1)}Z_{\text{w}} &= \frac{\delta^{(1)}e}{e} - \frac{\delta^{(1)}M_{W}}{M_{W}} - \frac{\delta^{(1)}s_{\text{w}}}{s_{\text{w}}}\ ;
\end{align}
where, however, in the gauge-less limit~$\delta^{(1)}e$ is equal to zero. 
The other elements of $\delta^{(2)}\mathbf{M}_{hHAG}$ not listed are determined
by symmetry (see App.~\ref{app:genuine2L}), or they
involve mixing with the Goldstone boson~$G$,
which is not needed for Eq.~\eqref{eq:higgspoles}.
Furthermore, in the gaugeless limit
the mass relations in Eq.~\eqref{eq:treelevelmasses} simplify to
$ m_h^2 = 0, \, m_H^2 = m_A^2,\,  m_{H^{\pm}}^2 = m_A^2$,
together with $m_G^2 = 0,\,  m_{G^{\pm}}^2 = 0$,
and  the mixing angle $\alpha$ is restricted by  the relations
\mbox{$s_{\alpha} = -c_{\beta}$} and \mbox{$c_{\alpha} = s_{\beta}$}.

In Eqs.~\eqref{eq:higgs2Lct} also several one-loop mass counterterms are needed, 
which in the present approximations are given by the following
expressions (symmetric in the neutral indices $h,H,A,G$), 
{\allowdisplaybreaks\vspace{-2ex}
\begin{subequations}\label{eq:atatMassCT_1L}
\begin{align}
  \delta^{(1)}m_h^2 &= -\frac{e}{2\,s_{\text{w}}\,M_W}\,\delta^{(1)}T_h\ ,\\
  \delta^{(1)}m_H^2 &= \delta^{(1)}m_{H^{\pm}}^2\ ,\\
  \delta^{(1)}m_{A}^{2} &= \delta^{(1)}m_{H^{\pm}}^{2}\ ,\label{eq:dmA_1L}\\
  \delta^{(1)}m_{hH}^2 &= -\frac{e}{2\,s_{\text{w}}\,M_W}\,\delta^{(1)}T_H + m_{H^{\pm}}^{2}\,c_{\beta}^2\,\delta^{(1)}t_{\beta}\ ,\\
  \delta^{(1)}m_{hA}^2 &= -\frac{e}{2\,s_{\text{w}}\,M_W}\,\delta^{(1)}T_A\ ,\\
  \delta^{(1)}m_{hG}^2 &= 0\ ,\\
  \delta^{(1)}m_{HA}^2 &= 0\ ,\\
  \delta^{(1)}m_{HG}^2 &= \delta^{(1)}m_{hA}^2\ ,\\
  \delta^{(1)}m_{AG}^2 &=  -\frac{e}{2\,s_{\text{w}}\,M_W}\,\delta^{(1)}T_H  -m_A^2\,  c_\beta\,  s_\beta \, \delta^{(1)}t_\beta  \ ,\\
  \delta^{(1)}m_{H^-G^+}^2 &= \frac{e}{2\,s_{\text{w}}\,M_W} \left[\delta^{(1)}T_H - \I \delta^{(1)}T_A\right] - m_{H^{\pm}}^2\,c_{\beta}^2\,\delta^{(1)}t_{\beta}\ ,\\
  \delta^{(1)}m_{G^-H^+}^2 &= \left(\delta^{(1)}m_{H^-G^+}^2\right)^*\ .
\end{align}
\end{subequations}
}%

\medskip
The renormalization constants in Eqs.~\eqref{eq:tadpolct}--\eqref{eq:countertermscharged} are
determined via renormalization conditions that are extended from the
one-loop level, as specified in Ref.~\cite{Frank:2006yh}, to two-loop
order; explicit expressions for
the renormalization constants are given in App.~\ref{sec:renconst}:


\begin{itemize}
  \item 
The tadpole counterterms $\delta^{(k)} T_i $ are fixed by requiring that
the minimum of the Higgs potential is not shifted, which means that the  tadpole coefficients 
have to vanish at each order,\footnote{The counterterms $\delta^{(k)}T_G$ are not
  independent and do not need separate renormalization conditions}
   \begin{align}
    T_{i}^{(1)} + \delta^{(1)}T_{i}  &= 0\ , \quad
    T_{i}^{(2)} + \delta^{(2)}\TZ_{i} = 0\ , \quad
    i = h,\,H,\,A\ , 
\end{align}
where
\begin{subequations}
  \begin{align}
    \left(\delta^{(2)}\TZ_h,\ \delta^{(2)}\TZ_H \right) &= 
    \left(\delta^{(1)} T_h,\ \delta^{(1)} T_H\right) \delta^{(1)}\mathbf{Z}_{hH} + \left(\delta^{(2)}T_h,\ \delta^{(2)}T_H\right),\\
    \left(\delta^{(2)}\TZ_A,\ \delta^{(2)}\TZ_G\right) &= 
    \left(\delta^{(1)} T_A,\ \delta^{(1)} T_G \right) \delta^{(1)}\mathbf{Z}_{AG} + \left(\delta^{(2)}T_A,\ \delta^{(2)}T_G\right).
  \end{align}
\end{subequations}
$T_{i}^{(k)}$ denote the unrenormalized one-point vertex functions at one- and two-loop order; the
two-loop diagrams contributing to $T_i^{(2)}$ are displayed in Fig.~\ref{fig:tadpoles} and 
written down in App.~\ref{sec:tadpole2L} and App.~\ref{sec:tadpoleCT}. 
The relation for the mixing angles \mbox{$\beta_{n} = \beta_{c} = \beta$}
is a consequence of the tapole conditions \mbox{$T_i = 0$} at lowest order.

\begin{figure}[t]

  \centering
  \includegraphics[width=0.85\textwidth]{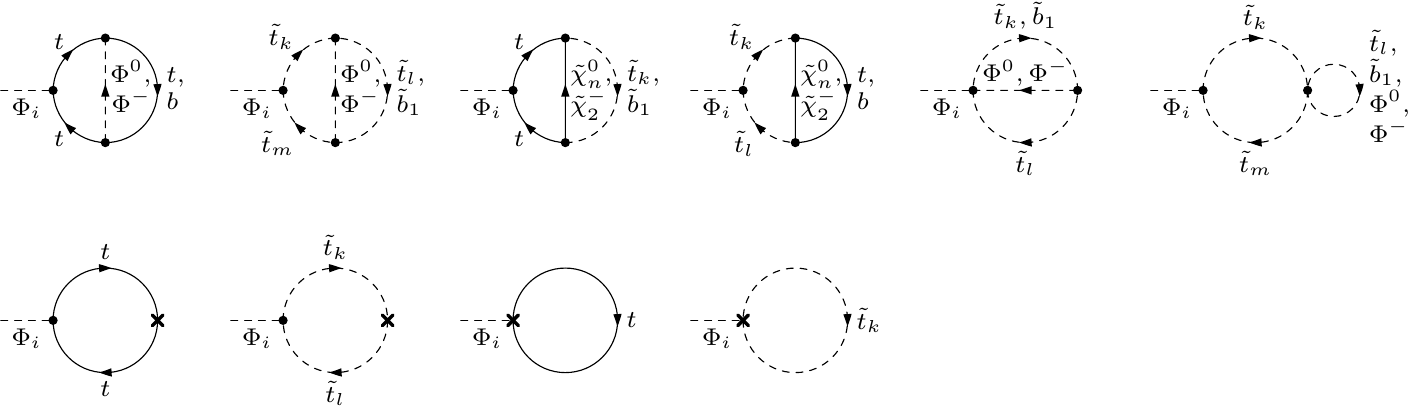}
  \caption{\label{fig:tadpoles} List of two-loop tadpole diagrams 
  contributing to $T^{(2)}_i$.
  One-loop counterterm insertions are denoted by a cross.
  \mbox{$\Phi_{i} = h,\,H,\,A$}; \mbox{$\;\Phi^0 = h,\,H,\,A,\,G$};  \mbox{$\;\Phi^- =H^-,\,G^-$}. }

\end{figure}

 %
  \item 
The charged Higgs-boson mass $m_{H^\pm}$ is the only independent mass
parameter of the Higgs sector and is used as an input
quantity. Accordingly, the corresponding mass counterterm is fixed by
an independent renormalization condition, chosen as on-shell
condition, which in the~\mbox{$p^2=0$} approximation is given by
\mbox{$\Real{\hat{\Sigma}_{H^{\pm}}^{(k)}(0)} = 0$} for the renormalized
charged-Higgs self-energy, at the two-loop level specified in terms of
the unrenormalized charged self-energies~(the contributing Feynman
diagrams are shown in Fig.~\ref{fig:selfenergiescharged}) and
respective counterterms,
\begin{align}
\label{eq:renchargedselfenergy}
\hat{\Sigma}_ {H^\pm} ^{(2)}  &=
\left( \mathbf{\hat{\Sigma}}_{H^\pm G^\pm}^{(2)}  \right) _{11}\ , \quad 
 \mathbf{\hat{\Sigma}}_{H^\pm G^\pm}^{(2)} (0) = \mathbf{\Sigma}_{H^\pm G^\pm} ^{(2)} (0) -\delta^{(2)} \MZ_{H^\pm G^\pm}\ , 
\end{align}
with
\begin{align}
\begin{split}
  \delta^{(2)}\MZ_{H^{\pm}G^{\pm}} &=
  \delta^{(2)}\mathbf{Z}_{H^{\pm}G^{\pm}}^{T}\mathbf{M}_{H^{\pm}G^{\pm}}^{(0)} + \mathbf{M}_{H^{\pm}G^{\pm}}^{(0)}\delta^{(2)}\mathbf{Z}_{H^{\pm}G^{\pm}}\\
  &\quad + \delta^{(1)}\mathbf{Z}_{H^{\pm}G^{\pm}}^{T} \delta^{(1)}\mathbf{M}_{H^{\pm}G^{\pm}} + \delta^{(1)}\mathbf{M}_{H^{\pm}G^{\pm}}\delta^{(1)}\mathbf{Z}_{H^{\pm}G^{\pm}}\\
  &\quad + \delta^{(1)}\mathbf{Z}_{H^{\pm}G^{\pm}}^{T}\mathbf{M}_{H^{\pm}G^{\pm}}^{(0)}\delta^{(1)}\mathbf{Z}_{H^{\pm}G^{\pm}} + \delta^{(2)}\mathbf{M}_{H^{\pm}G^{\pm}}\ .
\end{split}
\end{align}
From the on-shell condition, the independent mass counterterm 
\mbox{$\delta^{(2)} m_{H^{\pm}}^{2} = \left( \delta^{(2)} \mathbf{M}_{H^\pm G^\pm} \right)_{11}$}
can be extracted, yielding
\begin{align}
  \begin{split}
  \delta^{(2)}m_{H^{\pm}}^{2} &= \Real{\Sigma_{H^\pm}^{(2)}{\left(0\right)}} - \Real{\delta^{(1)}m_{H^{\pm}H^{\pm}}\,\delta^{(1)}Z_{H^{\pm}H^{\pm}} + \delta^{(1)}m_{H^{\pm}G^{\pm}}\,\delta^{(1)}Z_{H^{\pm}G^{\pm}}}\\
  &\quad - \Real{m_{H^{\pm}}^{2}\left(\left(\frac{1}{2}\delta^{(1)}Z_{H^{\pm}H^{\pm}}\right)^{2} + \delta^{(2)}Z_{H^{\pm}H^{\pm}}\right)}.
  \end{split}
\end{align}
The result for the charged Higgs-boson self-energy can be found in App.~\ref{sec:self2L} and App.~\ref{sec:selfCT}.
%
\begin{figure}[t]
  \centering
  \includegraphics[width=0.85\textwidth]{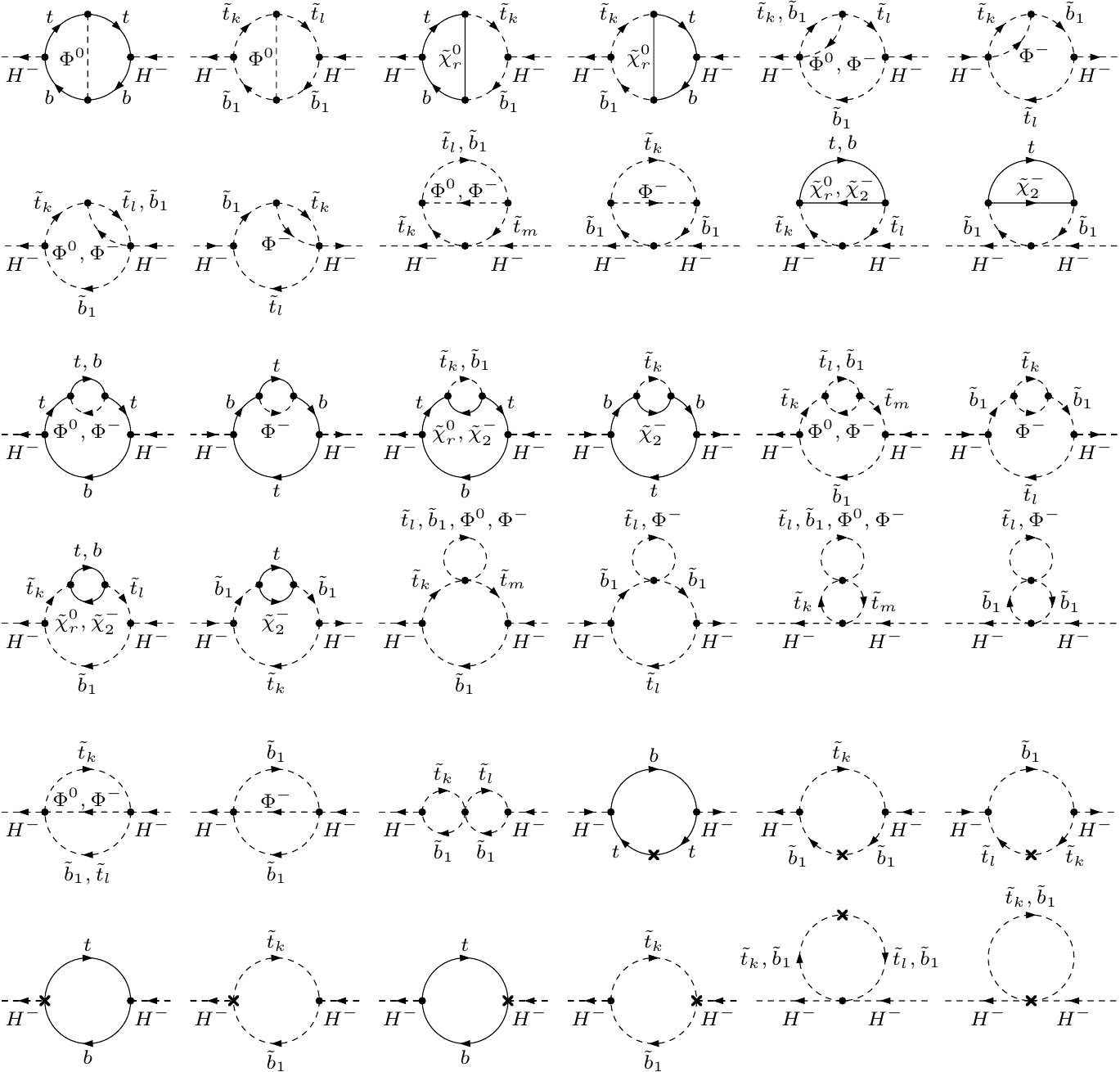}
  \caption{\label{fig:selfenergiescharged} List of two-loop self-energy diagrams for the charged Higgs bosons.
   One-loop counterterm insertions are denoted by a cross.
 \mbox{$\;\Phi^0 = h,\,H,\,A,\,G$};  \mbox{$\;\Phi^- =H^-,\,G^-$}.  }
\end{figure}

  \item The field-renormalization constants of the Higgs mass eigenstates in Eq.~\eqref{eq:neutralhiggsfieldren} 
        are combinations of the basic doublet-field renormalization constants
        $\delta^{(k)}Z_{\mathcal{H}_{1}}$ and $\delta^{(k)}Z_{\mathcal{H}_{2}}$ $(k=1,2)$, 
        which are fixed by the~$\overline{\text{DR}}$-conditions for the derivatives
        of the corresponding self-energies,
        \begin{alignat}{4}
          \delta^{(k)}Z_{\mathcal{H}_{1}} &= 
          -\left[ \Sigma^{(k)\, '}_{HH}(0) \right]_{\alpha=0}^{\rm div} , &\quad
          \delta^{(k)}Z_{\mathcal{H}_{2}} &=
          -\left[ \Sigma^{(k)\, '}_{hh}(0) \right]_{\alpha=0}^{\rm div} .
        \end{alignat}
%
  \item 
        $t_\beta \equiv \tan\beta$ is
        renormalized in the~$\overline{\text{DR}}$-scheme, 
        which has been shown to be a very convenient choice~\cite{Freitas:2002um} 
        (alternative process-dependent definitions and renormalization
        of  $t_{\beta}$ can be found in Ref.~\cite{Baro:2008bg}). 
        It has been clarified in Refs.~\cite{Sperling:2013eva} 
        that the counterterm for \mbox{$k=1,2$} can be written as
      \begin{align}
          \delta^{(k)}t_{\beta}^2 &= t_{\beta}^2 \left(\delta^{(k)}Z_{\mathcal{H}_{2}} - \delta^{(k)}Z_{\mathcal{H}_{1}}\right),
        \end{align}
      which at the two-loop level, however, is a special feature of
      our approximation and not generally valid.
  \item 
In the on-shell scheme, also the counterterms 
$\left.\delta^{(1)}M_W^2\middle/M_W^2\right.$ and $\left.\delta^{(1)}M_Z^2\middle/M_Z^2\right.$ 
appear, which are required for renormalization of the top Yukawa coupling 
    \mbox{$h_{t} = \left.\left(e\,m_{t}\right)\middle/\left(\sqrt{2}\,s_{\beta}\,s_{\rm w}\,M_{W}\right)\right.$}.
Also in the gauge-less limit these ratios have finite and divergent contributions 
arising from the Yukawa couplings and thus  have to be included as one-loop quantities
$\sim h_t^2$; they are evaluated from the~$W$ and~$Z$ self-energies yielding 
    \begin{alignat}{6}
       \frac{\delta^{(1)}M_W^2}{M_W^2} &=  \frac{\Sigma^{(1)}_W(0)}{M_W^2} ,  &\quad \frac{\delta^{(1)}M_Z^2}{M_Z^2}  &=  \frac{\Sigma^{(1)}_Z(0)}{M_Z^2} ,  &\quad
        \delta^{(1)}s_{\rm w}^2  &= c_{\rm w}^{2}  \left(\frac{\delta^{(1)}M_{Z}^2}{M_{Z}^2} - \frac{\delta^{(1)}M_{W}^2}{M_{W}^2}\right).
    \end{alignat}
In the Yukawa approximation, $\delta^{(1)}s_{\rm w}^2 $ is finite. 
The corresponding Feynman graphs are contained in Fig.~\ref{fig:RCCha}.
\end{itemize}

\medskip
The appearance of the finite
quantity~$\delta^{(1)} s_{\rm w}^2$ in the~$\mathcal{O}{\left(\alpha_{t}^{2}\right)}$ terms 
is a consequence of the on-shell scheme
where the top-Yukawa coupling~\mbox{$h_{t} =\left.m_{t}\middle/v_{2}\right. =\left.m_{t}\middle/(v\,s_{\beta})\right.$} 
is expressed in terms of the parameters
\begin{align}
  \frac{1}{v}  &= \frac{g_{\text{w}}}{\sqrt{2}\,M_W} = \frac{e}{\sqrt{2}\,s_{\rm w}\,M_{W}} .
\end{align}
Accordingly, also the one-loop self-energies have to be parametrized in terms
of this representation for~$h_t$ when added to the two-loop
self-energies in Eq.~\eqref{eq:masscorr}. On the other hand, it may be
convenient to use the Fermi constant~$G_{\rm F}$
for parametrization of the one-loop self-energies; in that case the relation
\begin{align}
  \sqrt{2}\,G_{\text{F}} &= \frac{e^2}{4\,s^2_{\rm w}\,M^2_{W}}\left(1 + \Delta^{(1)}r\right) ,
\end{align}
has to be applied, which is affected by loop contributions also in the gaugeless limit,
described by the one-loop quantity 
\begin{align}
  \Delta^{(1)} r = - \frac{c^2_{\rm w}}{s^2_{\rm w}} 
  \left(\frac{\delta^{(1)}M_{Z}^2}{M_{Z}^2} - \frac{\delta^{(1)}M_{W}^2}{M_{W}^2} \right)
  \, =\, - \frac{\delta^{(1)}s_{\rm w}^2}{s^2_{\rm w}} \, .
\end{align}
This finite shift in the one-loop self-energies induces two-loop terms
of~$\mathcal{O}{\left(\alpha_t^2\right)}$, 
effectively cancelling all other occurrences of $\delta^{(1)}s_{\rm w}^2$.

\section{Subrenormalization}
\label{sec:subren}
The two-loop top Yukawa coupling contributions to the self-energies and tadpoles 
involve one-loop diagrams with insertions of one-loop counterterms.
This subrenormalization concerns masses and couplings in the colored sector and in the  
chargino--neutralino sector.

\subsection{The third-generation quark--squark sector}
The required one-loop counterterms for subrenormalization arise from 
the top and scalar top~$\big(\st\big)$ as well as scalar bottom~$\big(\sbottom\big)$ sectors.
The stop and sbottom mass matrices in the 
$\big(\tilde{t}_{\text{L}},\tilde{t}_{\text{R}}\big)$ and
$\big(\tilde{b}_{\text{L}},\tilde{b}_{\text{R}}\big)$ bases are given by
\begin{align}
  \label{eq:squarks}
    \mathbf{M}_{\tilde{q}} &= 
    \begin{pmatrix}
     m_{\tilde{q}_{\text{L}}}^{2} + m_{q}^{2} + M_Z^2\,c_{2\beta} (T_q^3 - Q_q\,s^2_{\rm w}) & 
     m_{q}\left(A_{q}^{*} - \mu\,\kappa_q \right)\\[0.1cm]
     m_{q}\left(A_{q} - \mu^{*}\,\kappa_q \right) & 
     m_{\tilde{q}_{\text{R}}}^{2} + m_{q}^{2} + M_Z^2\,c_{2\beta}\,Q_q\,s^2_{\rm w}
   \end{pmatrix}, &
     \kappa_t &= \frac{1}{t_{\beta}},& \kappa_b &= t_{\beta} ,
\end{align}
with $Q_q$ and $T^3_q$ denoting charge and isospin of $q = t,\,b$.
$SU(2)$-invariance requires
\mbox{$m_{\tilde{t}_{\text{L}}}^{2} = m_{\tilde{b}_{\text{L}}}^{2} \equiv m_{\tilde{q}_{3}}^{2}$}. 
In the gaugeless approximation the $D$-terms vanish in 
both the $\st$ and $\sbottom$ matrices.
Moreover, in our approximation the $b$-quark is treated as massless; 
hence, the off-diagonal entries of the sbottom matrix are zero and the
mass eigenvalues can be read off directly,
\mbox{$m_{\tilde{b}_{1}}^{2} = m_{\sbottom_{\rm L}}^2 = m_{\tilde{q}_{3}}^{2}$},
\mbox{$m_{\tilde{b}_{2}}^{2} = m_{\tilde{b}_{\text{R}}}^{2}$}. 
The stop mass eigenvalues 
can be obtained by performing a unitary transformation,
\begin{align}
  \label{eq:squarkdiag}
  \mathbf{U}_{\tilde{t}}\mathbf{M}_{\tilde{t}}\mathbf{U}_{\tilde{t}}^{\dagger}  &= 
  \mathrm{diag}{\left(m_{\tilde{t}_{1}}^{2},\, m_{\tilde{t}_{2}}^{2}\right)}.
\end{align}
Since $A_{t}$ and $\mu$ are complex parameters in general, the unitary
matrix $\mathbf{U}_{\tilde{t}}$  consists of one 
mixing angle $\theta_{\st}$ and one phase $\varphi_{\st}$.

\begin{figure}[t]
  \centering
  \includegraphics[width=\textwidth]{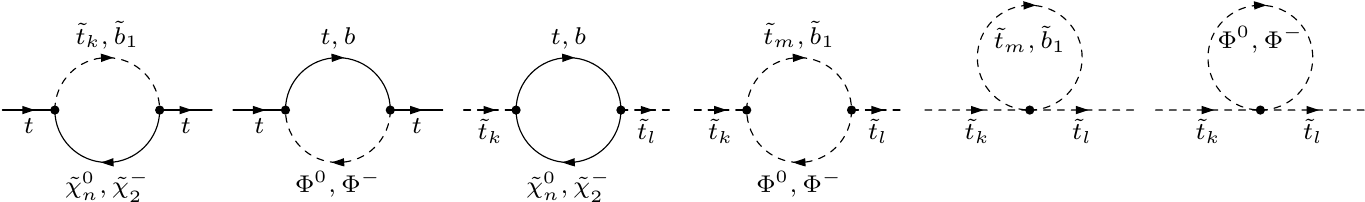}
  \caption{\label{fig:RCTop}
    Feynman diagrams for renormalization of the quark--squark sector.
   \mbox{$\Phi^{0} = h,\,H,\,A,\,G$}; 
   \mbox{$\Phi^{-} = H^{-},\,G^{-}$.} }
  \vspace{2ex}
\end{figure}

\medskip
Five independent parameters are introduced by the quark--squark sector,
which enter the two-loop calculation in addition to those of the previous section: 
the top mass $m_{t}$, the soft SUSY-breaking parameters 
$m_{\tilde{q}_{3}}$ and $m_{\tilde{t}_{\text{R}}}$ ($m_{\tilde{b}_{\text{R}}}$  decouples for $m_{b} = 0$), 
and the complex mixing parameter~\mbox{$A_{t} = \lvert A_{t}\rvert\E^{i\phi_{A_{t}}}$}. 
On top, $\mu$ enters as another free parameter related to the Higgsino sector.  
These parameters have to be renormalized at the one-loop level, 
\begin{subequations}
\begin{align}
  m_t &\rightarrow m_t + \delta^{(1)}m_t,\\
  \mathbf{M}_{\st}  &\rightarrow \mathbf{M}_{\st} + \delta^{(1)} \mathbf{M}_{\st} .
\end{align}
\end{subequations}
The independent renormalization conditions for the colored sector
are formulated in the following way:
\begin{itemize}
  \item 
The mass of the top quark is defined on-shell,
\IE\footnote{$\tilde{\Re\hspace{2pt}}\hspace{-3pt}\mathfrak{e}$ denotes the real part of all loop integrals, 
but leaves the couplings unaffected.}
    \begin{align}
      \delta^{(1)}m_{t} &=
      m_{t} \, \Realtilde{\frac{1}{2}\left(\Sigma_{t}^{(1),\,\text{L}}{\left(m_{t}^{2}\right)} 
    + \Sigma_{t}^{(1),\,\text{R}}{\left(m_{t}^{2}\right)}\right) + \Sigma_{t}^{(1),\,\text{S}}{\left(m_{t}^{2}\right)}},
    \end{align}
 according to the Lorentz decomposition of the self-energy of the top quark 
 (the contributing Feynman diagrams are depicted in Fig.~\ref{fig:RCTop})
\begin{align}
\label{eq:Lorentz} 
\Sigma_t (p) & =\,  \not{\! p}\, \omega_-\,  \Sigma_t^{\rm L} (p^2) +
                        \not{\! p}\, \omega_+\,   \Sigma_t^{\rm R}(p^2) 
                        + m_t \,\Sigma_t^{\rm S}(p^2)  
                        + m_t \gamma_5\, \Sigma_t^{\rm PS}(p^2)  .
\end{align}
%
  \item $m_{\tilde{q}_{3}}^{2}$ and $m_{\tilde{t}_{\text{R}}}^{2}$ are
    traded for $m_{\tilde{t}_{1}}^{2}$ and $m_{\tilde{t}_{2}}^{2}$, 
    which are then fixed by on-shell conditions for the top-squarks,
    \begin{align}
      \delta^{(1)}m_{\tilde{t}_{i}}^{2} &= \Realtilde{\, \Sigma_{\tilde{t}_{ii}}^{(1)}{\left(m_{\tilde{t}_{i}}^{2}\right)}}, 
      \quad  i=1,\,2 \, ,
   \end{align}
  involving the diagonal $\st_1$ and $\st_2$  self-energies 
  (diagrammatically visualized in Fig.~\ref{fig:RCTop}).
  These on-shell conditions determine the diagonal entries
  of the counterterm matrix
  \begin{align}
   \label{eq:stopcountermmatrix}     
     \mathbf{U}_{\tilde{t}}\, \delta\mathbf{M}_{\tilde{t}}\, \mathbf{U}_{\tilde{t}}^{\dagger} & =
     \begin{pmatrix} 
         \delta^{(1)}m_{\tilde{t}_{1}}^{2} & \delta^{(1)}m_{\tilde{t}_{1}\tilde{t}_{2}}^{2}  \\ 
         \delta^{(1)}m_{\tilde{t}_{1}\tilde{t}_{2}}^{2\,*} & \delta^{(1)}m_{\tilde{t}_{2}}^{2} 
    \end{pmatrix} .
    \end{align}
  \item 
   The mixing parameter $A_{t}$ is correlated with the $\st$-mass
    eigenvalues, $t_\beta$, and $\mu$, through Eq.~\eqref{eq:squarkdiag}. 
   Exploiting Eq.~\eqref{eq:stopcountermmatrix}
   and the unitarity of $\mathbf{U}_{\tilde{t}}$ yields the expression
    \begin{multline}
    \label{eq:Atrenormalization}
      \left(A_{t} - \frac{\mu^{*}}{t_{\beta}}\right)\delta^{(1)}m_{t} + m_{t} \left(\delta^{(1)}A_{t} - 
      \frac{\delta^{(1)}\mu^{*}}{t_{\beta}} + \frac{\mu^{*}\,\delta^{(1)}t_{\beta}}{t_{\beta}^{2}}\right) =\\
      \mathbf{U}_{\tilde{t}\,11}\mathbf{U}_{\tilde{t}\,12}^{*}\left(\delta^{(1)}m_{\tilde{t}_{1}}^{2} - \delta^{(1)}m_{\tilde{t}_{2}}^{2}\right)
      + \mathbf{U}_{\tilde{t}\,21}\mathbf{U}_{\tilde{t}\,12}^{*}\,\delta^{(1)}m_{\tilde{t}_{1}\tilde{t}_{2}}^{2}
      + \mathbf{U}_{\tilde{t}\,22}\mathbf{U}_{\tilde{t}\,11}^{*}\,\delta^{(1)}m_{\tilde{t}_{1}\tilde{t}_{2}}^{2\,*}\ .
    \end{multline}
    For the non-diagonal entry of Eq.~\eqref{eq:stopcountermmatrix},
    the renormalization condition
    \begin{align}
      \delta^{(1)}m_{\tilde{t}_{1}\tilde{t}_{2}}^{2} &= 
      \frac{1}{2}\Realtilde{\Sigma_{\tilde{t}_{12}}^{(1)}{\left(m_{\tilde{t}_{1}}^{2}\right)}
          + \Sigma_{\tilde{t}_{12}}^{(1)}{\left(m_{\tilde{t}_{2}}^{2}\right)}} 
    \end{align}
   is imposed, as in Ref.~\cite{Heinemeyer:2007aq}, which involves the 
   non-diagonal $\st_1$--$\st_2$ self-energy (Fig.~\ref{fig:RCTop}). 
   By means of Eq.~\eqref{eq:Atrenormalization} the counterterm $\delta^{(1)}A_t$ 
   is then determined. Actually this yields two conditions, 
   for $\lvert A_t \rvert $ and for the phase $\phi_{A_t}$ separately.
   The additionally required  
   mass counterterm $\delta^{(1)}\mu$ is obtained as described below 
   in section~\ref{sec:higgsinos}.
  \item 
As already mentioned, the relevant sbottom mass is not an independent parameter, 
and hence its counterterm is a derived quantity that can be obtained 
from Eq.~\eqref{eq:stopcountermmatrix},
\begin{align}
  \begin{split}
  \delta^{(1)}m_{\tilde{b}_{1}}^{2} \,\equiv\, \delta^{(1)}m_{\tilde{q}_{3}}^{2} &=
  \lvert\mathbf{U}_{\tilde{t}\,11}\rvert^{2}\,\delta^{(1)}m_{\tilde{t}_{1}}^{2} + \lvert\mathbf{U}_{\tilde{t}\,12}\rvert^{2}\,\delta^{(1)}m_{\tilde{t}_{2}}^{2}\\
  &\quad - \mathbf{U}_{\tilde{t}\,22}\mathbf{U}_{\tilde{t}\,12}^{*}\,\delta^{(1)}m_{\tilde{t}_{1}\tilde{t}_{2}}^{2} - \mathbf{U}_{\tilde{t}\,12}\mathbf{U}_{\tilde{t}\,22}^{*}\,\delta^{(1)}m_{\tilde{t}_{1}\tilde{t}_{2}}^{2\,*} 
  - 2\,m_{t}\,\delta^{(1)}m_{t}\ .
  \end{split}
\end{align}

\end{itemize}

\subsection{The chargino--neutralino sector \label{sec:higgsinos}}

For the calculation of the~$\mathcal{O}{\left(\alpha_{t}^{2}\right)}$ contributions 
to the Higgs-boson self-energies and tadpoles,
also the neutralino and chargino sectors have to be considered. 
Chargino/neutralino vertices and propagators enter
only at the two-loop level and thus do not need renormalization;
in the one-loop terms, however, the Higgsino-mass 
parameter~$\mu$ enters via the couplings of Higgs bosons to stops and the counterterm~$\delta^{(1)}\mu$ 
is required for the one-loop subrenormalization.
The mass matrices in the bino/wino/higgsino bases
are given by
\begin{alignat}{4}
  \mathbf{Y} &= 
  \begin{pmatrix}
   M_{1} & 0 & -M_{Z}\,s_{\rm w}\,c_{\beta} & M_{Z}\,s_{\rm w}\,s_{\beta}\\ 0 & M_{2} & M_{Z}\,c_{\rm w}\,c_{\beta} & M_{Z}\,c_{\rm w}\,s_{\beta}\\
   -M_{Z}\,s_{\rm w}\,c_{\beta} & M_{Z}\,c_{\rm w}\,c_{\beta} & 0 & -\mu\\ M_{Z}\,s_{\rm w}\,s_{\beta} & M_{Z}\,c_{\rm w}\,s_{\beta} & -\mu & 0 
   \end{pmatrix}, &\quad
  \mathbf{X} &= 
  \begin{pmatrix}
    M_{2} & \sqrt{2}\,M_{W}\,s_{\beta}\\
    \sqrt{2}\,M_{W}\,c_{\beta} & \mu
   \end{pmatrix} .
\end{alignat}
Diagonal matrices with real and positive entries are obtained with the
help of unitary matrices $ \mathbf{N}, \mathbf{U}, \mathbf{V}$
by the singular value decompositions
\begin{alignat}{4}
 \label{eq:ewdiag}
  \mathbf{N}^{*}\mathbf{Y}\mathbf{N}^{\dagger} &= 
  \mathrm{diag}{\left(m_{\tilde{\chi}^{0}_{1}},\, m_{\tilde{\chi}^{0}_{2}},\, m_{\tilde{\chi}^{0}_{3}},\, m_{\tilde{\chi}^{0}_{4}}\right)}\ , &\quad
  \mathbf{U}^{*}\mathbf{X}\mathbf{V}^{\dagger} &= \mathrm{diag}{\left(m_{\tilde{\chi}^{\pm}_{1}},\, m_{\tilde{\chi}^{\pm}_{2}}\right)}\ .
\end{alignat}
In the gaugeless limit the off-diagonal $(2\times 2)$-blocks of
$\mathbf{Y}$ and the off-diagonal entries of $\mathbf{X}$ vanish. 
For this special case the transformation matrices and diagonal entries in Eq.~\eqref{eq:ewdiag} simplify,
\begin{subequations}\begin{align}&\begin{aligned}
  \mathbf{N} &= 
  \begin{pmatrix}
    \begin{matrix}\E^{\frac{\I}{2} \phi_{M_{1}}} & 0\\ 0 & \E^{\frac{\I}{2} \phi_{M_{2}}}\end{matrix} & 
     \zero \\  \zero  & \frac{1}{\sqrt{2}}\E^{\frac{\I}{2}\phi_{\mu}}
        \begin{pmatrix} 1 & -1\\ \I & \I \end{pmatrix} 
  \end{pmatrix}, &
  \mathbf{U} &= \begin{pmatrix}\E^{\I\phi_{M_{2}}} & 0\\ 0 & \E^{\I\phi_{\mu}}\end{pmatrix}, &
 \mathbf{V} &= \unity \ ;\end{aligned}\\
 &\begin{aligned}
   m_{\tilde{\chi}^{0}_{1}} &= \lvert M_{1}\rvert\ , & 
   m_{\tilde{\chi}^{0}_{2}} &= \lvert M_{2}\rvert\ , & 
   m_{\tilde{\chi}^{0}_{3}} &= \lvert\mu\rvert\ , &
   m_{\tilde{\chi}^{0}_{4}} &= \lvert\mu\rvert\ , &
   m_{\tilde{\chi}^{\pm}_{1}} &= \lvert M_{2}\rvert\ , &
   m_{\tilde{\chi}^{\pm}_{2}} &= \lvert\mu\rvert \ ;
\end{aligned}\end{align}\end{subequations}
and only the Higgsinos $\tilde{\chi}^{0}_{3},\,\tilde{\chi}^{0}_{4},\, \tilde{\chi}^{\pm}_{2}$ 
remain in the $\mathcal{O}{\left(\alpha_{t}^{2}\right)}$ contributions.

\medskip
The Higgsino mass parameter $\mu$ is an independent input quantity and
has to be renormalized accordingly, $\mu \rightarrow \mu + \delta^{(1)}\mu$,
fixing the counterterm $\delta^{(1)}\mu$ by an independent renormalization
condition, which renders the one-loop subrenormalization complete. 
Together with the soft-breaking parameters $M_{1}$ and $M_{2}$,  $\mu$ can
be defined  in the neutralino/chargino sector by requiring on-shell
conditions for the two charginos and one neutralino.

\begin{figure}[tb]
  \centering
  \includegraphics[width=\textwidth]{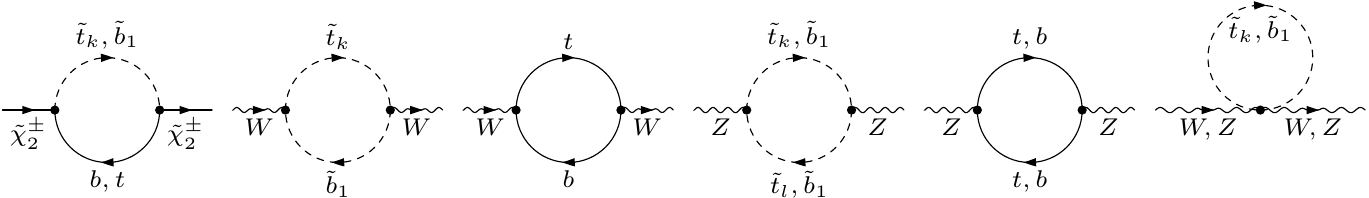}
  \caption{\label{fig:RCCha}
   Feynman diagrams for the counterterms 
  $\delta^{(1)}\mu$, $\left.\delta^{(1)}M_{W}^2\middle/M_W^2\right.$, and  $\left.\delta^{(1)}M_{Z}^2\middle/M_Z^2\right.$.}
\end{figure}

However, since only $\delta^{(1)}\mu$ is required here, it is sufficient to impose 
a renormalization condition for $\tilde{\chi}^{\pm}_{2}$ only;
the appropriate on-shell condition reads,
\begin{subequations}\label{eq:murenormalization}
\begin{align}
  \delta^{(1)}\mu &= \E^{\I \phi_{\mu}}\,\delta^{(1)}\lvert\mu\rvert\ ,
\end{align}
\vspace{-6ex}
\begin{align}
  \begin{split}
    \delta^{(1)}\lvert\mu\rvert &= \lvert\mu\rvert \left\{\tfrac{1}{2}\left[\Real{\Sigma^{(1),\,\text{L}}_{\tilde{\chi}^\pm_2}(\lvert\mu\rvert^{2}) + \Sigma^{(1),\,\text{R}}_{\tilde{\chi}^\pm_2}(\lvert\mu\rvert^{2})}\right]
    + \Real{\Sigma^{(1),\,\text{S}}_{\tilde{\chi}^\pm_2}(\lvert\mu\rvert^{2})}\right\} ,
  \end{split}
\end{align}
\end{subequations}
where the Lorentz decomposition of the self-energy 
for the Higgsino-like chargino $\tilde{\chi}^{\pm}_{2}$ 
(see Fig.~\ref{fig:RCCha})
has been applied, in analogy to Eq.~\eqref{eq:Lorentz}.

Another option is the~$\overline{\text{DR}}$-renormalization of $\mu$,
which defines the counterterm $\delta^{(1)}\mu$ in the~$\overline{\text{DR}}$-scheme,
 \IE~by the divergent part of the expression in Eq.~\eqref{eq:murenormalization}.

\section{Numerical results for masses and mixings}
\label{sec:numeric}

In this section we present numerical analyses for the masses of the neutral Higgs
bosons derived from Eq.~\eqref{eq:higgspoles} in various SUSY-parameter scenarios.
The complete one-loop results with the full dependence on the external squared-momentum 
$p^2$,  and the two-loop 
$\mathcal{O}{\left(\alpha_{t}\alpha_{s}\right)}$ terms are taken
from {\tt FeynHiggs}, while the
$\mathcal{O}{\left(\alpha_{t}^{2}\right)}$ terms are computed by means
of the corresponding two-loop self-energies as specified in the previous
sections. In our strategy, the new $\mathcal{O}{\left(\alpha_{t}^{2}\right)}$
self-energies are combined with the complementary self-energies
according to Eq.~\eqref{eq:masscorr} within {\tt FeynHiggs}, and the
masses are then derived via Eq.~\eqref{eq:higgspoles}, ordered as
$m_{h_1} < m_{h_2} < m_{h_3}$.

\smallskip
The Standard Model~(SM) parameters are collected in Tab.~\ref{tab:parameters}, as
well as those~MSSM~parameters that are kept for the analyses which are
performed in this section. The residual input parameters of the~MSSM
are shown in the figures or their captions.  The parameters~$\mu,\,
t_{\beta}$, and the Higgs field-renormalization constants are defined
in the~$\overline{\text{DR}}$~scheme at the scale~$m_{t}$.

\begin{table}[h]
  \centering
  \caption[Default input parameters]{Default input values of the MSSM and SM parameters.}
  \label{tab:parameters}
  \begin{tabular}{r@{$\;=\;$}lr@{$\;=\;$}l}
    \toprule
    \multicolumn{2}{c}{MSSM input} & \multicolumn{2}{c}{SM input}\\
    \midrule
    $M_2$ & $200$~GeV,  & $m_t$ & $173.2$~GeV,\\
    $M_1$ & $\left.\left(5s_{\rm w}^2\right)\middle/\left(3c_{\rm w}^2\right) M_2\right.$, & $m_b$ & $4.2$~GeV,\\
    $m_{\tilde{l}_1} = m_{\tilde{e}_{\rm R}}$ & $2000$~GeV, & $m_{\tau}$ & $1.77703$~GeV,\\
    $m_{\tilde{q}_1} = m_{\tilde{u}_{\rm R}} = m_{\tilde{d}_{\rm R}}$ & $2000$~GeV, & $M_W$ & $80.385$~GeV,\\
    $A_u = A_d = A_e$ & $0$~GeV, & $M_Z$ & $91.1876$~GeV,\\
    $m_{\tilde{l}_2} = m_{\tilde{\mu}_{\rm R}}$ & $2000$~GeV, & $G_{\text{F}}$ & $1.16639\cdot 10^{-5}$,\\
    $m_{\tilde{q}_2} = m_{\tilde{c}_{\rm R}} = m_{\tilde{s}_{\rm R}}$ & $2000$~GeV, & $\alpha_s(M_Z)$ & $0.118$,\\
    $A_c = A_s = A_\mu$ & $0$~GeV, & $\left.1\middle/\alpha\right.$ & $128.945$.\\
    \bottomrule
  \end{tabular}
\end{table}

\subsection*{Higgs-boson masses in the real MSSM}

In the case of the MSSM with real parameters, 
conventionally the mass $m_A$ of the $CP$-odd $A$ boson 
is chosen as an input parameter, and the masses of the two
$CP$-even neutral scalar bosons are predicted in terms of $m_A$ and 
the other SUSY parameters. In this special case, a
comparison of our diagrammatic result with those of the 
previously known $\mathcal{O}{\left(\alpha_{t}^{2}\right)}$
contributions~\cite{Brignole:2001jy}  obtained by
the effective-potential method is possible. 
In practice, this comparison is made by means of the default version
of {\tt FeynHiggs} which incorporates the $\mathcal{O}{\left(\alpha_{t}^{2}\right)}$ 
terms from~\cite{Brignole:2001jy}.
The beautiful agreement between the two independent 
calculations has been shown recently
for the mass of the lightest Higgs boson 
in Ref.~\cite{Hollik:2014wea};
similar good agreement has been 
found also for the heavier Higgs-boson mass.
The impact of the $\mathcal{O}{\left(\alpha_{t}^{2}\right)}$ terms 
in particular on the mass of the lightest Higgs boson is substantial,
yielding a mass shift of $\approx 5$~GeV, and demonstrates the importance
of the two-loop Yukawa contributions for a reliable prediction of the
Higgs-boson masses. For complex parameters, additional mass shifts of
several GeV can occur from the complex phases.

\subsection*{Higgs-boson masses in the complex MSSM}

\begin{figure}[tb]
  \centering
  \includegraphics[width=.97\textwidth]{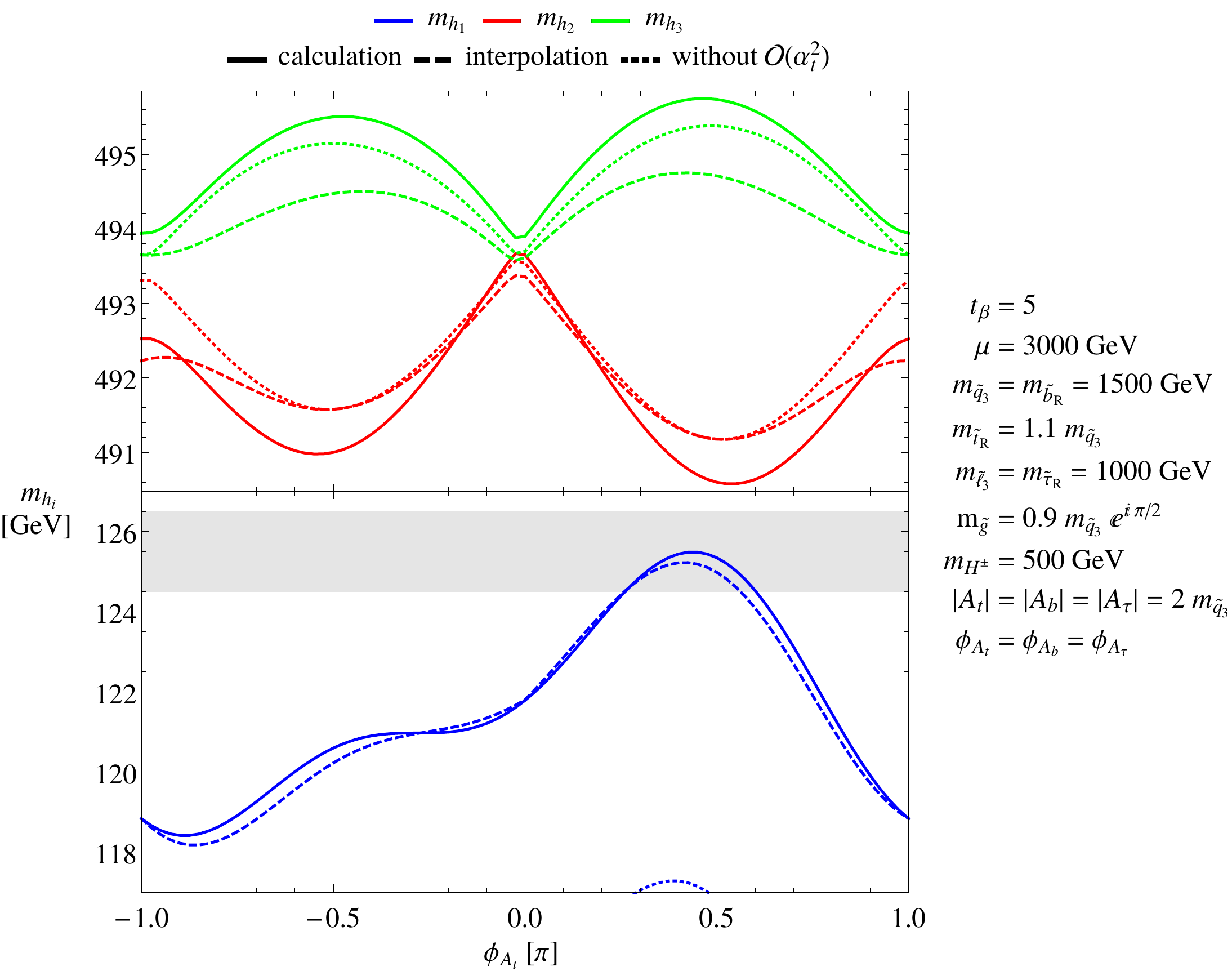}
  \caption{\label{fig:complexPhiAt}%
The result for the masses of the neutral Higgs bosons from the
diagrammatic calculation (full), in comparison with the
approximate result from interpolation between the phases
\mbox{$\phi_{A_{t}} = 0,\, \pm\pi$} (dashed). For reference the result
without the $\mathcal{O}{\left(\alpha_{t}^{2}\right)}$ contributions
is depicted (dotted). The gray area depicts the mass
range between~$124.5$~GeV and~$126.5$~GeV.}
\end{figure}

In the current public version of {\tt FeynHiggs} for complex parameters, the
dependence of the $\mathcal{O}{\left(\alpha_{t}^{2}\right)}$ terms on
the phases of $\phi_{A_{t}}$ and $\phi_{\mu}$ is approximated by an
interpolation between the real results for the phases~$0$
and~$\pm\pi$~\cite{Hahn:2009zz,Hahn:2007fq}. In
Ref.~\cite{Hollik:2014wea} a comparison with the full diagrammatic
calculation for the mass of the lightest Higgs boson was presented
showing notable deviations. Fig.~\ref{fig:complexPhiAt} contains 
the comparison for all three mass eigenvalues.
The dependence of the heavier~(upper plot) and the lightest~(lower plot) 
neutral Higgs-boson masses on the
phase~$\phi_{A_t}$ is illustrated. Rather large
deviations from the previous result of {\tt FeynHiggs} are found owing
to the~$\mathcal{O}{\left(\alpha_{t}^{2}\right)}$ contributions to the
charged-Higgs self-energy  which were not known before but which
are required for consistent renormalization of the self-energies of
the neutral Higgs bosons in the complex~MSSM. So far, {\tt FeynHiggs}
utilized the known~$\mathcal{O}{\left(\alpha_{t}^{2}\right)}$
contribution to the self-energy of the $A$-boson for renormalization;
the visible deviations at the real edges in
Fig.~\ref{fig:complexPhiAt} originate from the difference of these
renormalization schemes,
\IE~the~$\mathcal{O}{\left(\alpha_{t}^{2}\right)}$ terms
in the difference~\mbox{$\Sigma_A - \Sigma_{H^\pm}$}. 
A similar effect has also been found previously in
the~$\mathcal{O}{\left(\alpha_{t}\alpha_{s}\right)}$ corrections to the 
$m_A$--$m_{H^\pm}$ mass correlation in Ref.~\cite{Frank:2013hba}.

\subsection*{\textit{CP}-mixing}

\begin{figure}[tb]
  \begin{subfigure}{.705\textwidth}
  \centering
  \vspace{1pt}
  \includegraphics[width=\textwidth]{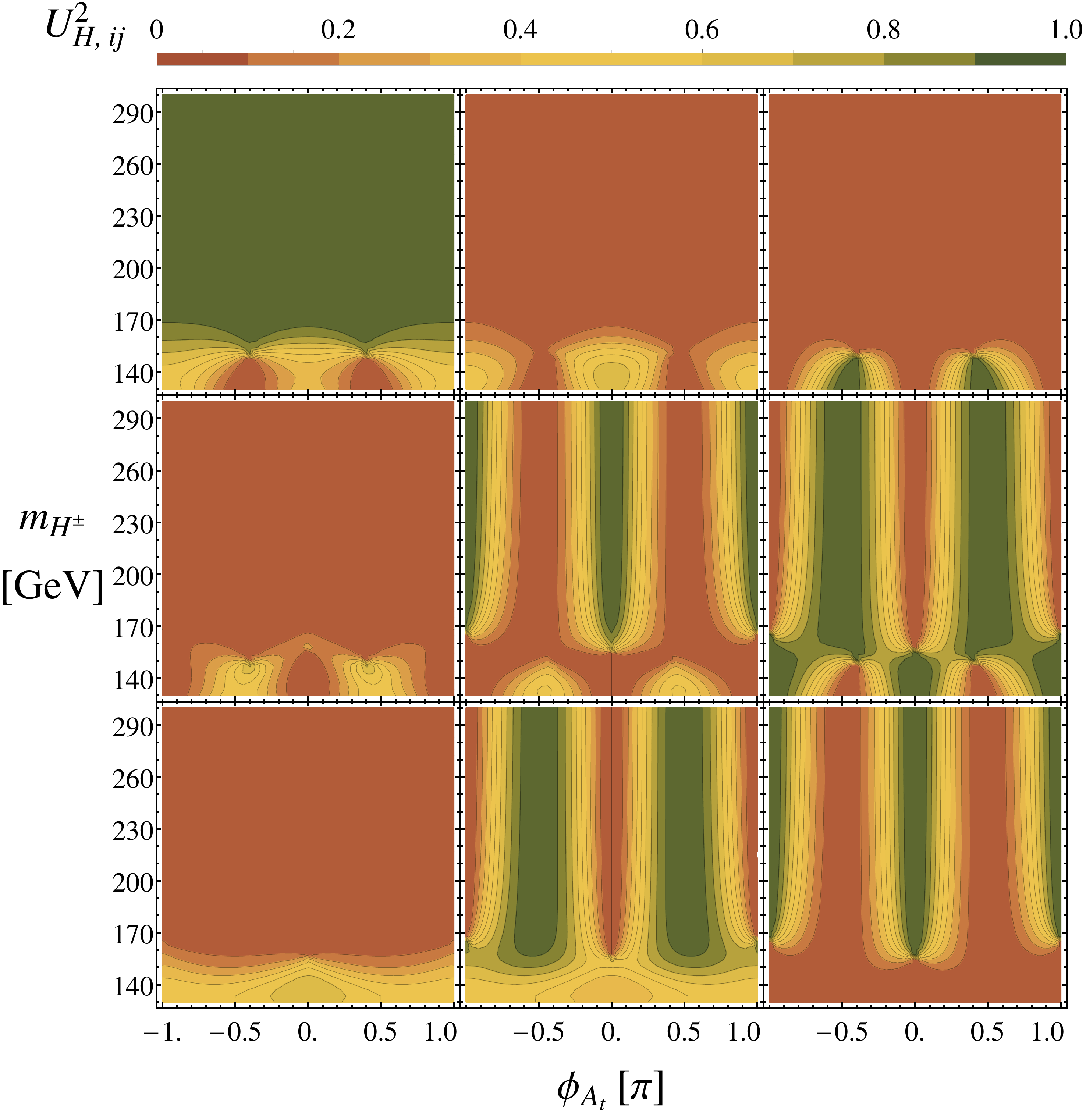}
  \end{subfigure}
  \begin{subfigure}{.283\textwidth}
  \centering
  \includegraphics[width=\textwidth]{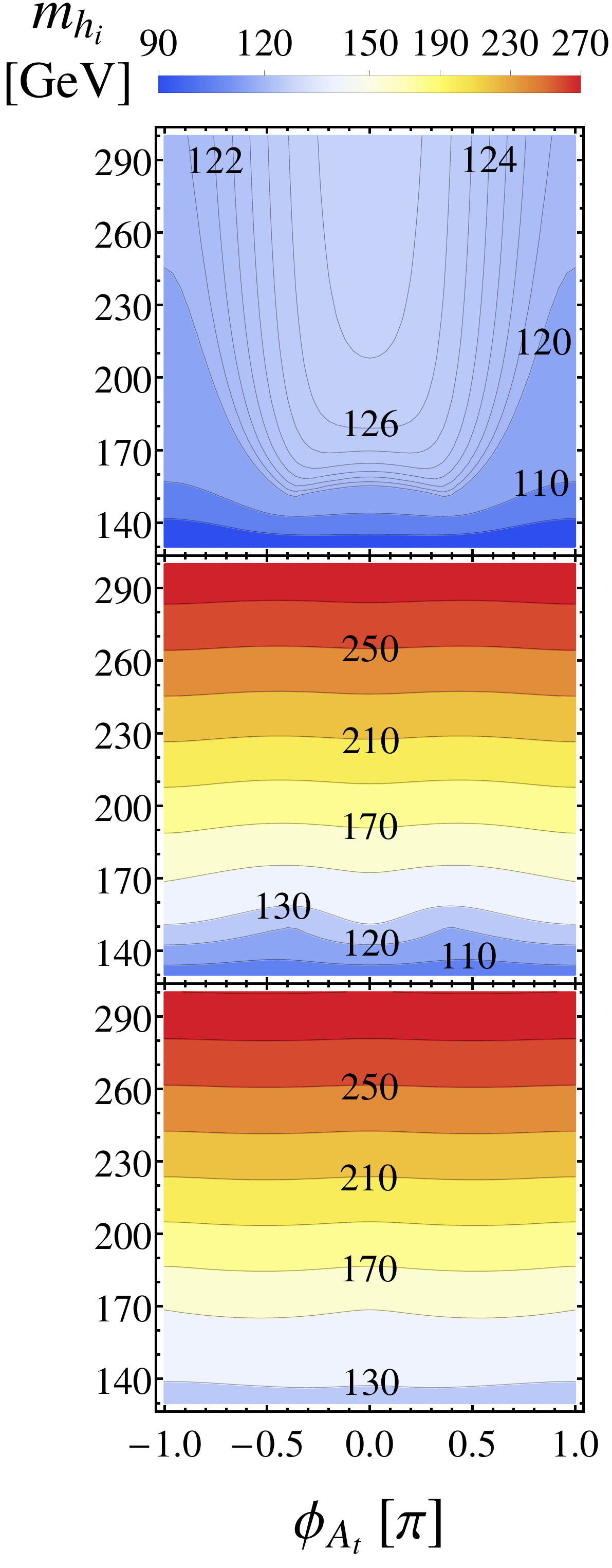}
  \end{subfigure}
  \caption[Dependence on~$m_{H^{\pm}}$ of the masses and mixing-matrix elements squared for complex parameters]{\label{fig:U}%
    The mixing-matrix elements squared~\mbox{$U_{H,\,ij}^2,\,i,j\in\{1,\,2,\,3\}$}~(left) and masses~$m_{h_i}$~(right), where~$i$ is the index of the row and~$j$ is the index of the column,
    are illustrated with the phase~$\phi_{A_t}$ and the input mass~$m_{H^{\pm}}$ being varied.
    The color coding is explained in the labels at the top of the figures; for convenience some contours of the mass plots are signed with their corresponding mass values. The input parameters are fixed at
    \mbox{$\mu = 2500$~GeV}, \mbox{$t_{\beta} = 10$}, \mbox{$m_{\tilde{\ell}_3} = m_{\tilde{\tau}_{\rm R}} = 1500$~GeV}, \mbox{$m_{\tilde{q}_3} = m_{\tilde{t}_{\text{R}}} = m_{\tilde{b}_{\text{R}}} = 1500$~GeV},
    \mbox{$\lvert A_t \rvert = \lvert A_b \rvert = \lvert A_{\tau} \rvert = 2\,m_{\tilde{q}_3}$}, \mbox{$m_{\tilde{g}} = 2000$~GeV}.}
\end{figure}

In the complex MSSM all three neutral Higgs bosons mix at higher
orders according to the off-diagonal entries of the mass matrix in
Eq.~\eqref{eq:higgspoles}, leading to violation of $CP$-symmetry.
Since the self-energies contributing to the mass matrix
at higher orders are momentum-dependent,  
this $CP$-mixing depends on~$p^2$ 
and hence it is not possible to describe $CP$-mixing
in terms of a constant mixing matrix.  
A convenient approximation for the discussion of $CP$-mixing is
given by setting~\mbox{$p^2 = 0$} in the renormalized self-energies of the
Higgs bosons at all orders. In this case, Eq.~\eqref{eq:higgspoles}
simplifies to the eigenvalue equation for the
matrix~$\mathbf{M}_{hHA}^{(2)}{\left(0\right)}$ in
Eq.~\eqref{eq:masscorr}. The real mixing matrix which
diagonalizes~$\mathbf{M}_{hHA}^{(2)}{\left(0\right)}$ is denoted
by~$\mathbf{U}_{H}$ in the following. It allows to define an
approximate mass-eigenstate basis according to
\begin{alignat}{4}\label{eq:CPviolation}
  \begin{pmatrix}h_{1}\\h_{2}\\h_{3}\end{pmatrix} &= \mathbf{U}_{H}\begin{pmatrix}h\\H\\A\end{pmatrix}, &\quad
    \mathbf{U}_{H} &= \begin{pmatrix}U_{H,\,11} & U_{H,\,12} & U_{H,\,13}\\U_{H,\,21} & U_{H,\,22} & U_{H,\,23}\\U_{H,\,31} & U_{H,\,32} & U_{H,\,33}\end{pmatrix}.
\end{alignat}
In general the~\mbox{$h_i,\,i\in\{1,\,2,\,3\}$}, are no
longer~$CP$-eigenstates since they are composed of the
$CP$-even ~$h$ and~$H$ and the $CP$-odd~$A$. 
The elements~$U_{H,\,i3}$ of~$\mathbf{U}_{H}$ in
Eq.~\eqref{eq:CPviolation} 
squared tell the amount of the~$A$~boson inside of
\begin{align}
  h_{i} = U_{H,\,i1}\, h + U_{H,\,i2}\, H + U_{H,\,i3}\, A
\end{align}
and thus the $CP$-odd admixture in~$h_i$.

The dependence of the mixing-matrix elements squared~$U_{H,\,ij}^2$,
in the approximation of Eq.~\eqref{eq:CPviolation}, 
on the charged Higgs-boson mass~$m_{H^{\pm}}$ 
and the basically unconstrained complex
phase~$\phi_{A_t}$~\cite{Falk:1996ni,Ibrahim:1997nc,Ibrahim:1997gj,Accomando:1999uj,Bartl:1999bc,Barger:2001nu}
is shown in the left part of  Fig.~\ref{fig:U}. 
Therein, the tiles are ordered according to the matrix array of~$U_{H,\,ij}^2$, 
with $i$ and~$j$ indicating row and column, respectively. 
The right part of Fig.~\ref{fig:U} displays 
the masses~$m_{h_i}$ in
ascending order from the first to the third row.

Whenever two masses~$m_{h_i}^2$ and~$m_{h_j}^2$ are close to each
other, the entries~$U_{H,\,ij}^2$ rapidly change from zero to unity,
\IE~$h_i$ and~$h_j$ interchange their meaning. 
For a large value of the charged Higgs-boson mass, the lightest Higgs~$h_1$ is
basically equal to~$h$. In contrast, the heavier Higgs bosons can be
composed of~$H$ and~$A$ in all possible variations, depending on the
phase, thus yielding the possibility of very large $CP$-mixing. At the
nodal points at the real values of~$A_t$ for~\mbox{$\phi_{A_t} =
  \pm\pi$} and~\mbox{$m_{H^{\pm}} \approx 165$~GeV} as well as
for~\mbox{$\phi_{A_t} = 0$} and~\mbox{$m_{H^{\pm}} \approx 155$~GeV}
the masses of~$h_2$ and~$h_3$ are equal; slightly above or below the
nodes,~$h_2$ and~$h_3$ can  be identified as~$H$
or~$A$. Between~\mbox{$m_{H^{\pm}}\approx 155$~GeV}
and~\mbox{$m_{H^{\pm}}\approx 150$~GeV} an extreme situation is
observed, where each~$h_i$ is almost equal to~$h,\,H$ or~$A$ for any
complex phase. However the situation changes for lower input values
of~$m_{H^{\pm}}$ where a large admixture of~$A$ to the lightest Higgs
boson~$h_1$ is predicted, depending on the complex
phase~$\phi_{A_t}$. In the same parameter range the heaviest
state~$h_3$ is nearly $CP$-even for any phase. In this scenario the
Higgs-like particle discovered at the LHC is interpreted as
the heaviest neutral Higgs boson of the~MSSM (so-called
low-$m_H$~scenario
). The strongest gradients for the mixing of~$h$
and~$A$ to~$h_1$ and~$h_2$ appear at~\mbox{$\phi_{A_t} \approx \pm
  0.4$} and \mbox{$m_{H^{\pm}} \approx 150$~GeV}; for the present
choice of parameters the masses~$m_{h_1}$ and~$m_{h_2}$ are equal 
at this point.

\begin{figure}[tb]
  \centering
  \includegraphics[width=.49\textwidth]{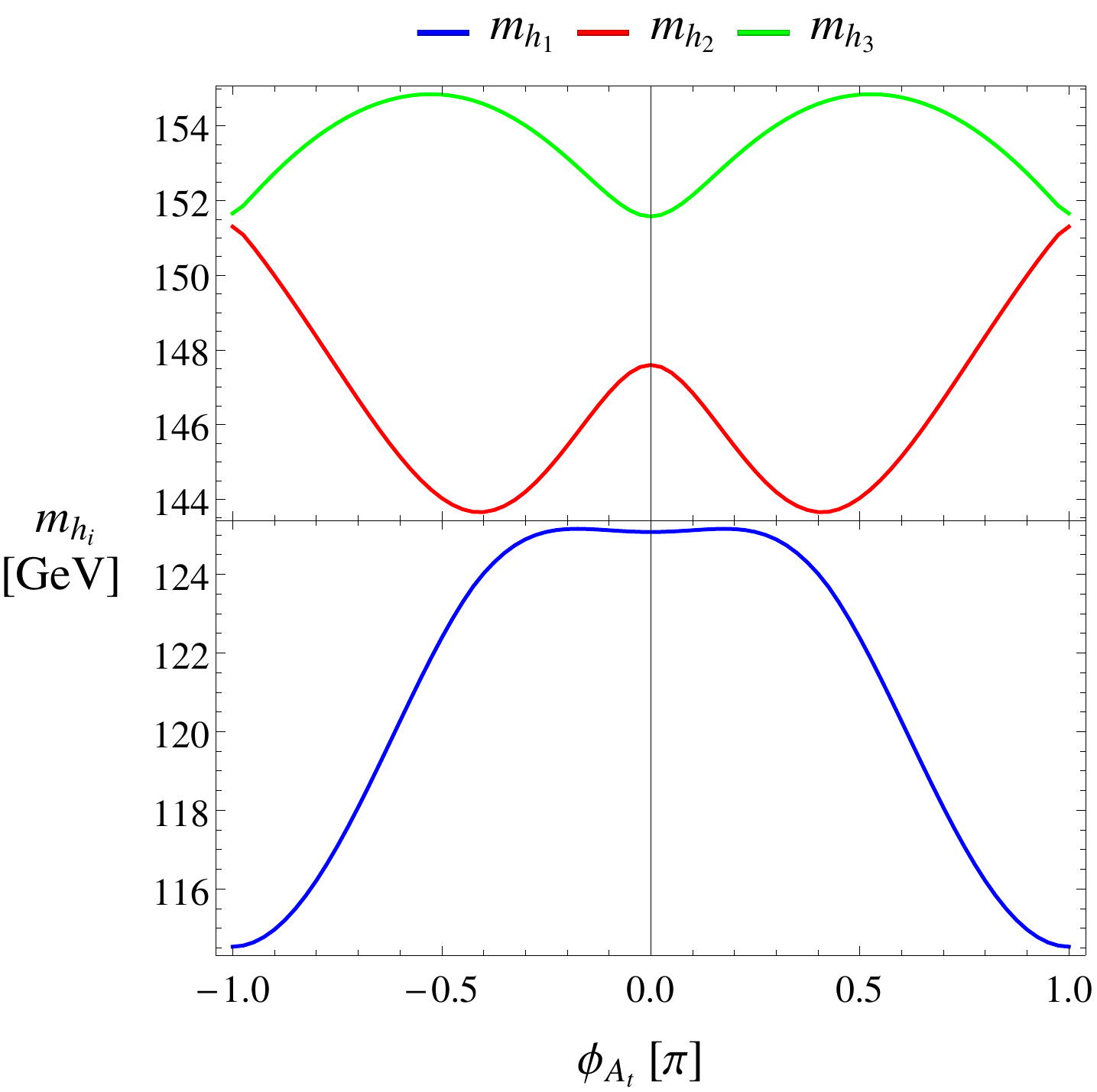}\hfill\includegraphics[width=.49\textwidth]{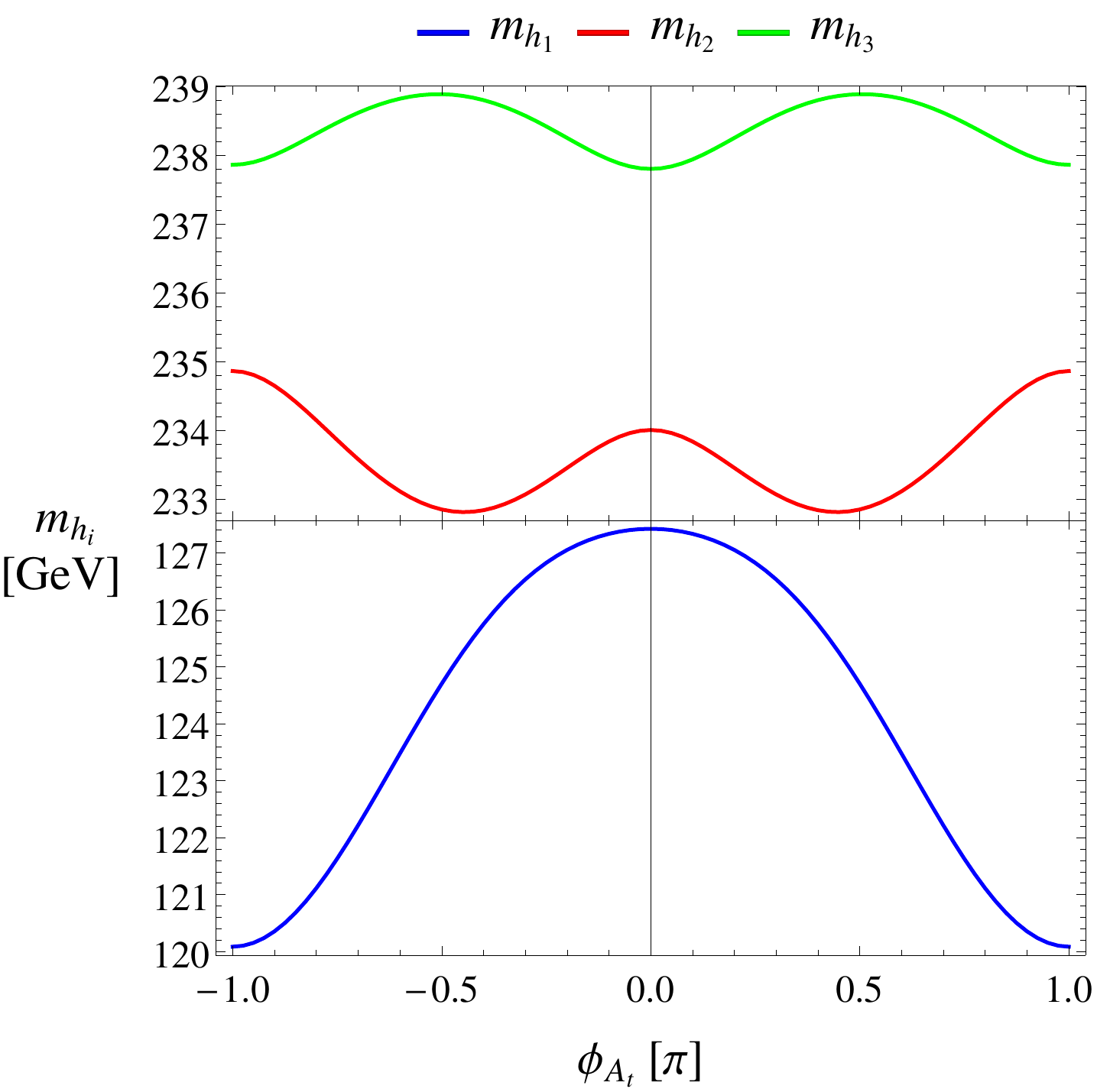}
  \caption{\label{fig:UMHp}%
    The dependence of the neutral Higgs-boson masses on the
    phase~$\phi_{A_t}$ is depicted for an input value of the charged
    Higgs-boson mass of~\mbox{$m_{H^\pm} = 170$~GeV}~(left)
    and~\mbox{$m_{H^\pm} = 250$~GeV}~(right). The other input
    parameters are the same as in Fig.~\ref{fig:U}. }
\end{figure}

For the same parameter set,
in Fig.~\ref{fig:UMHp} the Higgs-boson masses are depicted for the two
special cases of~\mbox{$m_{H^\pm} = 170$~GeV}~(left)
and~\mbox{$m_{H^\pm} = 250$~GeV}~(right). For larger~$m_{H^\pm}$ the
mass of the lightest Higgs boson can be lowered by an increasing
phase~$\phi_{A_t}$ to be in the mass range of the experimentally
discovered Higgs-like particle, remaining essentially $CP$-even,
whereas the heavier Higgs bosons get a larger mass splitting  
and a substantial~$CP$-mixing.
For the lower~$m_{H^\pm}$ value, the phase cannot be too large for the
right mass~$m_{h_1}$, and mass splitting
of the two heavy Higgs bosons shows a stronger variation.

\subsection*{Complex-valued $\boldsymbol{\mu}$}

\begin{figure}[tb]
  \centering
  \includegraphics[width=\textwidth]{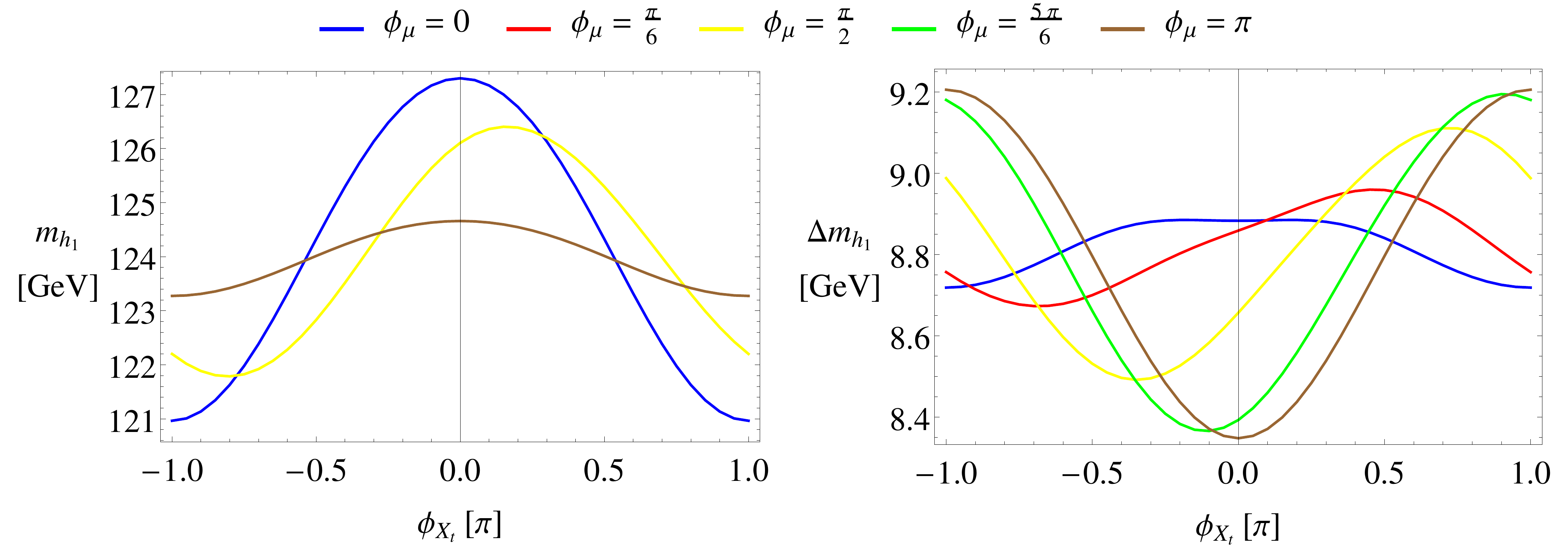}
  \caption{\label{fig:Mh1complexshift}%
The dependence of the lightest neutral Higgs-boson mass on the
phases~$\phi_{X_t}$ and~$\phi_\mu$ is depicted. Left: The value
of~$m_{h_1}$ including all available contributions, with the phase dependence
arising from one-loop, $\mathcal{O}{\left(\alpha_t \alpha_s\right)}$ and
$\mathcal{O}{\left(\alpha_t^2\right)}$ terms.
Right: The contribution~$\Delta m_{h_1}$ to~$m_{h_1}$ owing exclusively to
the~$\mathcal{O}{\left(\alpha_t^2\right)}$ terms, 
for different phases. The input parameters are~\mbox{$m_{H^\pm} = 200$~GeV}, \mbox{$\lvert\mu\rvert = 2500$~GeV}, \mbox{$t_{\beta} = 10$}, \mbox{$m_{\tilde{\ell}_3} = m_{\tilde{\tau}_{\rm R}} = 1000$~GeV}, \mbox{$m_{\tilde{q}_3} = m_{\tilde{t}_{\text{R}}} = m_{\tilde{b}_{\text{R}}} = 1500$~GeV}, \mbox{$\lvert X_t \rvert = 2\,m_{\tilde{q}_3}$}, \mbox{$A_b = A_{\tau} = 0$}, \mbox{$m_{\tilde{g}} = 2000$~GeV}.}
\end{figure}

Also the  coefficient $\mu$ of the bilinear term of the superpotential is in general a complex
quantity. The phase of $\mu$ is severely constrained by the experimental limits 
on the electric dipole moments of electron and neutron. These bounds can, however,
be circumvented in principle by a specific fine-tuning of the phases
of $\mu$ and of the non-universal SUSY 
parameters ~\cite{Masiero:1997bn,Brhlik:1998zn,Brhlik:1999ub,Ibrahim:1997gj,Bartl:1999bc},   
leaving room also for a non-vanishing phase $\phi_\mu$, and thus we
want to illustrate potential effects of $\phi_\mu \neq 0$ in terms of an example.
In Fig.~\ref{fig:Mh1complexshift} we display the  
influence of the phases~$\phi_\mu$ and $\phi_{X_t}$ 
(with~\mbox{$X_t = A_t^* - \left.\mu\middle/t_\beta\right.$}) 
on the mass of the lightest Higgs boson.
In the depicted scenario the variation of $m_{h_1}$ with $\phi_\mu$
is of the order of~$0.5$~GeV. 
Changing of the sign of~$\phi_\mu$ 
mirrors the graphs at the axis~$\phi_{X_t} = 0$.

As one can see in the expressions in App.~\ref{app:fullresults}, the off-diagonal
self-energies~$\Sigma_{hA}^{(2)}$ and~$\Sigma_{HA}^{(2)}$ are
proportional to~$\Imag{X_t\mu^*}$; thus no $CP$-mixing by the
$\mathcal{O}{\left(\alpha_t^2\right)}$~terms occurs
for~\mbox{$\phi_{X_t} - \phi_\mu = 0,\,\pm\pi$}. Nevertheless, large
mass shifts do occur.

\section{Conclusions}

We have presented the full results for the leading two-loop Yukawa
contributions of $\mathcal{O}{\left(\alpha_{t}^{2}\right)}$ from the
top--stop sector in the calculation of the Higgs-boson masses of the
MSSM with complex parameters. They generalize the previously known
result for the real MSSM to the case of complex phases entering at the
two-loop level.
The combination of the new terms with the hitherto available full one-loop
result and leading two-loop terms of
$\mathcal{O}{\left(\alpha_t\alpha_s\right)}$ yields an improved
prediction for the Higgs-boson mass spectrum also for complex
parameters that is equivalent in accuracy to that of the real~MSSM.

In the complex MSSM, the masses of the three neutral Higgs
are derived quantities whereas the mass of the charged Higgs boson
is chosen as an input parameter.  
The mass shifts that originate from the
$\mathcal{O}{\left(\alpha_{t}^{2}\right)}$ terms are significant, and
hence an adequate treatment also for complex parameters is a necessity.
Besides the mass shift of about $5$~GeV in the
real~MSSM, additional shifts of the same size 
can be induced by complex parameters.

Large $CP$-mixing among the heavy Higgs bosons is found
for~\mbox{$m_{H^\pm}\gtrsim 160$~GeV}. In this case the lightest Higgs boson is basically
$CP$-even and can be identified with the Higgs signal observed in the
LHC experiments ATLAS and CMS. 
At lower values of the charged Higgs-boson mass, 
the lighter Higgs bosons can be strongly $CP$-mixed, with low masses,
and with the heaviest
Higgs boson basically $CP$-even. In this case, the observed Higgs
particle can be interpreted as the heaviest neutral Higgs boson of the MSSM.
Such a scenario, however, might be ruled out by the experimental exclusion
of a light charged Higgs particle.

Our new results will become part of the code {\tt FeynHiggs}, 
where so far the complex phases are treated in an
approximate way by interpolating between the real results for phases
$0$ and $\pm \pi$. 
At the formal side, we have given the complete counterterm structure 
at the two-loop level for the renormalization of the self-energies
in the Higgs sector of the complex MSSM, which can be used  
for further two-loop calculations going beyond the Yukawa approximation.


\section*{Acknowledgments}
\sloppy{We thank Stefano Di Vita, Thomas Hahn, Sven Heinemeyer, Heidi Rzehak, Pietro Slavich, \mbox{Dominik Stöckinger},
Alexander Voigt, and Georg Weiglein for helpful discussions.}

\newpage

\begin{appendix}

\section{Two-loop mass counterterms\label{app:genuine2L}}

The genuine two-loop mass counterterms introduced in Eq.~\eqref{eq:dm2Lneutral} and Eq.~\eqref{eq:dm2Lcharged} are listed in the following. Thereby~$\delta^{(1)}e,\, \delta^{(1)}M_{W}$ and~$\delta^{(1)}s_{\text{w}}$ always appear in the combination~\mbox{$\delta^{(1)}Z_{\text{w}} = \left.\delta^{(1)}e\middle/e\right. - \left.\delta^{(1)}M_{W}\middle/M_{W}\right. - \left.\delta^{(1)}s_{\text{w}}\middle/s_{\text{w}}\right.$}:
{\allowdisplaybreaks
\begin{subequations}
\begin{align}
  \begin{split}
    \delta^{(2)}m_{h}^{2} &= c_{\alpha-\beta}^{2}\,\delta^{(2)}m_A^{2} + s_{\alpha+\beta}^{2}\,\delta^{(2)}m_Z^{2} + c_{\beta}^{2}\,\delta^{(2)}t_{\beta} \left(s_{2(\alpha-\beta)}\,m_A^{2} + s_{2(\alpha+\beta)}\,m_Z^{2}\right)\\
                           &\quad + c_{\beta}^{2}\,\delta^{(1)}t_{\beta} \left(s_{2(\alpha-\beta)}\,\delta^{(1)}m_A^{2} + s_{2(\alpha+\beta)}\,\delta^{(1)}m_Z^{2}\right)\\
                           &\quad + \frac{1}{2}\,c_{\beta}^{3}\left(\delta^{(1)}t_{\beta}\right)^{2} \left[s_{\alpha-\beta} \left(3\,s_{\alpha - 2\beta} - s_{\alpha}\right) m_A^2 + 2\,c_{2\alpha+3\beta}\,m_Z^2\right]\\
                           &\quad +
    \begin{aligned}[t]
      \frac{e\,s_{\alpha-\beta}}{2\,M_{W}\,s_{\text{w}}} \Biggl[
        &\left(1 + c_{\alpha-\beta}^{2}\right)\left(\delta^{(2)}T_{h} + \delta^{(1)}T_{h}\,\delta^{(1)}Z_{\text{w}}\right)\\
        &+ s_{\alpha-\beta}\,c_{\alpha-\beta}\left(\delta^{(2)}T_{H} + \delta^{(1)}T_{H}\,\delta^{(1)}Z_{\text{w}}\right)\\
        &+ s_{\alpha-\beta}\,c_{\beta}^{2}\,\delta^{(1)}t_{\beta}\left(c_{\alpha-\beta}\,\delta^{(1)}T_{h} + s_{\alpha-\beta}\,\delta^{(1)}T_{H}\right)\Biggr]\ ,
    \end{aligned}
  \end{split}\\
  \begin{split}
    \delta^{(2)}m_{H}^{2} &= s_{\alpha-\beta}^{2}\,\delta^{(2)}m_A^{2} + c_{\alpha+\beta}^{2}\,\delta^{(2)}m_Z^{2} - c_{\beta}^{2}\,\delta^{(2)}t_{\beta} \left(s_{2(\alpha-\beta)}\,m_A^{2} + s_{2(\alpha+\beta)}\,m_Z^{2}\right)\\
                           &\quad - c_{\beta}^{2}\,\delta^{(1)}t_{\beta} \left(s_{2(\alpha-\beta)}\,\delta^{(1)}m_A^{2} + s_{2(\alpha+\beta)}\,\delta^{(1)}m_Z^{2}\right)\\
                           &\quad + \frac{1}{2}\,c_{\beta}^{3} \left(\delta^{(1)}t_{\beta}\right)^{2} \left[c_{\alpha-\beta} \left(3\,c_{\alpha - 2\beta} - c_{\alpha}\right) m_A^2 - 2\,c_{2\alpha+3\beta}\,m_Z^2\right]\\
                           &\quad -
    \begin{aligned}[t]
      \frac{e\,c_{\alpha-\beta}}{2\,M_{W}\,s_{\text{w}}} \Biggl[
        &\left(1 + s_{\alpha-\beta}^{2}\right)\left(\delta^{(2)}T_{H} + \delta^{(1)}T_{H}\,\delta^{(1)}Z_{\text{w}}\right)\\
        &+ c_{\alpha-\beta}\,s_{\alpha-\beta}\left(\delta^{(2)}T_{h} + \delta^{(1)}T_{h}\,\delta^{(1)}Z_{\text{w}}\right)\\
        &- c_{\alpha-\beta}\,c_{\beta}^{2}\,\delta^{(1)}t_{\beta}\left(c_{\alpha-\beta}\,\delta^{(1)}T_{h} + s_{\alpha-\beta}\,\delta^{(1)}T_{H}\right)\Biggr]\ ,
    \end{aligned}
  \end{split}\\
  \begin{split}
    \delta^{(2)}m_{G}^{2} &= c_{\beta}^{4}\,m_A^2 \left(\delta^{(1)}t_{\beta}\right)^{2}\\
                           &\quad +
    \begin{aligned}[t]
      \frac{e}{2\,M_{W}\,s_{\text{w}}} \Biggl[
        & s_{\alpha-\beta}\left(\delta^{(2)}T_{h} + \delta^{(1)}T_{h}\,\delta^{(1)}Z_{\text{w}}\right)\\
        &- c_{\alpha-\beta}\left(\delta^{(2)}T_{H} + \delta^{(1)}T_{H}\,\delta^{(1)}Z_{\text{w}}\right)\\
        &+ c_{\beta}^{2}\,\delta^{(1)}t_{\beta}\left(c_{\alpha-\beta}\,\delta^{(1)}T_{h} + s_{\alpha-\beta}\,\delta^{(1)}T_{H}\right)\Biggr]\ ,
    \end{aligned}
  \end{split}\\
  \begin{split}
    \delta^{(2)}m_{hH}^{2} &= c_{\alpha-\beta}\,s_{\alpha-\beta}\,\delta^{(2)}m_A^{2} - c_{\beta}^{2}\,\delta^{(2)}t_{\beta} \left(c_{2(\alpha-\beta)}\,m_A^{2} + c_{2(\alpha+\beta)}\,m_Z^{2}\right)\\
                           &\quad - c_{\alpha+\beta}\,s_{\alpha+\beta}\,\delta^{(2)}m_Z^{2} - c_{\beta}^{2}\,\delta^{(1)}t_{\beta} \left(c_{2(\alpha-\beta)}\,\delta^{(1)}m_A^{2} + c_{2(\alpha+\beta)}\,\delta^{(1)}m_Z^{2}\right)\\
                           &\quad + \frac{1}{2}\,c_{\beta}^{3} \left(\delta^{(1)}t_{\beta}\right)^{2} \left[\left(-3\,c_{\beta}\,s_{2(\alpha - \beta)} + 2\,s_{2\alpha-\beta}\right) m_A^2 + 2\,s_{2\alpha+3\beta}\,m_Z^2\right]\\
                           &\quad +
    \begin{aligned}[t]
      \frac{e}{2\,M_{W}\,s_{\text{w}}} \Biggl[
        &-c_{\alpha-\beta}^{3}\left(\delta^{(2)}T_{h} + \delta^{(1)}T_{h}\,\delta^{(1)}Z_{\text{w}}\right)\\
        &+s_{\alpha-\beta}^{3}\left(\delta^{(2)}T_{H} + \delta^{(1)}T_{H}\,\delta^{(1)}Z_{\text{w}}\right)\\
        &-c_{\alpha-\beta}\,s_{\alpha-\beta}\,c_{\beta}^{2}\,\delta^{(1)}t_{\beta} \left(c_{\alpha-\beta}\,\delta^{(1)}T_{h} + s_{\alpha-\beta}\,\delta^{(1)}T_{H}\right)\Biggr]\ ,
    \end{aligned}
  \end{split}\\
  \delta^{(2)}m_{hA}^{2} &= \frac{e}{2\,M_{W}\,s_{\text{w}}}\,s_{\alpha-\beta}\left(\delta^{(2)}T_{A} + \delta^{(1)}T_{A}\,\delta^{(1)}Z_{\text{w}}\right)\ ,\\
  \delta^{(2)}m_{hG}^{2} &= \frac{e}{2\,M_{W}\,s_{\text{w}}}\,c_{\alpha-\beta}\left(\delta^{(2)}T_{A} + \delta^{(1)}T_{A}\,\delta^{(1)}Z_{\text{w}}\right)\ ,\\
  \delta^{(2)}m_{HA}^{2} &= -\delta^{(2)}m_{hG}^{2}\ ,\\
  \delta^{(2)}m_{HG}^{2} &= \delta^{(2)}m_{hA}^{2}\ ,\\
  \begin{split}
    \delta^{(2)}m_{AG}^{2} &= - c_{\beta}^{2}\,m_A^{2}\,\delta^{(2)}t_{\beta} - c_{\beta}^{2}\,\delta^{(1)}m_A^{2}\,\delta^{(1)}t_{\beta} + c_{\beta}^{3}\,s_{\beta}\,m_A^2 \left(\delta^{(1)}t_{\beta}\right)^{2}\\
                           &\quad -
    \begin{aligned}[t]
      \frac{e}{2\,M_{W}\,s_{\text{w}}} \Biggl[
        & c_{\alpha-\beta}\left(\delta^{(2)}T_{h} + \delta^{(1)}T_{h}\,\delta^{(1)}Z_{\text{w}}\right)\\
        &+s_{\alpha-\beta}\left(\delta^{(2)}T_{H} + \delta^{(1)}T_{H}\,\delta^{(1)}Z_{\text{w}}\right)\Biggr]\ ,
    \end{aligned}
  \end{split}\\
     \delta^{(2)} m_{H^\pm}^2 &= \delta^{(2)}m_A^2 +  \delta^{(2)}M_W^2  \,  ,  \\
 \begin{split}
    \delta^{(2)}m_{G^{\pm}}^{2} &= c_{\beta}^{4}\,m_{H^{\pm}}^2 \left(\delta^{(1)}t_{\beta}\right)^{2}\\
                           &\quad +
    \begin{aligned}[t]
      \frac{e}{2\,M_{W}\,s_{\text{w}}} \Biggl[
        & s_{\alpha-\beta}\left(\delta^{(2)}T_{h} + \delta^{(1)}T_{h}\,\delta^{(1)}Z_{\text{w}}\right)\\
        &-c_{\alpha-\beta}\left(\delta^{(2)}T_{H} + \delta^{(1)}T_{H}\,\delta^{(1)}Z_{\text{w}}\right)\\
        &+c_{\beta}^{2}\,\delta^{(1)}t_{\beta}\left(c_{\alpha-\beta}\,\delta^{(1)}T_{h} + s_{\alpha-\beta}\,\delta^{(1)}T_{H}\right)\Biggr]\ ,
    \end{aligned}
  \end{split}\\
  \begin{split}
    \delta^{(2)}m_{H^-G^+}^{2} &= - c_{\beta}^{2}\,m_{H^{\pm}}^2\,\delta^{(2)}t_{\beta} + c_{\beta}^{3}\,s_{\beta}\,m_{H^{\pm}}^2 \left(\delta^{(1)}t_{\beta}\right)^{2}\\
                           &\quad -
    \begin{aligned}[t]
      \frac{e}{2\,M_{W}\,s_{\text{w}}} \Biggl[
        & c_{\alpha-\beta}\left(\delta^{(2)}T_{h} + \delta^{(1)}T_{h}\,\delta^{(1)}Z_{\text{w}}\right)\\
        &+s_{\alpha-\beta}\left(\delta^{(2)}T_{H} + \delta^{(1)}T_{H}\,\delta^{(1)}Z_{\text{w}}\right)\\
        &+\I\left(\delta^{(2)}T_{A} + \delta^{(1)}T_{A}\,\delta^{(1)}Z_{\text{w}}\right)\Biggr]\ ,
    \end{aligned}
  \end{split}\\
  \delta^{(2)}m_{G^-H^+}^{2} &= \left(\delta^{(2)}m_{H^-G^+}^{2}\right)^* \, .
\end{align}
\end{subequations}
}%
The neutral counterterms are symmetric,
\IE~$\delta^{(2)} m^2_{ab} = \delta^{(2)} m^2_{ba}
\quad (a,b = h,H,A,G)$.

\section{Couplings and counterterm insertions\label{sec:couplings}}

\subsection{Tree-level vertices}

The tree-level vertices contain the top-Yukawa coupling~\mbox{$h_{t} =
  \left.m_t\middle/v_2\right.$}. In the case of the Higgs bosons,
their different couplings are accommodated by explicit
factors~$c\left(\,\dots\right)$. The symbols~$\Phi^0$ and~$\Phi^{\pm}$
are used as generic expressions for the Higgs bosons,
\IE~\mbox{$\Phi^0 \in \{h,\,H,\,A,\,G\}$} and~\mbox{$\Phi^{\pm} \in
  \{H^{\pm},\,G^{\pm}\}$}. For fermion couplings, the left-chiral part
is the first and the right-chiral part the second entry of the
column. The present approximations have already been applied to this
expressions, leaving only those parts proportional to~$h_{t}$
or~$h_{t}^{2}$. The mixing matrix~$\mathbf{U}$ of the charginos does
not appear in the following;~\mbox{$\mathbf{V} = \unity$} is already
inserted. \\

{\allowdisplaybreaks
\begin{subequations}
\begin{align}
\begin{minipage}[t][.04\textwidth][b]{.2\textwidth}\includegraphics[width=\textwidth]{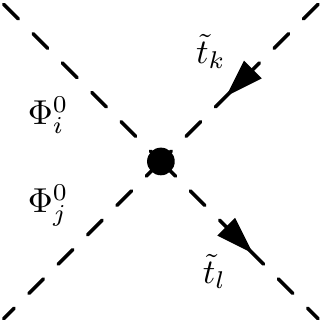}\end{minipage}\quad
\begin{split}
  &= -\I\,C{\left(\Phi^{0}_{i},\,\Phi^{0}_{j},\,\tilde{t}_{k},\,\tilde{t}_{l}\right)}\\
  &= -\I\,h_{t}^{2}\Big[c_{4}{\left(\Phi^{0}_{i},\, \Phi^{0}_{j}\right)} \left(\mathbf{U}_{\tilde{t}\,k1}^{*}\,\mathbf{U}_{\tilde{t}\,l1} + \mathbf{U}_{\tilde{t}\,k2}^{*}\,\mathbf{U}_{\tilde{t}\,l2}\right)\Big]\ ,\\
  &\quad\begin{aligned}[t]
    c_{4}{\left(h,\, h\right)} &= c_{\alpha}^{2}\ ,\\
    c_{4}{\left(H,\, H\right)} &= s_{\alpha}^{2}\ ,\\
    c_{4}{\left(A,\, A\right)} &= c_{\beta}^{2}\ ,\\
    c_{4}{\left(G,\, G\right)} &= s_{\beta}^{2}\ ,\\
    c_{4}{\left(h,\, H\right)} &= c_{\alpha}\,s_{\alpha}\ ,\\
    c_{4}{\left(A,\, G\right)} &= c_{\beta}\,s_{\beta}\ .
  \end{aligned}
\end{split}\\
\quad\notag\\
\begin{minipage}[t][.06\textwidth][b]{.2\textwidth}\includegraphics[width=\textwidth]{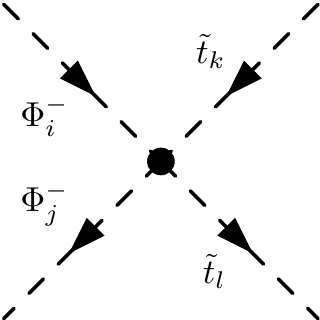}\end{minipage}\quad
\begin{split}
  &= -\I\,C{\left(\Phi^{-}_{i},\,\Phi^{+}_{j},\,\tilde{t}_{k},\,\tilde{t}_{l}\right)}\\
  &= -\I\,h_{t}^{2}\Big[c_{4}^{\tilde{t}}{\left(\Phi^{-}_{i},\, \Phi^{+}_{j}\right)}\,\mathbf{U}_{\tilde{t}\,k2}^{*}\,\mathbf{U}_{\tilde{t}\,l2}\Big]\ ,\\
  &\quad\begin{aligned}[t]
    c_{4}^{\tilde{t}}{\left(H^{-},\, H^{+}\right)} &= c_{\beta}^{2}\ ,\\
    c_{4}^{\tilde{t}}{\left(G^{-},\, G^{+}\right)} &= s_{\beta}^{2}\ ,\\
    c_{4}^{\tilde{t}}{\left(H^{-},\, G^{+}\right)} &= c_{\beta}\,s_{\beta}\ ,\\
    c_{4}^{\tilde{t}}{\left(G^{-},\, H^{+}\right)} &= c_{\beta}\,s_{\beta}\ .
  \end{aligned}
\end{split}\\
\quad\notag\\
\begin{minipage}{.2\textwidth}\includegraphics[width=\textwidth]{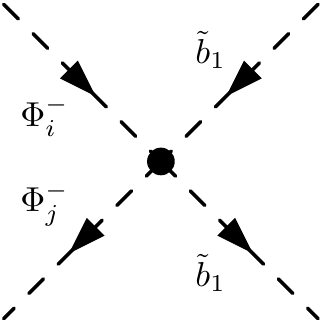}\end{minipage}\quad
\begin{split}
  &= -\I\,C{\left(\Phi^{-}_{i},\,\Phi^{+}_{j},\,\tilde{b}_{1},\,\tilde{b}_{1}\right)} =
     -\I\,h_{t}^{2}\Big[c_{4}^{\tilde{b}}{\left(\Phi^{-}_{i},\, \Phi^{+}_{j}\right)}\Big]\ ,\\
  &\quad\begin{aligned}[t]
    c_{4}^{\tilde{b}}{\left(H^{-},\, H^{+}\right)} &= c_{\beta}^{2}\ ,\\
    c_{4}^{\tilde{b}}{\left(G^{-},\, G^{+}\right)} &= s_{\beta}^{2}\ ,\\
    c_{4}^{\tilde{b}}{\left(H^{-},\, G^{+}\right)} &= c_{\beta}\,s_{\beta}\ ,\\
    c_{4}^{\tilde{b}}{\left(G^{-},\, H^{+}\right)} &= c_{\beta}\,s_{\beta}\ .
  \end{aligned}
\end{split}\\
\quad\notag\\
\begin{minipage}[t][.05\textwidth][b]{.2\textwidth}\includegraphics[width=\textwidth]{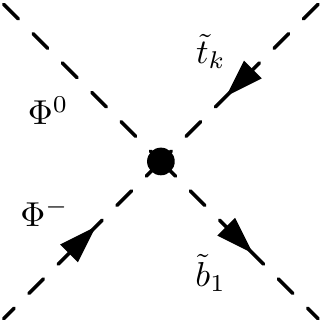}\end{minipage}\quad
\begin{split}
  &= -\I\,C{\left(\Phi^{0},\,\Phi^{-},\,\tilde{t}_{k},\,\tilde{b}_{1}\right)} = -\I\,C{\left(\Phi^{0},\,\Phi^{+},\,\tilde{t}_{k},\,\tilde{b}_{1}\right)}^{*}\\
  &= -\I\,h_{t}^{2}\Big[c_{4}{\left(\Phi^{0},\, \Phi^{-}\right)}\,\mathbf{U}_{\tilde{t}\,k1}^{*}\Big]\ ,\\
  &\quad\begin{aligned}[t]
    c_{4}{\left(h,\, H^{-}\right)} &= \tfrac{-c_{\alpha}\,c_{\beta}}{\sqrt{2}}\ , & c_{4}{\left(H,\, H^{-}\right)} &= \tfrac{-s_{\alpha}\,c_{\beta}}{\sqrt{2}}\ ,\\
    c_{4}{\left(h,\, G^{-}\right)} &= \tfrac{-c_{\alpha}\,s_{\beta}}{\sqrt{2}}\ , & c_{4}{\left(H,\, G^{-}\right)} &= \tfrac{-s_{\alpha}\,s_{\beta}}{\sqrt{2}}\ ,\\
    c_{4}{\left(A,\, H^{-}\right)} &= \tfrac{-\I\,c_{\beta}^{2}}{\sqrt{2}}\ , & c_{4}{\left(G,\, H^{-}\right)} &= \tfrac{-\I\,c_{\beta}\,s_{\beta}}{\sqrt{2}}\ ,\\
    c_{4}{\left(A,\, G^{-}\right)} &= \tfrac{-\I\,c_{\beta}\,s_{\beta}}{\sqrt{2}}\ , & c_{4}{\left(G,\, G^{-}\right)} &= \tfrac{-\I\,s_{\beta}^{2}}{\sqrt{2}}\ .
  \end{aligned}
\end{split}\\
\quad\notag\\
\begin{minipage}{.2\textwidth}\includegraphics[width=\textwidth]{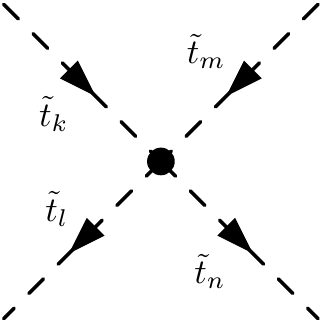}\end{minipage}\quad
\begin{split}
  &= -\I\,C{\left(\tilde{t}_{k},\,\tilde{t}_{l},\,\tilde{t}_{m},\,\tilde{t}_{n}\right)}\\
  &= -\I\,h_{t}^{2}\begin{aligned}[t]\Big[
        &\left(\mathbf{U}_{\tilde{t}\,k1}^{*}\,\mathbf{U}_{\tilde{t}\,m2}^{*} + \mathbf{U}_{\tilde{t}\,k2}^{*}\,\mathbf{U}_{\tilde{t}\,m1}^{*}\right)\\
        &\left(\mathbf{U}_{\tilde{t}\,l1}\,\mathbf{U}_{\tilde{t}\,n2} + \mathbf{U}_{\tilde{t}\,l2}\,\mathbf{U}_{\tilde{t}\,n1}\right)\Big]\ .
  \end{aligned}
\end{split}\\
\quad\notag\\
\begin{minipage}{.2\textwidth}\includegraphics[width=\textwidth]{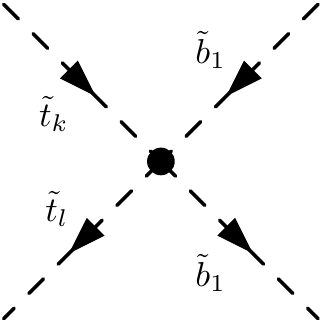}\end{minipage}\quad &
  = -\I\,C{\left(\tilde{t}_{k},\,\tilde{t}_{l},\,\tilde{b}_{1},\,\tilde{b}_{1}\right)} =
    -\I\,h_{t}^{2}\Big[\mathbf{U}_{\tilde{t}\,k2}^{*}\,\mathbf{U}_{\tilde{t}\,l2}\Big]\ .\\
\quad\notag\\[-1ex]
\begin{minipage}[t][2ex][b]{.2\textwidth}\includegraphics[width=\textwidth]{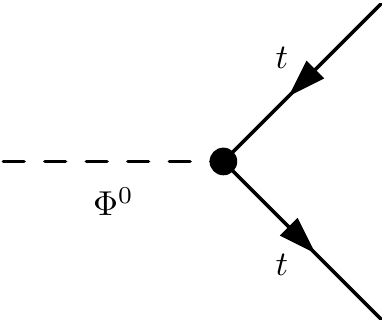}\end{minipage}\quad
\begin{split}
  &= -\I\,C{\left(\Phi^{0},\,t,\,t\right)} =
     -\I\,h_{t} \Big[c_{3}{\left(\Phi^{0}\right)}\Big] \begin{pmatrix} 1\\ \mathrm{sign}{\left(\Phi^{0}\right)}\end{pmatrix},\\
  &\quad\begin{aligned}[t]
    \left\{c_{3}{\left(h\right)},\, \mathrm{sign}{\left(h\right)}\right\} &= \left\{\tfrac{c_{\alpha}}{\sqrt{2}},\, 1\right\}\ ,\\
    \left\{c_{3}{\left(H\right)},\, \mathrm{sign}{\left(H\right)}\right\} &= \left\{\tfrac{s_{\alpha}}{\sqrt{2}},\, 1\right\}\ ,\\
    \left\{c_{3}{\left(A\right)},\, \mathrm{sign}{\left(A\right)}\right\} &= \left\{\tfrac{\I c_{\beta}}{\sqrt{2}},\, -1\right\}\ ,\\
    \left\{c_{3}{\left(G\right)},\, \mathrm{sign}{\left(G\right)}\right\} &= \left\{\tfrac{\I s_{\beta}}{\sqrt{2}},\, -1\right\}\ .
  \end{aligned}
\end{split}\\
\quad\notag\\
\begin{minipage}{.2\textwidth}\includegraphics[width=\textwidth]{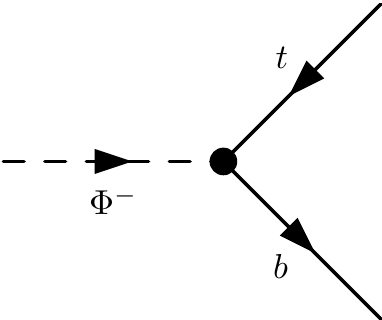}\end{minipage}\quad
\begin{split}
  &= -\I\,C{\left(\Phi^{-},\,t,\,\bar{b}\right)} =
     -\I\,h_{t} \Big[c_{3}{\left(\Phi^{-}\right)}\Big] \begin{pmatrix} 0\\ 1\end{pmatrix},\\
  &\quad\begin{aligned}[t]
    c_{3}{\left(H^{-}\right)} &= -c_{\beta}\ ,\\
    c_{3}{\left(G^{-}\right)} &= -s_{\beta}\ .
  \end{aligned}
\end{split}\\
\quad\notag\\
\begin{minipage}{.2\textwidth}\includegraphics[width=\textwidth]{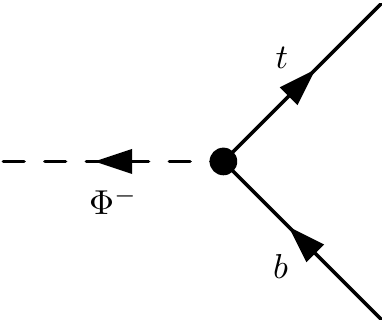}\end{minipage}\quad
\begin{split}
  &= -\I\,C{\left(\Phi^{+},\,\bar{t},\,b\right)} =
     -\I\,h_{t} \Big[\bar{c}_{3}{\left(\Phi^{+}\right)}\Big] \begin{pmatrix} 1\\ 0\end{pmatrix},\\
  &\quad\begin{aligned}[t]
    \bar{c}_{3}{\left(H^{+}\right)} &= -c_{\beta}\ ,\\
    \bar{c}_{3}{\left(G^{+}\right)} &= -s_{\beta}\ .
  \end{aligned}
\end{split}\\
\quad\notag\\
\begin{minipage}[t][-.03\textwidth][b]{.2\textwidth}\includegraphics[width=\textwidth]{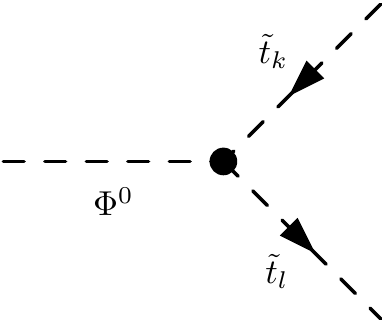}\end{minipage}\quad
\begin{split}
  &= -\I\,C{\left(\Phi^{0},\,\tilde{t}_{k},\,\tilde{t}_{l}\right)}\\
  &= -\I\,h_{t}\Big[
    \begin{aligned}[t]
      & c_{m}{\left(\Phi^{0}\right)} \left(\mathbf{U}_{\tilde{t}\,k1}^{*}\,\mathbf{U}_{\tilde{t}\,l1} + \mathbf{U}_{\tilde{t}\,k2}^{*}\,\mathbf{U}_{\tilde{t}\,l2}\right) m_{t}\\
      &+ c_{\mu}{\left(\Phi^{0}\right)} \left(\mathrm{sign}{\left(\Phi^{0}\right)}\,\mathbf{U}_{\tilde{t}\,k2}^{*}\,\mathbf{U}_{\tilde{t}\,l1}\,\mu + \mathbf{U}_{\tilde{t}\,k1}^{*}\,\mathbf{U}_{\tilde{t}\,l2}\,\mu^{*}\right)\\
      &+ c_{A}{\left(\Phi^{0}\right)} \left(\mathrm{sign}{\left(\Phi^{0}\right)}\,\mathbf{U}_{\tilde{t}\,k2}^{*}\,\mathbf{U}_{\tilde{t}\,l1}\,A_{t}^{*} + \mathbf{U}_{\tilde{t}\,k1}^{*}\,\mathbf{U}_{\tilde{t}\,l2}\,A_{t}\right)\Big]\ ,
    \end{aligned}\\
  &\begin{aligned}[t]
    \left\{c_{m}{\left(h\right)},\, c_{\mu}{\left(h\right)},\, c_{A}{\left(h\right)},\, \mathrm{sign}{\left(h\right)}\right\} &= \left\{\sqrt{2}\,c_{\alpha},\, \tfrac{s_{\alpha}}{\sqrt{2}},\, \tfrac{c_{\alpha}}{\sqrt{2}},\, 1\right\}\ ,\\
    \left\{c_{m}{\left(H\right)},\, c_{\mu}{\left(H\right)},\, c_{A}{\left(H\right)},\, \mathrm{sign}{\left(H\right)}\right\} &= \left\{\sqrt{2}\,s_{\alpha},\, \tfrac{-c_{\alpha}}{\sqrt{2}},\, \tfrac{s_{\alpha}}{\sqrt{2}},\, 1\right\}\ ,\\
    \left\{c_{m}{\left(A\right)},\, c_{\mu}{\left(A\right)},\, c_{A}{\left(A\right)},\, \mathrm{sign}{\left(A\right)}\right\} &= \left\{0,\, \tfrac{\I\,s_{\beta}}{\sqrt{2}},\, \tfrac{\I\,c_{\beta}}{\sqrt{2}},\, -1\right\}\ ,\\
    \left\{c_{m}{\left(G\right)},\, c_{\mu}{\left(G\right)},\, c_{A}{\left(G\right)},\, \mathrm{sign}{\left(G\right)}\right\} &= \left\{0,\, \tfrac{-\I\,c_{\beta}}{\sqrt{2}},\, \tfrac{\I\,s_{\beta}}{\sqrt{2}},\, -1\right\}\ .
  \end{aligned}
\end{split}\\
\quad\notag\\
\begin{minipage}{.2\textwidth}\includegraphics[width=\textwidth]{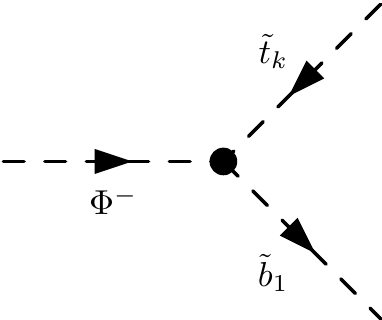}\end{minipage}\quad
\begin{split}
  &= -\I\,C{\left(\Phi^{-},\,\tilde{t}_{k},\,\tilde{b}_{1}\right)} = -\I\,C{\left(\Phi^{+},\,\tilde{t}_{k},\,\tilde{b}_{1}\right)}^{*}\\
  &= -\I\,h_{t}\Big[c_{m}{\left(\Phi^{-}\right)}\,\mathbf{U}_{\tilde{t}\,k1}^{*}\,m_{t}
                + c_{\mu}{\left(\Phi^{-}\right)}\,\mathbf{U}_{\tilde{t}\,k2}^{*}\,\mu + c_{A}{\left(\Phi^{-}\right)}\,\mathbf{U}_{\tilde{t}\,k2}^{*}\,A_{t}^{*}\Big]\ ,\\
  &\quad\begin{aligned}[t]
    \left\{c_{m}{\left(H^{-}\right)},\, c_{\mu}{\left(H^{-}\right)},\, c_{A}{\left(H^{-}\right)}\right\} &= \left\{-c_{\beta},\, -s_{\beta},\, -c_{\beta}\right\}\ ,\\
    \left\{c_{m}{\left(G^{-}\right)},\, c_{\mu}{\left(G^{-}\right)},\, c_{A}{\left(G^{-}\right)}\right\} &= \left\{-s_{\beta},\, c_{\beta},\, -s_{\beta}\right\}\ .
  \end{aligned}
\end{split}\\
\quad\notag\\
\begin{minipage}{.2\textwidth}\includegraphics[width=\textwidth]{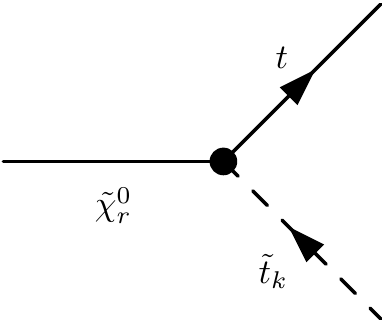}\end{minipage}\quad &
  = -\I\,C{\left(\tilde{\chi}^{0}_{r},\,\bar{t},\,\tilde{t}_{k}\right)} =
    -\I\,h_{t} \begin{pmatrix} \mathbf{U}_{\tilde{t}\,k1}^{*}\,\mathbf{N}_{r4}^{*}\\ \mathbf{U}_{\tilde{t}\,k2}^{*}\,\mathbf{N}_{r4}\end{pmatrix}.\\
\quad\notag\\
\begin{minipage}{.2\textwidth}\includegraphics[width=\textwidth]{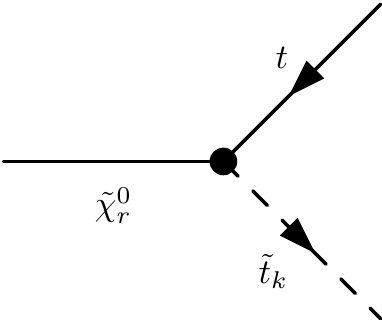}\end{minipage}\quad &
  = -\I\,C{\left(\tilde{\chi}^{0}_{r},\,t,\,\tilde{t}_{k}\right)} =
    -\I\,h_{t} \begin{pmatrix} \mathbf{U}_{\tilde{t}\,k2}\,\mathbf{N}_{r4}^{*}\\ \mathbf{U}_{\tilde{t}\,k1}\,\mathbf{N}_{r4}\end{pmatrix}.\\
\quad\notag\\
\begin{minipage}{.2\textwidth}\includegraphics[width=\textwidth]{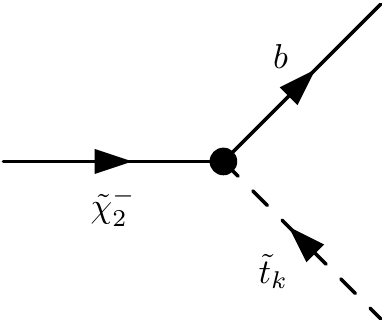}\end{minipage}\quad &
  = -\I\,C{\left(\tilde{\chi}^{-}_{2},\,\bar{b},\,\tilde{t}_{k}\right)} =
    -\I\,h_{t} \begin{pmatrix} 0\\ -\mathbf{U}_{\tilde{t}\,k2}^{*}\end{pmatrix}.\\
\quad\notag\\
\begin{minipage}{.2\textwidth}\includegraphics[width=\textwidth]{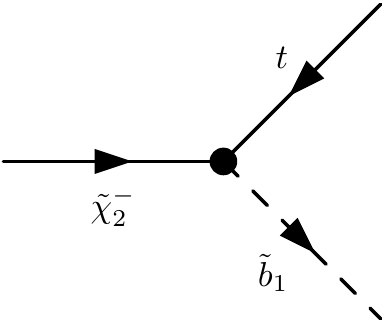}\end{minipage}\quad &
  = -\I\,C{\left(\tilde{\chi}^{-}_{2},\,t,\,\tilde{b}_{1}\right)} =
    -\I\,h_{t} \begin{pmatrix} 0\\ -1\end{pmatrix}.\\
\quad\notag\\
\begin{minipage}{.2\textwidth}\includegraphics[width=\textwidth]{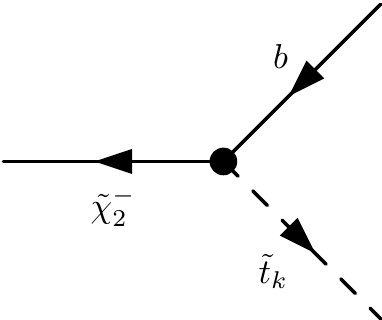}\end{minipage}\quad &
  = -\I\,C{\left(\tilde{\chi}^{+}_{2},\,b,\,\tilde{t}_{k}\right)} =
    -\I\,h_{t} \begin{pmatrix} -\mathbf{U}_{\tilde{t}\,k2}\\ 0\end{pmatrix}.\\
\quad\notag\\
\begin{minipage}{.2\textwidth}\includegraphics[width=\textwidth]{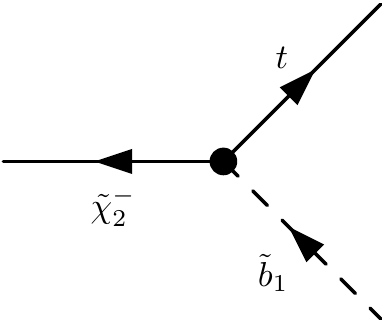}\end{minipage}\quad &
  = -\I\,C{\left(\tilde{\chi}^{+}_{2},\,\bar{t},\,\tilde{b}_{1}\right)} =
    -\I\,h_{t} \begin{pmatrix} -1\\ 0\end{pmatrix}.
\end{align}
\end{subequations}
}%

\vspace*{0.5cm}

\subsection{Counterterm vertices}

The following one-loop counterterms for the two, three- and four-point vertices 
appear as insertions in the two-loop diagrams with subrenormalization
for masses and couplings. For the two-point vertices we have

\begin{subequations}
\begin{align}
  \begin{minipage}{.2\textwidth}\includegraphics[width=\textwidth]{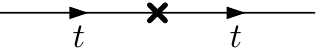}\end{minipage} 
   &  \quad = \I\,\delta^{(1)}m_t\ .\\
  \quad\notag\\
  \begin{minipage}{.2\textwidth}\includegraphics[width=\textwidth]{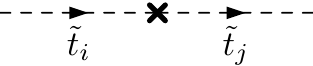}\end{minipage} 
   & \quad  = \I\,\delta^{(1)}m_{\tilde{t}_i\tilde{t}_j}^2\ .\\
  \quad\notag\\
  \begin{minipage}{.2\textwidth}\includegraphics[width=\textwidth]{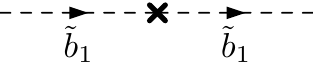}\end{minipage} 
   & \quad  = \I\,\delta^{(1)}m_{\tilde{b}_1}^2 \equiv \I\,\delta^{(1)}m_{\tilde{q}_{3}}^2\ .
\end{align}
\end{subequations}

\vspace*{0.5cm}
\noindent
To shorten the notation for the three- and four-point vertices, 
the previously defined tree-level couplings~$C{\left(\,\dots\right)}$ are re-utilized;  
their corresponding one-loop counterterms are named~$\delta^{(1)}C{\left(\,\dots\right)}$. 
Since each of the vertices contains the top-Yukawa coupling~$h_t$, 
its renormalization constant~$\delta^{(1)}h_{t}$, given by
  \begin{align}
  \delta^{(1)}h_{t} & = h_{t} \left(\frac{\delta^{(1)}m_{t}}{m_{t}} 
                             - \frac{\delta^{(1)}M_{W}}{M_{W}} 
                             - \frac{\delta^{(1)}s_{\text{w}}}{s_{\text{w}}} 
                              -\frac{\delta^{(1)}s_{\beta}}{s_{\beta}}\right)  ,
  \end{align}
is part of each vertex counterterm
[the renormalization constant~\mbox{$\left.\delta^{(1)}e\middle/e\right.$} 
is omitted since there are no contributions of~$\mathcal{O}{\left(\alpha_t\right)}$].
Also the field-renormalization constants of the Higgs bosons are kept; 
all other field-renormalization constants cancel out in the sum of the full set of Feynman diagrams, 
since the corresponding particles exclusively appear in internal propagators.\\

\vspace*{0.3cm}
{\allowdisplaybreaks
\begin{subequations}
\begin{align}
\begin{minipage}{.2\textwidth}\includegraphics[width=\textwidth]{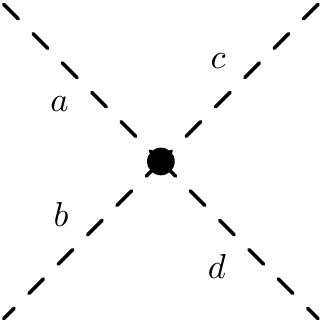}\end{minipage} &= -\I\,C{\left(a,\, b,\, c,\, d\right)}\ , &
\begin{minipage}{.2\textwidth}\includegraphics[width=\textwidth]{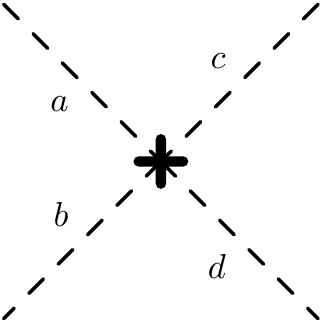}\end{minipage} &= -\I\,\delta^{(1)}C{\left(a,\, b,\, c,\, d\right)}\ ,\\
\quad\notag\\
\begin{minipage}{.2\textwidth}\includegraphics[width=\textwidth]{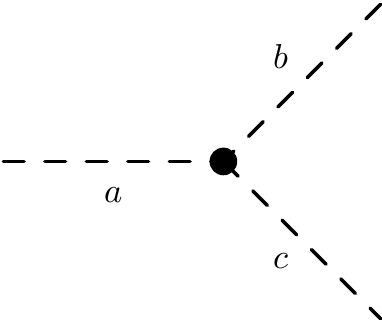}\end{minipage} &= -\I\,C{\left(a,\, b,\, c\right)}\ , &
\begin{minipage}{.2\textwidth}\includegraphics[width=\textwidth]{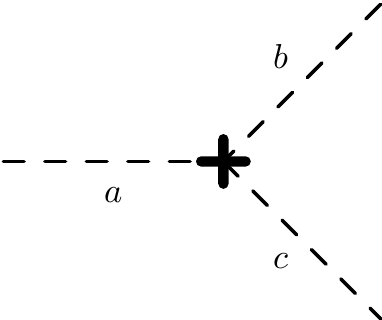}\end{minipage} &= -\I\,\delta^{(1)}C{\left(a,\, b,\, c\right)}\ ,
\end{align}\vspace{-2ex}\nextParentEquation
\begin{align}
\delta^{(1)}C{\left(\Phi^{0}_{i},\,\Phi^{0}_{j},\,\tilde{t}_{k},\,\tilde{t}_{l}\right)} &= C{\left(\Phi^{0}_{i},\,\Phi^{0}_{j},\,\tilde{t}_{k},\,\tilde{t}_{l}\right)}
                                                                            \left(\tfrac{2\,\delta^{(1)}h_{t}}{h_{t}} + \delta^{(1)}Z_{\mathcal{H}_{2}}\right)\ ,\\
\delta^{(1)}C{\left(\Phi^{-}_{i},\,\Phi^{+}_{j},\,\tilde{t}_{k},\,\tilde{t}_{l}\right)} &= C{\left(\Phi^{-}_{i},\,\Phi^{+}_{j},\,\tilde{t}_{k},\,\tilde{t}_{l}\right)}
                                                                            \left(\tfrac{2\,\delta^{(1)}h_{t}}{h_{t}} + \delta^{(1)}Z_{\mathcal{H}_{2}}\right)\ ,\\
\delta^{(1)}C{\left(\Phi^{-}_{i},\,\Phi^{+}_{j},\,\tilde{b}_{1},\,\tilde{b}_{1}\right)} &= C{\left(\Phi^{-}_{i},\,\Phi^{+}_{j},\,\tilde{b}_{1},\,\tilde{b}_{1}\right)}
                                                                            \left(\tfrac{2\,\delta^{(1)}h_{t}}{h_{t}} + \delta^{(1)}Z_{\mathcal{H}_{2}}\right)\ ,\\
\delta^{(1)}C{\left(\Phi^{0},\,\Phi^{-},\,\tilde{t}_{k},\,\tilde{b}_{1}\right)} &= C{\left(\Phi^{0},\,\Phi^{-},\,\tilde{t}_{k},\,\tilde{b}_{1}\right)}
                                                                            \left(\tfrac{2\,\delta^{(1)}h_{t}}{h_{t}} + \delta^{(1)}Z_{\mathcal{H}_{2}}\right)\ ,\\
\delta^{(1)}C{\left(\tilde{t}_{k},\,\tilde{t}_{l},\,\tilde{t}_{m},\,\tilde{t}_{n}\right)} &= C{\left(\tilde{t}_{k},\,\tilde{t}_{l},\,\tilde{t}_{m},\,\tilde{t}_{n}\right)}\left(\tfrac{2\,\delta^{(1)}h_{t}}{h_{t}}\right)\ ,\\
\delta^{(1)}C{\left(\tilde{t}_{k},\,\tilde{t}_{l},\,\tilde{b}_{1},\,\tilde{b}_{1}\right)} &= C{\left(\tilde{t}_{k},\,\tilde{t}_{l},\,\tilde{b}_{1},\,\tilde{b}_{1}\right)}\left(\tfrac{2\,\delta^{(1)}h_{t}}{h_{t}}\right)\ ,\\
\delta^{(1)}C{\left(\Phi^{0},\,t,\,t\right)} &= C{\left(\Phi^{0},\,t,\,t\right)} \left(\tfrac{\delta^{(1)}h_{t}}{h_{t}} + \tfrac{1}{2}\,\delta^{(1)}Z_{\mathcal{H}_{2}}\right)\ ,\\
\delta^{(1)}C{\left(\Phi^{-},\,t,\,\bar{b}\right)} &= C{\left(\Phi^{-},\,t,\,\bar{b}\right)} \left(\tfrac{\delta^{(1)}h_{t}}{h_{t}} + \tfrac{1}{2}\,\delta^{(1)}Z_{\mathcal{H}_{2}}\right)\ ,\\
\delta^{(1)}C{\left(\Phi^{+},\,\bar{t},\,b\right)} &= C{\left(\Phi^{+},\,\bar{t},\,b\right)} \left(\tfrac{\delta^{(1)}h_{t}}{h_{t}} + \tfrac{1}{2}\,\delta^{(1)}Z_{\mathcal{H}_{2}}\right)\ ,\\
\begin{split}\label{eq:nofactct1}
  \delta^{(1)}C{\left(\Phi^{0},\,\tilde{t}_{k},\,\tilde{t}_{l}\right)} &= C{\left(\Phi^{0},\,\tilde{t}_{k},\,\tilde{t}_{l}\right)} \left(\tfrac{\delta^{(1)}h_{t}}{h_{t}}\right)\\
                                                            &\quad +
    \begin{aligned}[t]
      h_{t}\Bigg\{
      & c_{m}{\left(\Phi^{0}\right)} \left(\mathbf{U}_{\tilde{t}\,k1}^{*}\,\mathbf{U}_{\tilde{t}\,l1} + \mathbf{U}_{\tilde{t}\,k2}^{*}\,\mathbf{U}_{\tilde{t}\,l2}\right)
      m_{t} \left(\tfrac{\delta^{(1)}m_{t}}{m_{t}} + \tfrac{1}{2}\,\delta^{(1)}Z_{\mathcal{H}_{2}}\right)\\
      & + \begin{aligned}[t]
        c_{A}{\left(\Phi^{0}\right)} \bigg[&\mathrm{sign}{\left(\Phi^{0}\right)}\,\mathbf{U}_{\tilde{t}\,k2}^{*}\,\mathbf{U}_{\tilde{t}\,l1}\,A_{t}^{*}
          \left(\tfrac{\delta^{(1)}A_{t}^{*}}{A_{t}^{*}} + \tfrac{1}{2}\,\delta^{(1)}Z_{\mathcal{H}_{2}}\right)\\
          &+ \mathbf{U}_{\tilde{t}\,k1}^{*}\,\mathbf{U}_{\tilde{t}\,l2}\,A_{t} \left(\tfrac{\delta^{(1)}A_{t}}{A_{t}} + \tfrac{1}{2}\,\delta^{(1)}Z_{\mathcal{H}_{2}}\right)\bigg]
      \end{aligned}\\
      & + \begin{aligned}[t]
          c_{\mu}{\left(\Phi^{0}\right)} \bigg[& \mathrm{sign}{\left(\Phi^{0}\right)}\,\mathbf{U}_{\tilde{t}\,k2}^{*}\,\mathbf{U}_{\tilde{t}\,l1}\,\mu \left(\tfrac{\delta^{(1)}\mu}{\mu} + \tfrac{1}{2}\,\delta^{(1)}Z_{\mathcal{H}_{1}}\right)\\
            &+ \mathbf{U}_{\tilde{t}\,k1}^{*}\,\mathbf{U}_{\tilde{t}\,l2}\,\mu^{*} \left(\tfrac{\delta^{(1)}\mu^{*}}{\mu^{*}} + \tfrac{1}{2}\,\delta^{(1)}Z_{\mathcal{H}_{1}}\right)\bigg]
        \end{aligned}\Bigg\}\ ,
    \end{aligned}
\end{split}\\
\begin{split}\label{eq:nofactct2}
  \delta^{(1)}C{\left(\Phi^{-},\,\tilde{t}_{k},\,\tilde{b}_{1}\right)} &= C{\left(\Phi^{-},\,\tilde{t}_{k},\,\tilde{b}_{1}\right)}\left(\tfrac{\delta^{(1)}h_{t}}{h_{t}}\right)\\
                                                            &\quad +
    \begin{aligned}[t]
      h_{t}\Bigg[& c_{\mu}{\left(\Phi^{-}\right)}\,\mathbf{U}_{\tilde{t}\,k2}^{*}\,\mu \left(\tfrac{\delta^{(1)}\mu}{\mu} + \tfrac{1}{2}\,\delta^{(1)}Z_{\mathcal{H}_{1}}\right)\\
                &+ c_{A}{\left(\Phi^{-}\right)}\,\mathbf{U}_{\tilde{t}\,k2}^{*}\,A_{t}^{*} \left(\tfrac{\delta^{(1)}A_{t}^{*}}{A_{t}^{*}} + \tfrac{1}{2}\,\delta^{(1)}Z_{\mathcal{H}_{2}}\right)\\
                &+ c_{m}{\left(\Phi^{-}\right)}\,\mathbf{U}_{\tilde{t}\,k1}^{*}\,m_{t} \left(\tfrac{\delta^{(1)}m_{t}}{m_{t}} + \tfrac{1}{2}\,\delta^{(1)}Z_{\mathcal{H}_{2}}\right)\Bigg]\ ,\\
    \end{aligned}
\end{split}\\
\delta^{(1)}C{\left(\tilde{\chi}^{0}_{r},\,\bar{t},\,\tilde{t}_{k}\right)} &= C{\left(\tilde{\chi}^{0}_{r},\,\bar{t},\,\tilde{t}_{k}\right)} \left(\tfrac{\delta^{(1)}h_{t}}{h_{t}}\right)\ ,\\
\delta^{(1)}C{\left(\tilde{\chi}^{0}_{r},\,t,\,\tilde{t}_{k}\right)} &= C{\left(\tilde{\chi}^{0}_{r},\,t,\,\tilde{t}_{k}\right)} \left(\tfrac{\delta^{(1)}h_{t}}{h_{t}}\right)\ ,\\
\delta^{(1)}C{\left(\tilde{\chi}^{-}_{2},\,\bar{b},\,\tilde{t}_{k}\right)} &= C{\left(\tilde{\chi}^{-}_{2},\,\bar{b},\,\tilde{t}_{k}\right)} \left(\tfrac{\delta^{(1)}h_{t}}{h_{t}}\right)\ ,\\
\delta^{(1)}C{\left(\tilde{\chi}^{-}_{2},\,t,\,\tilde{t}_{k}\right)} &= C{\left(\tilde{\chi}^{-}_{2},\,t,\,\tilde{t}_{k}\right)} \left(\tfrac{\delta^{(1)}h_{t}}{h_{t}}\right)\ ,\\
\delta^{(1)}C{\left(\tilde{\chi}^{+}_{2},\,b,\,\tilde{t}_{k}\right)} &= C{\left(\tilde{\chi}^{+}_{2},\,b,\,\tilde{t}_{k}\right)} \left(\tfrac{\delta^{(1)}h_{t}}{h_{t}}\right)\ ,\\
\delta^{(1)}C{\left(\tilde{\chi}^{+}_{2},\,\bar{t},\,\tilde{b}_{1}\right)} &= C{\left(\tilde{\chi}^{+}_{2},\,\bar{t},\,\tilde{b}_{1}\right)} \left(\tfrac{\delta^{(1)}h_{t}}{h_{t}}\right)\ .
\end{align}
\end{subequations}
}%
The counterterms of Eq.~\eqref{eq:nofactct1} and
Eq.~\eqref{eq:nofactct2}  should be emphasized, 
because they are the only ones 
which cannot be simply expressed as a product 
of a tree-level coupling and the counterterm of the 
top-Yukawa coupling and field renormalization.

\section{Loop integrals\label{app:loops}}

The analytical evaluation of the~$\mathcal{O}{\left(\alpha_t^2\right)}$~contributions requires the following explicit expressions for one-loop and two-loop integrals.

\subsection{One-loop functions\label{app:loops_1L}}

In the following all required one-loop integrals are listed up to~$\mathcal{O}{\left(\epsilon^{1}\right)}$, where $\epsilon = \left.\left(4 - D\right)\middle/2\right.$ parametrizes the divergent parts. $D$~is the dimension of the integrated momentum and~$\mu_{\text{D}}$ depicts the regularization parameter, so that
\begin{align}
  \int \mathrm{d}^{4}q &\rightarrow \mu_{\text{D}}^{4 - D}\int \mathrm{d}^{D}q\ .
\end{align}
The reduction to scalar integrals as described first by Ref.~\cite{Passarino:1978jh} has been used. The scalar integrals have been re-evaluated by using the technique of Feynman parameters.
{\allowdisplaybreaks\footnotesize
\begin{subequations}
\begin{align}
  \Anull{0} &= 0,\\
  \Anull{m^{2}} &= \frac{m^{2}}{\epsilon} - m^{2} \left\{\Log{m^{2}}\right\} + m^{2} \epsilon\left\{\frac{1}{2} + \frac{\pi^{2}}{12} + \frac{1}{2} \left[\Log{m^{2}}\right]^{2}\right\},\\\nextParentEquation
  \quad\notag\\
  \begin{split}
  \Bnull{p^{2},0,0} &= \frac{1}{\epsilon} + \left\{1 + C + \log{\left(-\tfrac{\mu_{\text{D}}}{p^2 - \I \epsilon^{\prime}}\right)}\right\}\\
  &\quad+ \epsilon\left\{2 - \frac{\pi^{2}}{12} + \frac{1}{2}\left[1 + C + \log{\left(-\tfrac{\mu_{\text{D}}}{p^2 - \I \epsilon^{\prime}}\right)}\right]^2\right\},
  \end{split}\\
  \Bnull{0,0,m^{2}} &= \Bnull{0,m^{2},0} = \frac{\Anull{m^{2}}}{m^{2}},\\
  \Bnull{0,m^{2},m^{2}} &= \left(1 - \epsilon\right) \frac{\Anull{m^{2}}}{m^{2}},\\
  \Bnull{0,m_{1}^{2},m_{2}^{2}} &= \frac{\Anull{m_{1}^{2}} - \Anull{m_{2}^{2}}}{m_{1}^{2} - m_{2}^{2}},\\
  \Bnull{m^{2},0,m^{2}} &= \frac{1}{\epsilon} + \left\{1 - \Log{m^{2}}\right\} + \epsilon\left\{2 + \frac{\pi^{2}}{12} + \frac{1}{2} \left[1 - \Log{m^{2}}\right]^{2}\right\},\\
  \Bnull{m^{2},m^{2},0} &= \Bnull{m^{2},0,m^{2}},\\
  \begin{split}
    \Bnull{m_{1}^{2},0,m_{2}^{2}} &= \frac{1}{\epsilon} + \left\{\frac{m_{2}^{2}}{m_{1}^{2}}\left[1 - \Log{m_{2}^{2}}\right] + \frac{m_{1}^{2} - m_{2}^{2}}{m_{1}^{2}}\left[1 + C + \log{\left(\tfrac{\mu_{\text{D}}}{m_{2}^{2} - m_{1}^{2}}\right)}\right]\right\}\\
    &\quad +\begin{aligned}[t]
    \epsilon\,\Biggl\{
    & \frac{m_{1}^{2} - m_{2}^{2}}{2m_{1}^{2}}\left[\left(1 + C + \log{\left(\tfrac{\mu_{\text{D}}}{m_{2}^{2} - m_{1}^{2}}\right)}\right)^{2} - 2 \dilog{\tfrac{-m_{1}^{2}}{m_{2}^{2} - m_{1}^{2}}}\right]\\
    & + 2 + \frac{\pi^{2}}{12} + \frac{m_{2}^{2}}{2m_{1}^{2}}\left[1 - \Log{m_{2}^{2}}\right]^{2}\Biggr\},
    \end{aligned}
  \end{split}\\
  \Bnull{m_{1}^{2},m_{2}^{2},0} &= \Bnull{m_{1}^{2},0,m_{2}^{2}},\\
  \begin{split}
    B_{0}{\Big((m_{1} \pm m_{2}}&)^{2}, m_{1}^{2},m_{2}^{2}\Big) =\\
    &\,\quad \frac{1}{\epsilon} + \left\{\frac{m_{1}\left[1 - \Log{m_{1}^{2}}\right] \pm m_{2}\left[1 - \Log{m_{2}^{2}}\right]}{m_{1} \pm m_{2}}\right\}\\
    &\quad + \epsilon\left\{2 + \frac{\pi^{2}}{12} + \frac{m_{1}\left[1 - \Log{m_{1}^{2}}\right]^{2} \pm m_{2}\left[1 - \Log{m_{2}^{2}}\right]^{2}}{2\left(m_{1} \pm m_{2}\right)}\right\},
  \end{split}\\
  \begin{split}
    \Bnull{p^{2},m_{1}^{2},m_{2}^{2}} &= \frac{1}{\epsilon} + \begin{aligned}[t] \biggl\{
      & \frac{m_{1}^{2} - m_{2}^{2} + p^{2}}{2 p^{2}}\left[1 - \Log{m_{1}^{2}}\right] + \frac{m_{2}^{2} - m_{1}^{2} + p^{2}}{2 p^{2}}\left[1 - \Log{m_{2}^{2}}\right]\\
      & + \frac{R}{2 p^{2}}\left[\log{\scriptstyle\left(m_{1}^{2} + m_{2}^{2} - p^{2} + R\right)} + \log{\left(\tfrac{1}{m_{1}^{2} + m_{2}^{2} - p^{2} - R}\right)}\right]\biggr\}
    \end{aligned}\\
    &\quad + \begin{aligned}[t] \epsilon\,\biggl\{
      & 2 + \frac{\pi^{2}}{12} + \tfrac{m_{1}^{2} - m_{2}^{2} + p^{2}}{4p^{2}}\left[1 - \Log{m_{1}^{2}}\right]^{2} + \tfrac{m_{2}^{2} - m_{1}^{2} + p^{2}}{4p^{2}}\left[1 - \Log{m_{2}^{2}}\right]^{2}\\
      & -\begin{aligned}[t] \frac{R}{4p^{2}}\biggl[
        & \left(1 - \Log{m_{1}^{2}}\right)\left(\log{\scriptstyle\left(m_{1}^{2} - m_{2}^{2} + p^{2} + R\right)} - \log{\scriptstyle\left(m_{2}^{2} - m_{1}^{2} - p^{2} + R\right)}\right)\\
        & + \left(1 - \Log{m_{2}^{2}}\right)\left(\log{\scriptstyle\left(m_{2}^{2} - m_{1}^{2} + p^{2} + R\right)} - \log{\scriptstyle\left(m_{1}^{2} - m_{2}^{2} - p^{2} + R\right)}\right)\\
        & + \left(-1 - \Log{-p^{2}} + 2 \Log{R}\right)\\
        &\quad \left(\log{\scriptstyle\left(m_{1}^{2} + m_{2}^{2} - p^{2} + R\right)} + \log{\left(\tfrac{1}{m_{1}^{2} + m_{2}^{2} - p^{2} - R}\right)}\right)\\
        & + 2\left(\dilog{\tfrac{m_{1}^{2} - m_{2}^{2} - p^{2} + R}{2 R}} - \dilog{\tfrac{m_{2}^{2} - m_{1}^{2} + p^{2} + R}{2 R}}\right)\\
        & + 2\left(\dilog{\tfrac{m_{2}^{2} - m_{1}^{2} - p^{2} + R}{2 R}} - \dilog{\tfrac{m_{1}^{2} - m_{2}^{2} + p^{2} + R}{2 R}}\right)\biggr]\biggr\},
      \end{aligned}
    \end{aligned}
  \end{split}\\\nextParentEquation
  \Beins{0,m_{1}^{2},m_{2}^{2}} &= -\frac{1}{2} \Bnull{0,m_{1}^{2},m_{2}^{2}} + \frac{m_{2}^{2} - m_{1}^{2}}{2} \DBnull{0,m_{1}^{2},m_{2}^{2}},\\
  \Beins{p^{2},m_{1}^{2},m_{2}^{2}} &= \frac{1}{2 p^{2}} \left[\Anull{m_{1}^{2}} - \Anull{m_{2}^{2}} - \left(p^{2} - m_{2}^{2} + m_{1}^{2}\right) \Bnull{p^{2},m_{1}^{2},m_{2}^{2}}\right],\\
  \begin{split}
    \Bnnull{p^{2},m_{1}^{2},m_{2}^{2}} &=
    \begin{aligned}[t]
      \frac{1}{2\left(3 - 2 \epsilon\right)} \Big[
      & \Anull{m_{2}^{2}} + 2 m_{1}^{2} \Bnull{p^{2},m_{1}^{2},m_{2}^{2}}\\
      &+ \left(p^{2} - m_{2}^{2} + m_{1}^{2}\right) \Beins{p^{2},m_{1}^{2},m_{2}^{2}}\Big],
    \end{aligned}
  \end{split}\\\nextParentEquation
  \quad\notag\\
  \DBnull{0,0,0} &= 0,\\
  \DBnull{0,0,m^{2}} &= \frac{1}{2 m^{2}} + \frac{\epsilon}{2 m^{2}} \left\{\frac{1}{2} - \Log{m^{2}}\right\},\\
  \DBnull{0,m^{2},0} &= \DBnull{0,0,m^{2}},\\
  \DBnull{0,m^{2},m^{2}} &= \frac{1}{6 m^{2}} + \frac{\epsilon}{6 m^{2}} \left\{-1 - \Log{m^{2}}\right\},\\
  \begin{split}
    \DBnull{0,m_{1}^{2},m_{2}^{2}} &= \frac{1}{2 \left(m_{1}^{2} - m_{2}^{2}\right)^{3}} \left\{m_{1}^{4} - m_{2}^{4} + 2 m_{1}^{2} m_{2}^{2} \log{\left(\tfrac{m_{2}^{2}}{m_{1}^{2}}\right)}\right\}\\
    &\quad + \begin{aligned}[t]
      \frac{\epsilon}{2 \left(m_{1}^{2} - m_{2}^{2}\right)^{3}} \biggl\{
      & m_{1}^{4} \left[\frac{1}{2} - \Log{m_{1}^{2}}\right] - m_{2}^{4} \left[\frac{1}{2} - \Log{m_{2}^{2}}\right]\\
      & + m_{1}^{2} m_{2}^{2} \left[\left(\Log{m_{1}^{2}}\right)^{2} - \left(\Log{m_{2}^{2}}\right)^{2}\right]\biggr\},
    \end{aligned}
  \end{split}\\\nextParentEquation
  \quad\notag\\
  \Cnull{0,0,0,m^{2},m^{2},m^{2}} &= -\frac{\epsilon}{2 m^{2}} \Bnull{0,m^{2},m^{2}},\\\nextParentEquation
  \quad\notag\\
  \Log{m^{2}} &= \log{\left(\tfrac{m^{2}}{\mu_{\text{D}}}\right)} - C,\\
  C &= 1 - \gamma_{\text{E}} + \log{\left(4\pi\right)},\\
  R &= \sqrt{m_{1}^{4} + m_{2}^{4} + p^{4} - 2m_{1}^{2}m_{2}^{2} - 2m_{2}^{2}p^{2} - 2p^{2}m_{1}^{2}}.
\end{align}
\end{subequations}}

\subsection{Two-loop functions\label{app:loops_2L}}
The notation of the two-loop integrals follows the conventions which have been introduced by Refs.~\cite{Weiglein:1993hd,Weiglein:1995qs}.
After reducing the appearing two-loop integrals to a set of master integrals and applying the approximation of a vanishing external momentum, only the following function is left which cannot be completely expressed in terms of one-loop functions. The result is taken from Ref.~\cite{Berends:1994ed} and reordered in the given way. Up to~$\mathcal{O}{\left(\epsilon^{0}\right)}$ it reads:
{\allowdisplaybreaks\footnotesize
\begin{subequations}
\begin{align}
  \begin{split}\label{eq:TVacuumMaster}
    \mathrm{T}_{134}{\left(m_{1}^{2},m_{2}^{2},m_{3}^{2}\right)} &= \frac{1 - \epsilon}{2\left(1 - 2 \epsilon\right)} \left\{\frac{\left[\Anull{m_{1}^{2}}\right]^{2}}{m_{1}^{2}} + \frac{\left[\Anull{m_{2}^{2}}\right]^{2}}{m_{2}^{2}} + \frac{\left[\Anull{m_{3}^{2}}\right]^{2}}{m_{3}^{2}}\right\}\\
     &\quad + \Phi^{\text{cyc}}{\left(m_{1}^{2}, m_{2}^{2}, m_{3}^{2}\right)}\ ,
  \end{split}\\
  \Phi^{\text{cyc}}{\left(m^{2}, 0, 0\right)} &= m^{2} \frac{\pi^{2}}{6},\label{eq:TVacuumMasterPhiStart}\\
  \Phi^{\text{cyc}}{\left(m_{1}^{2}, m_{2}^{2}, 0\right)} &= m_{1}^{2} \dilog{\tfrac{m_{1}^{2} - m_{2}^{2}}{m_{1}^{2}}} + m_{2}^{2} \dilog{\tfrac{m_{2}^{2} - m_{1}^{2}}{m_{2}^{2}}}\ ,\\
  \begin{split}
    \Phi^{\text{cyc}}{\left(m_{1}^{2}, m_{2}^{2}, m_{3}^{2}\right)} &= 
      -\frac{m_{1}^{2}}{2} \log{\left(\tfrac{m_{1}^{2}}{m_{2}^{2}}\right)}\log{\left(\tfrac{m_{1}^{2}}{m_{3}^{2}}\right)}
      -\frac{m_{2}^{2}}{2} \log{\left(\tfrac{m_{2}^{2}}{m_{3}^{2}}\right)}\log{\left(\tfrac{m_{2}^{2}}{m_{1}^{2}}\right)}
      -\frac{m_{3}^{2}}{2} \log{\left(\tfrac{m_{3}^{2}}{m_{1}^{2}}\right)}\log{\left(\tfrac{m_{3}^{2}}{m_{2}^{2}}\right)}\\
      &\quad +
      \begin{aligned}[t]
        R\biggl[
        & \frac{\pi^{2}}{6} - \frac{1}{2}\log{\left(\tfrac{m_{1}^{2}}{m_{3}^{2}}\right)}\log{\left(\tfrac{m_{2}^{2}}{m_{3}^{2}}\right)} + \log{\left(\tfrac{m_{1}^{2} - m_{2}^{2} + m_{3}^{2} - R}{2m_{3}^{2}}\right)}\log{\left(\tfrac{m_{2}^{2} - m_{1}^{2} + m_{3}^{2} - R}{2m_{3}^{2}}\right)}\\
        & - \dilog{\tfrac{m_{1}^{2} - m_{2}^{2} + m_{3}^{2} - R}{2m_{3}^{2}}} - \dilog{\tfrac{m_{2}^{2} - m_{1}^{2} + m_{3}^{2} - R}{2m_{3}^{2}}}\biggr]\ ,
      \end{aligned}
  \end{split}\\
  R &= \sqrt{m_{1}^{4} + m_{2}^{4} + m_{3}^{4} - 2m_{1}^{2}m_{2}^{2} - 2m_{2}^{2}m_{3}^{2} - 2m_{3}^{2}m_{1}^{2}}\ .\label{eq:TVacuumMasterPhiEnd}
\end{align}
\end{subequations}}
The function $\Phi^{\text{cyc}}$ is cyclic in its arguments and contains only finite parts.

During the reduction to master integrals some terms can be expressed as products of one-loop integrals:
{\footnotesize
\begin{subequations}\label{eq:TOneLoop}
\begin{align}
  \mathrm{T}_{ab}{\left(m_{1}^{2},m_{2}^{2}\right)} &= \Anull{m_{1}^{2}}\Anull{m_{2}^{2}},\\
  \quad\notag\\
  \mathrm{T}_{a^{x}b^{y}}{\Big(\underbrace{m_{1}^{2},\dots,m_{1}^{2}}_{x},\underbrace{m_{2}^{2},\dots,m_{2}^{2}}_{y}\Big)} &= \frac{x - 3 + \epsilon}{\left(1 - x\right) m_{1}^{2}}\mathrm{T}_{a^{x - 1}b^{y}}{\Big(\underbrace{m_{1}^{2},\dots,m_{1}^{2}}_{x - 1},\underbrace{m_{2}^{2},\dots,m_{2}^{2}}_{y}\Big)}\\
  &\quad\ \text{for $x > 1,\, y \ge 1,$}\nonumber\\
  \quad\notag\\
  \mathrm{T}_{a^{x}b^{y}}{\Big(\underbrace{m_{1}^{2},\dots,m_{1}^{2}}_{x},\underbrace{m_{2}^{2},\dots,m_{2}^{2}}_{y}\Big)} &= \frac{y - 3 + \epsilon}{\left(1 - y\right) m_{2}^{2}}\mathrm{T}_{a^{x}b^{y - 1}}{\Big(\underbrace{m_{1}^{2},\dots,m_{1}^{2}}_{x},\underbrace{m_{2}^{2},\dots,m_{2}^{2}}_{y - 1}\Big)}\\
  &\quad\ \text{for $x \ge 1,\, y > 1$}\nonumber,
\end{align}
\end{subequations}
}%
with~\mbox{$a\neq b$} and~{$a,\,b \in \{1,\,3,\,4\}$}.

All other appearing integrals can be reduced to Eq.~\eqref{eq:TVacuumMaster} or Eq.~\eqref{eq:TOneLoop} by using the following formulas:
{\allowdisplaybreaks\footnotesize
\begin{subequations}
\begin{align}
  \begin{split}
  \mathrm{T}_{11334}{\left(m_{1}^{2},m_{1}^{2},m_{1}^{2},m_{1}^{2},0\right)} &= \frac{2}{m_{1}^{2}}\mathrm{T}_{1113}{\left(m_{1}^{2},m_{1}^{2},m_{1}^{2},m_{1}^{2}\right)} - 2\mathrm{T}_{11134}{\left(m_{1}^{2},m_{1}^{2},m_{1}^{2},m_{1}^{2},0\right)},
  \end{split}\\
  \begin{split}
    \mathrm{T}_{11334}{\left(m_{1}^{2},m_{1}^{2},m_{2}^{2},m_{2}^{2},m_{3}^{2}\right)} &=
    -\tfrac{\left(m_{1}^{2} - m_{2}^{2}\right)^{2} - m_{3}^{4}}{\left(m_{1}^{2} - m_{2}^{2}\right)^{2} + m_{3}^{4}}
    \begin{aligned}[t] \biggl[
        & \mathrm{T}_{11134}{\left(m_{1}^{2},m_{1}^{2},m_{1}^{2},m_{2}^{2},m_{3}^{2}\right)}\\
        &+ \mathrm{T}_{11134}{\left(m_{2}^{2},m_{2}^{2},m_{2}^{2},m_{1}^{2},m_{3}^{2}\right)}\biggr]
    \end{aligned}\\
    &\quad+\tfrac{m_{3}^{2}}{\left(m_{1}^{2} - m_{2}^{2}\right)^{2} + m_{3}^{4}}\\
    &\qquad\times\biggl[\mathrm{T}_{1134}{\left(m_{1}^{2},m_{1}^{2},m_{2}^{2},m_{3}^{2}\right)} + \mathrm{T}_{1134}{\left(m_{2}^{2},m_{2}^{2},m_{1}^{2},m_{3}^{2}\right)}\biggr]\\
    &\quad-\tfrac{m_{3}^{2}}{\left(m_{1}^{2} - m_{2}^{2}\right)^{2} + m_{3}^{4}} \mathrm{T}_{1133}{\left(m_{1}^{2},m_{1}^{2},m_{2}^{2},m_{2}^{2}\right)}\\
    &\quad+\tfrac{m_{1}^{4}\left(m_{1}^{2} - m_{2}^{2} + m_{3}^{2}\right) + m_{2}^{4}\left(m_{2}^{2} - m_{1}^{2} + m_{3}^{2}\right)}{2\left[\left(m_{1}^{2} - m_{2}^{2}\right)^{2} + m_{3}^{4}\right]}\\
    &\qquad\times\left[\frac{\mathrm{T}_{1113}{\left(m_{1}^{2},m_{1}^{2},m_{1}^{2},m_{2}^{2}\right)}}{m_{2}^{4}} + \frac{\mathrm{T}_{1113}{\left(m_{2}^{2},m_{2}^{2},m_{2}^{2},m_{1}^{2}\right)}}{m_{1}^{4}}\right]\\
    &\quad-\tfrac{m_{2}^{2} - m_{1}^{2} + m_{3}^{2}}{\left(m_{1}^{2} - m_{2}^{2}\right)^{2} + m_{3}^{4}} \mathrm{T}_{1114}{\left(m_{1}^{2},m_{1}^{2},m_{1}^{2},m_{3}^{2}\right)}\\
    &\quad-\tfrac{m_{1}^{2} - m_{2}^{2} + m_{3}^{2}}{\left(m_{1}^{2} - m_{2}^{2}\right)^{2} + m_{3}^{4}} \mathrm{T}_{1114}{\left(m_{2}^{2},m_{2}^{2},m_{2}^{2},m_{3}^{2}\right)},
  \end{split}\\
  \quad\notag\\
  \mathrm{T}_{11134}{\left(m_{1}^{2},m_{1}^{2},m_{1}^{2},m_{1}^{2},0\right)} &= -\frac{3}{1 + 2\epsilon}\mathrm{T}_{11113}{\left(m_{1}^{2},m_{1}^{2},m_{1}^{2},m_{1}^{2},m_{1}^{2}\right)},\\
  \begin{split}
    \mathrm{T}_{11134}{\left(m_{1}^{2},m_{1}^{2},m_{1}^{2},m_{2}^{2},m_{3}^{2}\right)} &=
    \tfrac{\left(m_{2}^{2} - m_{1}^{2} + m_{3}^{2}\right)\epsilon}{\left[\left(m_{1} + m_{2}\right)^{2} - m_{3}^{2}\right]\left[\left(m_{1} - m_{2}\right)^{2} - m_{3}^{2}\right]} \mathrm{T}_{1134}{\left(m_{1}^{2},m_{1}^{2},m_{2}^{2},m_{3}^{2}\right)}\\
    &\quad+\tfrac{m_{2}^{2}}{\left[\left(m_{1} + m_{2}\right)^{2} - m_{3}^{2}\right]\left[\left(m_{1} - m_{2}\right)^{2} - m_{3}^{2}\right]}\\
    &\qquad\times\biggl[\mathrm{T}_{1134}{\left(m_{1}^{2},m_{1}^{2},m_{2}^{2},m_{3}^{2}\right)} + \mathrm{T}_{1134}{\left(m_{2}^{2},m_{2}^{2},m_{1}^{2},m_{3}^{2}\right)}\biggr]\\
    &\quad-\tfrac{m_{1}^{2} - m_{2}^{2} + m_{3}^{2}}{\left[\left(m_{1} + m_{2}\right)^{2} - m_{3}^{2}\right]\left[\left(m_{1} - m_{2}\right)^{2} - m_{3}^{2}\right]} \mathrm{T}_{1113}{\left(m_{1}^{2},m_{1}^{2},m_{1}^{2},m_{2}^{2}\right)}\\
    &\quad-\tfrac{m_{1}^{2} + m_{2}^{2} - m_{3}^{2}}{\left[\left(m_{1} + m_{2}\right)^{2} - m_{3}^{2}\right]\left[\left(m_{1} - m_{2}\right)^{2} - m_{3}^{2}\right]} \mathrm{T}_{1114}{\left(m_{1}^{2},m_{1}^{2},m_{1}^{2},m_{3}^{2}\right)}\\
    &\quad-\tfrac{1}{\left[\left(m_{1} + m_{2}\right)^{2} - m_{3}^{2}\right]\left[\left(m_{1} - m_{2}\right)^{2} - m_{3}^{2}\right]} \mathrm{T}_{113}{\left(m_{1}^{2},m_{1}^{2},m_{2}^{2}\right)},
  \end{split}\\
  \quad\notag\\
  \mathrm{T}_{1134}{\left(m_{1}^{2},m_{1}^{2},m_{1}^{2},0\right)} &= \frac{1}{2m_{1}^{2}}\mathrm{T}_{113}{\left(m_{1}^{2},m_{1}^{2},m_{1}^{2}\right)},\\
  \begin{split}
    \mathrm{T}_{1134}{\left(m_{1}^{2},m_{1}^{2},m_{2}^{2},m_{3}^{2}\right)} &=
    \tfrac{\left(m_{2}^{2} - m_{1}^{2} + m_{3}^{2}\right)\left(-1 + 2\epsilon\right)}{\left[\left(m_{1} + m_{2}\right)^{2} - m_{3}^{2}\right]\left[\left(m_{1} - m_{2}\right)^{2} - m_{3}^{2}\right]} \mathrm{T}_{134}{\left(m_{1}^{2},m_{2}^{2},m_{3}^{2}\right)}\\
    &\quad-\tfrac{m_{1}^{2} - m_{2}^{2} + m_{3}^{2}}{\left[\left(m_{1} + m_{2}\right)^{2} - m_{3}^{2}\right]\left[\left(m_{1} - m_{2}\right)^{2} - m_{3}^{2}\right]} \mathrm{T}_{113}{\left(m_{1}^{2},m_{1}^{2},m_{2}^{2}\right)}\\
    &\quad-\tfrac{m_{1}^{2} + m_{2}^{2} - m_{3}^{2}}{\left[\left(m_{1} + m_{2}\right)^{2} - m_{3}^{2}\right]\left[\left(m_{1} - m_{2}\right)^{2} - m_{3}^{2}\right]} \mathrm{T}_{114}{\left(m_{1}^{2},m_{1}^{2},m_{3}^{2}\right)}\\
    &\quad+\tfrac{2 m_{2}^{2}}{\left[\left(m_{1} + m_{2}\right)^{2} - m_{3}^{2}\right]\left[\left(m_{1} - m_{2}\right)^{2} - m_{3}^{2}\right]} \mathrm{T}_{334}{\left(m_{2}^{2},m_{2}^{2},m_{3}^{2}\right)}.
  \end{split}
\end{align}
\end{subequations}
}%
Integrals with multiple denominators of the same loop-momentum structure and different masses are simplified by partial fractioning beforehand:
{\footnotesize
\begin{align}
  \mathrm{T}_{aa\dots}{\left(m_{1}^{2},m_{2}^{2},\dots\right)} &= \frac{1}{m_{1}^{2} - m_{2}^{2}}\left[\mathrm{T}_{a\dots}{\left(m_{1}^{2},\dots\right)} - \mathrm{T}_{a\dots}{\left(m_{2}^{2},\dots\right)}\right]\\
  &\quad\ \text{for $m_{1}^{2} \neq m_{2}^{2}$ and $a \in \{1,\,3,\,4\}$}.\nonumber
\end{align}
}%
All displayed integrals are symmetric under exchange of different loop-momentum structures:
{\footnotesize
\begin{align}
  \mathrm{T}_{a^{x}b^{y}\dots}{\Big(\underbrace{m_{1}^{2},\dots,m_{1}^{2}}_{x},\underbrace{m_{2}^{2},\dots,m_{2}^{2}}_{y},\dots\Big)} &=
  \mathrm{T}_{b^{x}a^{y}\dots}{\Big(\underbrace{m_{1}^{2},\dots,m_{1}^{2}}_{x},\underbrace{m_{2}^{2},\dots,m_{2}^{2}}_{y},\dots\Big)}\\
  &\quad\ \text{for $a,\,b \in \{1,\,3,\,4\}$}.\nonumber
\end{align}
}%

\section[Analytical \texorpdfstring{\boldmath{$\mathcal{O}{(\alpha_t^2)}$}}{order alpha top squared} results]{Analytical \boldmath{$\mathcal{O}{\scalebox{1.2}[1.3]{(}\alpha_t^2\scalebox{1.2}[1.3]{)}}$} results\label{app:fullresults}}

The analytical expressions for the~$\mathcal{O}{(\alpha_t^2)}$~contributions to the Higgs tadpoles and self-energies that are described in Section~\ref{sec:HiggsSect} are listed in the following.

\subsection{Symbols and abbreviations}

The following symbols and abbreviations are used to express the analytical results in a compact way. To shorten the notation the absolute-value bars of~\mbox{$\lvert X_t\rvert^2,\,\lvert Y_t\rvert^2$} and~$\lvert\mu\rvert^2$ are suppressed in the following terms:
{\footnotesize
\begin{subequations}
\begin{alignat}{4}
  \Delta_{a,b} &= m_{a}^{2} - m_{b}^{2}\ ,\\
  X_{t} &= A_{t}^{*} - \tfrac{\mu}{t_{\beta}}\ , &\quad Y_{t} &= A_{t}^{*} + \mu\,t_{\beta}\ ,\\
  X_{t}^{2} &\equiv \lvert X_t\rvert^2 = X_{t}\,X_{t}^{*}\ ,\quad x_{t}^{2} = \tfrac{X_{t}^{2}}{\Delta_{\tilde{t}_i\tilde{t}_j}}\ , &\quad Y_{t}^{2} &\equiv \lvert Y_t\rvert^2= Y_{t}\,Y_{t}^{*}\ ,\quad y_{t}^{2} = \tfrac{Y_{t}^{2}}{\Delta_{\tilde{t}_i\tilde{t}_j}}\ ,\\
  \mu^2 &\equiv \lvert\mu\rvert^2= \mu\,\mu^*\ , &\quad \eta &= \tfrac{\mu^2}{s_{\beta}^{2}c_{\beta}^{2}\Delta_{\tilde{t}_i\tilde{t}_j}} - x_{t}^{2} - y_{t}^{2}\ ,\\
  U_- &= \mathbf{U}_{\tilde{t}\,1i}\,\mathbf{U}_{\tilde{t}\,1i}^{*} - \mathbf{U}_{\tilde{t}\,1j}\,\mathbf{U}_{\tilde{t}\,1j}^{*}\ , &\quad U_{\times} &= \tfrac{1 - U_-^{2}}{4} = \tfrac{m_{t}^{2}}{\Delta_{\tilde{t}_i\tilde{t}_j}}\,x_{t}^{2}\ ,\\
  h_t &= \tfrac{e m_t}{\sqrt{2}s_{\beta}s_{\text{w}}M_W}\ , &\quad \delta^{(1)}h_t &= h_{t} \left(\tfrac{\delta^{(1)}m_{t}}{m_{t}} - \tfrac{\delta^{(1)}M_{W}}{M_{W}} - \tfrac{\delta^{(1)}s_{\text{w}}}{s_{\text{w}}} - \tfrac{\delta^{(1)}s_{\beta}}{s_{\beta}}\right)\ ,\\
  \tfrac{\delta^{(1)}x_{t}^{2}}{x_{t}^{2}} &= \tfrac{\delta^{(1)}X_{t}}{X_{t}} + \tfrac{\delta^{(1)}X_{t}^{*}}{X_{t}^{*}}\ , &\quad \delta^{(1)}\phi_{X} &= -\tfrac{\I}{2}\left(\tfrac{\delta^{(1)}X_{t}}{X_{t}} - \tfrac{\delta^{(1)}X_{t}^{*}}{X_{t}^{*}}\right)\ ,\\
  \delta^{(1)}X_{t} &= \delta^{(1)}A_{t}^{*} - \tfrac{\delta^{(1)}\mu}{t_{\beta}} + \tfrac{\mu\,\delta^{(1)}t_{\beta}}{t_{\beta}^2}\ , &\quad \delta^{(1)}X_{t}^{*} &= \delta^{(1)}A_{t} - \tfrac{\delta^{(1)}\mu^{*}}{t_{\beta}} + \tfrac{\mu^*\,\delta^{(1)}t_{\beta}}{t_{\beta}^2}\ .
\end{alignat}
\end{subequations}}%

\subsection{Genuine two-loop self-energies\label{sec:self2L}}

The explicit expressions of the genuine two-loop integrals contributing to the Higgs-boson self-energies are depicted in the following.
{\allowdisplaybreaks\footnotesize
\begin{subequations}
\begin{align}
  \begin{split}
  \Sigma^{(2)\,\text{gen}}_{hh} &= \frac{N_{c} s_{\beta}^{2} h_{t}^{4}}{256 \pi^{4}}\Big\{s_{A} + 4 m_{t}^{2} s_{B}\Big\}\\
  &\quad + \sum_{\ontop{\,i\;=\;1}{j\;\neq\;i}}^{2}\frac{N_{c} s_{\beta}^{2} h_{t}^{4}}{256 \pi^{4}}
  \begin{aligned}[t]\Big\{
    & -x_{t}^{2} \left(1 - 12 U_{\times}\right) s_{D} - 2 m_{t}^{2} \left(1 + x_{t}^{2}\right)^{2} s_{E} + 4 m_{t}^{2} \left(1 + x_{t}^{2}\right) s_{F}\\
    & - \left(1 - 16 U_{\times}\right) x_{t}^{2} s_{G} + \left(1 - 4 U_{\times}\right) x_{t}^{2} s_{H} + \left(1 + x_{t}^{2}\left(1 - 4 U_{\times}\right)\right) s_{I_{1}}\\
    & + \left(1 + x_{t}^{2} + 4 U_{\times}\right) s_{I_{2}} + 2 \left(m_{t}^{2} - U_{\times} X_{t}^{2}\right) s_{J} + \tfrac{1}{2} s_{K_{1}}
  \Big\},\end{aligned}
  \end{split}\\
  \begin{split}
  \Sigma^{(2)\,\text{gen}}_{HH} &= \frac{N_{c} c_{\beta}^{2} h_{t}^{4}}{256 \pi^{4}}\Big\{s_{A} + 4 m_{t}^{2} s_{B}\Big\}\\
  &\quad + \sum_{\ontop{\,i\;=\;1}{j\;\neq\;i}}^{2}\frac{N_{c} c_{\beta}^{2} h_{t}^{4}}{256 \pi^{4}}
  \begin{aligned}[t]\bigg\{
    & \tfrac{3 m_{t}^{2} \eta^{2} - Y_{t}^{2}}{\Delta_{\tilde{t}_i\tilde{t}_j}} s_{D}
      - \tfrac{m_{t}^{2}}{2} \left(\eta - 2\right)^{2} s_{E} - 2 m_{t}^{2} \left(\eta - 2\right) s_{F} + \tfrac{4 m_{t}^{2} \eta^{2} - Y_{t}^{2}}{\Delta_{\tilde{t}_i\tilde{t}_j}} s_{G}\\
    & - \tfrac{\left(m_{t}^{2} \eta^{2} - Y_{t}^{2}\right)}{\Delta_{\tilde{t}_i\tilde{t}_j}} s_{H} - \left(\tfrac{m_{t}^{2}\eta^{2} - Y_{t}^{2}}{\Delta_{\tilde{t}_i\tilde{t}_j}} - 1\right) s_{I_{1}}
      - \left(\tfrac{2 m_{t}^{2} \eta - Y_{t}^{2}}{\Delta_{\tilde{t}_i\tilde{t}_j}} - 1\right) s_{I_{2}}\\
    & - m_{t}^{2} \left(\tfrac{\eta^{2}}{2} - 2\right) s_{J} + \tfrac{1}{2} s_{L_{1}}
  \bigg\},\end{aligned}
  \end{split}\\
  \begin{split}
  \Sigma^{(2)\,\text{gen}}_{hH} &= -\frac{N_{c} s_{\beta}c_{\beta} h_{t}^{4}}{256 \pi^{4}}\Big\{s_{A} + 4 m_{t}^{2} s_{B}\Big\}\\
  &\quad - \sum_{\ontop{\,i\;=\;1}{j\;\neq\;i}}^{2}\frac{N_{c} s_{\beta}c_{\beta} h_{t}^{4}}{256 \pi^{4}}
  \begin{aligned}[t]\bigg\{
    & - \tfrac{\eta}{2} \left(1 - 12 U_{\times}\right) s_{D} + m_{t}^{2} \left(\eta - 2\right) \left(1 + x_{t}^{2}\right) s_{E}\\
    & - 2 m_{t}^{2} \left[\tfrac{\eta}{2} - 1 - \left(1 + x_{t}^{2}\right)\right] s_{F} + \left(1 - 16 U_{\times}\right) \tfrac{\eta}{2} s_{G} + \left(1 - 4 U_{\times}\right) \tfrac{\eta}{2} s_{H}\\
    & + \left[1 - \tfrac{\eta}{2} \left(1 - 4 U_{\times}\right)\right] s_{I_{1}} + \left[1 + 2 U_{\times} - \tfrac{\eta}{2}\left(1 + \tfrac{2 m_{t}^{2}}{\Delta_{\tilde{t}_i\tilde{t}_j}}\right)\right] s_{I_{2}}\\
    & + m_{t}^{2} \left(\eta x_{t}^{2} + 2\right) s_{J} + \tfrac{1}{2} s_{K_{1}} + \tfrac{1}{2} \left(\tfrac{\eta}{2} + x_{t}^{2}\right) s_{K_{2}} - s_{K_{3}}
  \bigg\},\end{aligned}
  \end{split}\\
  \begin{split}
  \Sigma^{(2)\,\text{gen}}_{AA} &= \frac{N_{c} c_{\beta}^{2} h_{t}^{4}}{256 \pi^{4}}s_{A}\\
  &\quad + \sum_{\ontop{\,i\;=\;1}{j\;\neq\;i}}^{2}\frac{N_{c} c_{\beta}^{2} h_{t}^{4}}{256 \pi^{4}}
  \begin{aligned}[t]\bigg\{
    & - \tfrac{3 m_{t}^{2} \eta^{2} + \left(1 - 12 U_{\times}\right) Y_{t}^{2}}{\Delta_{\tilde{t}_i\tilde{t}_j}} s_{D}
      + \tfrac{m_{t}^{2}}{2} \left(\eta^{2} - 4 x_{t}^{2} y_{t}^{2}\right) s_{E} - \tfrac{4 m_{t}^{2} \eta^{2} + \left(1 - 16 U_{\times}\right) Y_{t}^{2}}{\Delta_{\tilde{t}_i\tilde{t}_j}} s_{G}\\
    & + \tfrac{m_{t}^{2} \eta^{2} + \left(1 - 4 U_{\times}\right) Y_{t}^{2}}{\Delta_{\tilde{t}_i\tilde{t}_j}} s_{H} + \left[1 + \tfrac{m_{t}^{2}\eta^{2}}{\Delta_{\tilde{t}_i\tilde{t}_j}} + \left(1 - 4 U_{\times}\right) y_{t}^{2}\right] s_{I_{1}}\\
    & + \left(1 + y_{t}^{2}\right) s_{I_{2}} + \left(\tfrac{m_{t}^{2}\eta^{2}}{2} - 2 U_{\times} Y_{t}^{2}\right) s_{J} + \tfrac{1}{2} s_{L_{1}} + \eta s_{L_{2}} - s_{L_{3}}
  \bigg\},\end{aligned}
  \end{split}\\
  \begin{split}
  \Sigma^{(2)\,\text{gen}}_{hA} &= \sum_{\ontop{\,i\;=\;1}{j\;\neq\;i}}^{2}\frac{N_{c} h_{t}^{4}}{256 \pi^{4}}\frac{\Imag{X_{t}\mu^{*}}}{s_{\beta}c_{\beta}\Delta_{\tilde{t}_i\tilde{t}_j}}
  \begin{aligned}[t]\bigg\{
    & \left(1 - 12 U_{\times}\right) s_{D} + 2 m_{t}^{2} \left(1 + x_{t}^{2}\right) s_{E} - 2 m_{t}^{2} s_{F} + \left(1 - 16 U_{\times}\right) s_{G}\\
    & - \left(1 - 4 U_{\times}\right)\left(s_{H} + s_{I_{1}}\right) - \left(1 + \tfrac{2 m_{t}^{2}}{\Delta_{\tilde{t}_i\tilde{t}_j}}\right) s_{I_{2}} + 2 m_{t}^{2} x_{t}^{2} s_{J} + \tfrac{1}{2} s_{K_{2}}
  \bigg\},\hspace{-1em}\end{aligned}
  \end{split}\\
  \begin{split}
  \Sigma^{(2)\,\text{gen}}_{HA} &= \sum_{\ontop{\,i\;=\;1}{j\;\neq\;i}}^{2}\frac{N_{c} c_{\beta} h_{t}^{4}}{256 \pi^{4} s_{\beta}}\frac{\Imag{X_{t}\mu^{*}}}{s_{\beta}c_{\beta}\Delta_{\tilde{t}_i\tilde{t}_j}}
  \begin{aligned}[t]\bigg\{
    & -\tfrac{6 m_{t}^{2} \eta}{\Delta_{\tilde{t}_i\tilde{t}_j}} s_{D} + m_{t}^{2} \left(\eta - 2\right) s_{E} + 2 m_{t}^{2} s_{F}
      - \tfrac{8 m_{t}^{2} \eta}{\Delta_{\tilde{t}_i\tilde{t}_j}} s_{G}\\
    & + \tfrac{2 m_{t}^{2} \eta}{\Delta_{\tilde{t}_i\tilde{t}_j}} s_{H} + \tfrac{2 m_{t}^{2} \eta}{\Delta_{\tilde{t}_i\tilde{t}_j}} s_{I_{1}} + \tfrac{2 m_{t}^{2}}{\Delta_{\tilde{t}_i\tilde{t}_j}} s_{I_{2}} + m_{t}^{2} \eta s_{J} + s_{L_{2}}
  \bigg\},\end{aligned}
  \end{split}\\
  \begin{split}
  \Sigma^{(2)\,\text{gen}}_{H^{\pm}H^{\pm}} &= \frac{N_{c} c_{\beta}^{2} h_{t}^{4}}{256 \pi^{4}}\Big\{s_{A} + \tfrac{\mu^{2}}{s_{\beta}^{2}c_{\beta}^{2}}\tfrac{m_t^2}{\Delta_{\tilde{t}_i\tilde{b}_1}\Delta_{\tilde{t}_j\tilde{b}_1}} s_{C}\Big\}\\
  &\quad + \sum_{\ontop{\,i\;=\;1}{j\;\neq\;i}}^{2}\frac{N_{c} c_{\beta}^{2} h_{t}^{4}}{256 \pi^{4}}
  \begin{aligned}[t]\bigg\{
    & \tfrac{\Delta_{\tilde{t}_i\tilde{t}_j}}{2 \Delta_{\tilde{t}_i\tilde{b}_1}} \left[\left(y_t^2 + 1\right) \left(\tfrac{2 m_t^2}{\Delta_{\tilde{t}_i\tilde{t}_j}} + 1 - U_-\right) - \tfrac{2 \mu^2 m_t^2}{c_{\beta}^2s_{\beta}^2\Delta_{\tilde{t}_i\tilde{t}_j}^{2}}\right] \left(s_{I_{1}} + s_{I_{2}}\right)\\
    & + \tfrac{m_{t}^{2}\mu^{2}\Delta_{\tilde{t}_i\tilde{t}_j}}{2s_{\beta}^{2}c_{\beta}^{2}\Delta_{\tilde{t}_i\tilde{b}_1}\Delta_{\tilde{t}_j\tilde{b}_1}} s_{M} + s_{N} + \tfrac{\Delta_{\tilde{t}_i\tilde{t}_j}}{\Delta_{\tilde{t}_i\tilde{b}_1}} s_{O} \bigg\}.\end{aligned}
  \end{split}
\end{align}
\end{subequations}
%
%
%
%
\begin{subequations}
\begin{align}
  \begin{split}
  s_{A} &= \mathrm{T}_{134}{\left(m_t^2,m_{\tilde{b}_1}^2,\mu^2\right)} - 3\mathrm{T}_{113}{\left(m_t^2,m_t^2,\mu^2\right)} - 2 \mathrm{T}_{113}{\left(m_t^2,m_t^2,m_t^2\right)} + \mathrm{T}_{113}{\left(m_t^2,m_t^2,m_{\tilde{b}_1}^2\right)}\hspace{-1em}\\
  &\quad+ \left(m_t^2 - m_{\tilde{b}_1}^2 + \mu^2\right) \mathrm{T}_{1134}{\left(m_t^2,m_t^2,m_{\tilde{b}_1}^2,\mu^2\right)} + \sum_{i=1}^{2}\mathrm{T}_{113}{\left(m_t^2,m_t^2,m_{\tilde{t}_i}^2\right)}\\
  &\quad + c_{\beta}^{2} s_{A}^{H} + s_{\beta}^{2} \left.s_{A}^{H}\right|_{m_{H^{\pm}} \rightarrow 0,\ Y_{t}^{2} \rightarrow X_{t}^{2},\ y_{t}^{2} \rightarrow x_{t}^{2},\ \eta \rightarrow -2 x_{t}^{2}}\ ,
  \end{split}\\
  \begin{split}
  s_{A}^{H} &= 3 \mathrm{T}_{113}{\left(m_t^2,m_t^2,m_{H^{\pm }}^2\right)} + \mathrm{T}_{134}{\left(m_t^2,0,m_{H^{\pm }}^2\right)} + 2 \mathrm{T}_{134}{\left(m_t^2,m_t^2,m_{H^{\pm}}^2\right)}\\
  &\quad + \left(m_t^2 - m_{H^{\pm }}^2\right) \mathrm{T}_{1134}{\left(m_t^2,m_t^2,0,m_{H^{\pm }}^2\right)} + 2 \left(2 m_t^2 - m_{H^{\pm }}^2\right) \mathrm{T}_{1134}{\left(m_t^2,m_t^2,m_t^2,m_{H^{\pm }}^2\right)},
  \end{split}\\\nextParentEquation
  \begin{split}
  s_{B} &= \mathrm{T}_{1113}{\left(m_t^2,m_t^2,m_t^2,m_{\tilde{b}_1}^2\right)} - 3\mathrm{T}_{1113}{\left(m_t^2,m_t^2,m_t^2,\mu^2\right)} - 2 \mathrm{T}_{1113}{\left(m_t^2,m_t^2,m_t^2,m_t^2\right)}\\
  &\quad- \mathrm{T}_{1133}{\left(m_t^2,m_t^2,m_t^2,m_t^2\right)} + \mathrm{T}_{1134}{\left(m_t^2,m_t^2,m_{\tilde{b}_1}^2,\mu^2\right)}\\
  &\quad+ \left(m_t^2 - m_{\tilde{b}_1}^2 + \mu^2\right) \mathrm{T}_{11134}{\left(m_t^2,m_t^2,m_t^2,m_{\tilde{b}_1}^2,\mu^2\right)}\\
  &\quad+ \sum_{i=1}^{2}\left[\left(m_t^2 - m_{\tilde{t}_i}^2 + \mu^2\right) \mathrm{T}_{11134}{\left(m_t^2,m_t^2,m_t^2,m_{\tilde{t}_i}^2,\mu ^2\right)} + \mathrm{T}_{1113}{\left(m_t^2,m_t^2,m_t^2,m_{\tilde{t}_i}^2\right)}\right]\\
  &\quad+ c_{\beta }^2 s_{B}^{H} + s_{\beta}^{2} \left.s_{B}^{H}\right|_{m_{H^{\pm}} \rightarrow 0,\ Y_{t}^{2} \rightarrow X_{t}^{2},\ y_{t}^{2} \rightarrow x_{t}^{2},\ \eta \rightarrow -2 x_{t}^{2}}\ ,
  \end{split}\\
  \begin{split}
  s_{B}^{H} &= 3 \mathrm{T}_{1113}{\left(m_t^2,m_t^2,m_t^2,m_{H^{\pm }}^2\right)} + \left(m_t^2 - m_{H^{\pm }}^2\right) \mathrm{T}_{11134}{\left(m_t^2,m_t^2,m_t^2,0,m_{H^{\pm }}^2\right)}\\
  &\quad+ \mathrm{T}_{1134}{\left(m_t^2,m_t^2,0,m_{H^{\pm }}^2\right)} + 4 \mathrm{T}_{1134}{\left(m_t^2,m_t^2,m_t^2,m_{H^{\pm }}^2\right)}\\
  &\quad+ \left(2 m_t^2 - m_{H^{\pm }}^2\right) \left(2 \mathrm{T}_{11134}{\left(m_t^2,m_t^2,m_t^2,m_t^2,m_{H^{\pm }}^2\right)} + \mathrm{T}_{11334}{\left(m_t^2,m_t^2,m_t^2,m_t^2,m_{H^{\pm}}^2\right)}\right),
  \end{split}\\\nextParentEquation
  \begin{split}
  s_{C} &= -\mathrm{T}_{113}{\left(m_{\tilde{b}_1}^2,m_{\tilde{b}_1}^2,m_t^2\right)} - \mathrm{T}_{113}{\left(m_{\tilde{b}_1}^2,m_{\tilde{b}_1}^2,\mu^2\right)} + \mathrm{T}_{134}{\left(m_t^2,m_{\tilde{b}_1}^2,\mu ^2\right)}\\
  &\quad- \left(m_t^2 - m_{\tilde{b}_1}^2 + \mu ^2\right) \mathrm{T}_{1134}{\left(m_{\tilde{b}_1}^2,m_{\tilde{b}_1}^2,m_t^2,\mu ^2\right)}\\
  &\quad +\tfrac{m_t^2}{\Delta_{\tilde{t}_i\tilde{b}_1}\Delta_{\tilde{t}_j\tilde{b}_1}}
  \begin{aligned}[t]\Big[
    & \mathrm{T}_{13}{\left(m_{\tilde{b}_1}^2,\mu ^2\right)} - 4 \mathrm{T}_{13}{\left(m_{\tilde{b}_1}^2,m_{\tilde{b}_1}^2\right)} - \left(m_{\tilde{b}_1}^2 - \mu^2\right) \mathrm{T}_{134}{\left(m_{\tilde{b}_1}^2,\mu ^2,0\right)}
  \Big]\end{aligned}\\
  &\quad +c_{\beta }^2
  \begin{aligned}[t]\Big\{
    \mathrm{T}_{113}{\left(m_{\tilde{b}_1}^2,m_{\tilde{b}_1}^2,m_{H^{\pm }}^2\right)} - \tfrac{m_t^2}{\Delta_{\tilde{t}_i\tilde{b}_1}\Delta_{\tilde{t}_j\tilde{b}_1}}
    \left[\mathrm{T}_{13}{\left(m_{\tilde{b}_1}^2,m_{H^{\pm}}^2\right)} + \mu^2 \mathrm{T}_{134}{\left(m_{\tilde{b}_1}^2,m_{\tilde{b}_1}^2,m_{H^{\pm }}^2\right)}\right]
  \Big\},\end{aligned}
  \end{split}\\\nextParentEquation
  s_{D} &= \tfrac{U_-}{\Delta_{\tilde{t}_i\tilde{t}_j}}\left[\mathrm{T}_{13}{\left(m_{\tilde{t}_i}^2,m_{\tilde{b}_1}^2\right)} - \mathrm{T}_{13}{\left(m_{\tilde{t}_i}^2,\mu^2\right)} + c_{\beta}^2 \mathrm{T}_{13}{\left(m_{\tilde{t}_i}^2,m_{H^{\pm}}^2\right)}\right],\\\nextParentEquation
  \begin{split}
  s_{E} &= \left(1 - U_-\right) \mathrm{T}_{1113}{\left(m_{\tilde{t}_i}^2,m_{\tilde{t}_i}^2,m_{\tilde{t}_i}^2,m_{\tilde{b}_1}^2\right)} - 2\left(8 U_{\times} - 1\right) \mathrm{T}_{1113}{\left(m_{\tilde{t}_i}^2,m_{\tilde{t}_i}^2,m_{\tilde{t}_i}^2,m_{\tilde{t}_j}^2\right)}\\
  &\quad- 2 \left(m_t^2 - m_{\tilde{t}_i}^2 + \mu ^2\right) \mathrm{T}_{11134}{\left(m_{\tilde{t}_i}^2,m_{\tilde{t}_i}^2,m_{\tilde{t}_i}^2,m_t^2,\mu^2\right)} + \left(U_- - 3\right) \mathrm{T}_{1113}{\left(m_{\tilde{t}_i}^2,m_{\tilde{t}_i}^2,m_{\tilde{t}_i}^2,\mu^2\right)}\\
  &\quad+ \left(1 - U_-\right) \left(m_{\tilde{t}_i}^2 - \mu^2\right) \mathrm{T}_{11134}{\left(m_{\tilde{t}_i}^2,m_{\tilde{t}_i}^2,m_{\tilde{t}_i}^2,0,\mu^2\right)} + 16 U_{\times} \mathrm{T}_{1113}{\left(m_{\tilde{t}_i}^2,m_{\tilde{t}_i}^2,m_{\tilde{t}_i}^2,m_{\tilde{t}_i}^2\right)}\\
  &\quad+ 8 U_{\times} \mathrm{T}_{1133}{\left(m_{\tilde{t}_i}^2,m_{\tilde{t}_i}^2,m_{\tilde{t}_i}^2,m_{\tilde{t}_i}^2\right)} - 2 \mathrm{T}_{1113}{\left(m_{\tilde{t}_i}^2,m_{\tilde{t}_i}^2,m_{\tilde{t}_i}^2,m_t^2\right)}\\
  &\quad+ c_{\beta}^{2} s_{E}^{H} + s_{\beta}^{2} \left.s_{E}^{H}\right|_{m_{H^{\pm}} \rightarrow 0,\ Y_{t}^{2} \rightarrow X_{t}^{2},\ y_{t}^{2} \rightarrow x_{t}^{2},\ \eta \rightarrow -2 x_{t}^{2}}\ ,
  \end{split}\\
  \begin{split}
  s_{E}^{H} &= 2 m_t^2 \left(1 + x_t^2 y_t^2 - \eta\right) \left[2 \mathrm{T}_{11134}{\left(m_{\tilde{t}_i}^2,m_{\tilde{t}_i}^2,m_{\tilde{t}_i}^2,m_{\tilde{t}_i}^2,m_{H^{\pm }}^2\right)} + \mathrm{T}_{11334}{\left(m_{\tilde{t}_i}^2,m_{\tilde{t}_i}^2,m_{\tilde{t}_i}^2,m_{\tilde{t}_i}^2,m_{H^{\pm }}^2\right)}\right]\\
  &\quad + 2 Y_t^2 \left(1 - 2 U_{\times}\right) \mathrm{T}_{11134}{\left(m_{\tilde{t}_i}^2,m_{\tilde{t}_i}^2,m_{\tilde{t}_i}^2,m_{\tilde{t}_j}^2,m_{H^{\pm}}^2\right)} + \left(3 - U_-\right) \mathrm{T}_{1113}{\left(m_{\tilde{t}_i}^2,m_{\tilde{t}_i}^2,m_{\tilde{t}_i}^2,m_{H^{\pm }}^2\right)}\\
  &\quad + \left(m_t^2 \left(1 + U_- - 2 \eta\right) + \left(1 - U_-\right) Y_t^2\right) \mathrm{T}_{11134}{\left(m_{\tilde{t}_i}^2,m_{\tilde{t}_i}^2,m_{\tilde{t}_i}^2,m_{\tilde{b}_1}^2,m_{H^{\pm }}^2\right)},
  \end{split}\\\nextParentEquation
  s_{F} &= \left(m_t^2 - m_{\tilde{t}_i}^2 + \mu^2\right) \mathrm{T}_{11334}{\left(m_{\tilde{t}_i}^2,m_{\tilde{t}_i}^2,m_t^2,m_t^2,\mu^2\right)} + \mathrm{T}_{1133}{\left(m_{\tilde{t}_i}^2,m_{\tilde{t}_i}^2,m_t^2,m_t^2\right)},\\\nextParentEquation
  \begin{split}
  s_{G} &= \tfrac{1 - 4 U_{\times}}{\Delta_{\tilde{t}_i\tilde{t}_j}} \begin{aligned}[t]\Big[
      & 4 \mathrm{T}_{13}{\left(m_{\tilde{t}_i}^2,m_{\tilde{t}_i}^2\right)} - 4 \mathrm{T}_{13}{\left(m_{\tilde{t}_i}^2,m_{\tilde{t}_j}^2\right)}\\
      & - c_{\beta }^2 Y_t^2 \mathrm{T}_{134}{\left(m_{\tilde{t}_i}^2,m_{\tilde{t}_j}^2,m_{H^{\pm }}^2\right)} - s_{\beta }^2 X_t^2 \mathrm{T}_{134}{\left(m_{\tilde{t}_i}^2,m_{\tilde{t}_j}^2,0\right)}\Big],
  \end{aligned}\end{split}\\\nextParentEquation
  s_{H} &= -\mathrm{T}_{134}{\left(m_{\tilde{t}_i}^2,m_t^2,\mu ^2\right)},\\\nextParentEquation
  \begin{split}
  s_{I_{1}} &= -\tfrac{1}{2}\left(1 - 3 U_-\right) \mathrm{T}_{113}{\left(m_{\tilde{t}_i}^2,m_{\tilde{t}_i}^2,m_{\tilde{b}_1}^2\right)} + \left(m_t^2 - m_{\tilde{t}_j}^2 + \mu^2\right) \mathrm{T}_{1134}{\left(m_{\tilde{t}_i}^2,m_{\tilde{t}_i}^2,m_t^2,\mu ^2\right)}\\
  &\quad- \left(9 - 40 U_{\times}\right) \mathrm{T}_{113}{\left(m_{\tilde{t}_i}^2,m_{\tilde{t}_i}^2,m_{\tilde{t}_j}^2\right)} + \tfrac{3}{2}\left(1 - U_-\right) \mathrm{T}_{113}{\left(m_{\tilde{t}_i}^2,m_{\tilde{t}_i}^2,\mu^2\right)}\\
  &\quad+ 8\left(1 - 5 U_{\times}\right) \mathrm{T}_{113}{\left(m_{\tilde{t}_i}^2,m_{\tilde{t}_i}^2,m_{\tilde{t}_i}^2\right)} + \mathrm{T}_{113}{\left(m_{\tilde{t}_i}^2,m_{\tilde{t}_i}^2,m_t^2\right)}\\
  &\quad+ c_{\beta}^2 s_{I_{1}}^{H} + s_{\beta}^{2} \left.s_{I_{1}}^{H}\right|_{m_{H^{\pm}} \rightarrow 0,\ Y_{t}^{2} \rightarrow X_{t}^{2},\ y_{t}^{2} \rightarrow x_{t}^{2},\ \eta \rightarrow -2 x_{t}^{2}}\ ,
  \end{split}\\
  \begin{split}
  s_{I_{1}}^{H} &= -Y_t^2 \left(3 - 10 U_{\times}\right) \mathrm{T}_{1134}{\left(m_{\tilde{t}_i}^2,m_{\tilde{t}_i}^2,m_{\tilde{t}_j}^2,m_{H^{\pm }}^2\right)} - \tfrac{3}{2} \left(1 - U_-\right) \mathrm{T}_{113}{\left(m_{\tilde{t}_i}^2,m_{\tilde{t}_i}^2,m_{H^{\pm}}^2\right)},
  \end{split}\\\nextParentEquation
  \begin{split}
  s_{I_{2}} &= U_-\left[\mathrm{T}_{113}{\left(m_{\tilde{t}_i}^2,m_{\tilde{t}_i}^2,\mu^2\right)} - \mathrm{T}_{113}{\left(m_{\tilde{t}_i}^2,m_{\tilde{t}_i}^2,m_{\tilde{b}_1}^2\right)}\right] - \Delta_{\tilde{t}_i\tilde{t}_j} \mathrm{T}_{1134}{\left(m_{\tilde{t}_i}^2,m_{\tilde{t}_i}^2,m_t^2,\mu^2\right)}\\
  &\quad + 8\left(1 - 4 U_{\times}\right)\left[\mathrm{T}_{113}{\left(m_{\tilde{t}_i}^2,m_{\tilde{t}_i}^2,m_{\tilde{t}_j}^2\right)} - \mathrm{T}_{113}{\left(m_{\tilde{t}_i}^2,m_{\tilde{t}_i}^2,m_{\tilde{t}_i}^2\right)}\right]\\
  &\quad+ c_{\beta}^2 s_{I_{2}}^{H} + s_{\beta}^{2} \left.s_{I_{2}}^{H}\right|_{m_{H^{\pm}} \rightarrow 0,\ Y_{t}^{2} \rightarrow X_{t}^{2},\ y_{t}^{2} \rightarrow x_{t}^{2},\ \eta \rightarrow -2 x_{t}^{2}}\ ,
  \end{split}\\
  \begin{split}
  s_{I_{2}}^{H} &= 2 Y_t^2 \left(1 - 4 U_{\times}\right) \mathrm{T}_{1134}{\left(m_{\tilde{t}_i}^2,m_{\tilde{t}_i}^2,m_{\tilde{t}_j}^2,m_{H^{\pm }}^2\right)} - U_- \mathrm{T}_{113}{\left(m_{\tilde{t}_i}^2,m_{\tilde{t}_i}^2,m_{H^{\pm }}^2\right)},
  \end{split}\\\nextParentEquation
  \begin{split}
  s_{J} &= -\left(1 - 8 U_{\times }\right) \mathrm{T}_{1133}{\left(m_{\tilde{t}_i}^2,m_{\tilde{t}_i}^2,m_{\tilde{t}_j}^2,m_{\tilde{t}_j}^2\right)}
  - c_{\beta}^2 s_{J}^{H} - s_{\beta}^{2} \left.s_{J}^{H}\right|_{m_{H^{\pm}} \rightarrow 0,\ Y_{t}^{2} \rightarrow X_{t}^{2},\ y_{t}^{2} \rightarrow x_{t}^{2},\ \eta \rightarrow -2 x_{t}^{2}}\ ,
  \end{split}\\
  \begin{split}
  s_{J}^{H} &= Y_t^2 \left(1 - 2 U_{\times }\right) \mathrm{T}_{11334}{\left(m_{\tilde{t}_i}^2,m_{\tilde{t}_i}^2,m_{\tilde{t}_j}^2,m_{\tilde{t}_j}^2,m_{H^{\pm }}^2\right)},
  \end{split}\\\nextParentEquation
  \begin{split}
  s_{K_{1}} &= \left\{U_- \left[\left(x_t^2 \left(2\tfrac{\mu^{2} - m_{\tilde{t}_i}^2}{\Delta_{\tilde{t}_i\tilde{t}_j}} + 1\right) \left(1 - 12 U_{\times}\right)\right) + 1 - 8 U_{\times}\right] - x_t^2 \left(1 - 4 U_{\times}\right) - 1\right\} \mathrm{T}_{134}{\left(m_{\tilde{t}_i}^2,\mu^2,0\right)}\\
  &\quad+ \left[-8 U_{\times} \Delta_{\tilde{t}_i\tilde{t}_j} + 2 m_t^2 + 2 \left(\mu^{2} - m_{\tilde{t}_i}^2\right)\right] \mathrm{T}_{1134}{\left(m_t^2,m_t^2,m_{\tilde{t}_i}^2,\mu^2\right)}\\
  &\quad- \begin{aligned}[t]\Big\{
    & \left(\mu^{2} - m_{\tilde{t}_i}^2\right) \left[\left(1 - 4 U_{\times }\right) + 1 - U_- \left(x_t^2\left(1 - 12 U_{\times}\right) + 1 - 8 U_{\times}\right)\right]\\
    & + 4 \left(1 - U_-\right) m_t^2 \left(x_t^2 + 1\right)^2\Big\} \mathrm{T}_{1134}{\left(m_{\tilde{t}_i}^2,m_{\tilde{t}_i}^2,0,\mu^2\right)}\end{aligned}\\
  &\quad+ c_{\beta}^2 s_{K_{1}}^{H} + s_{\beta}^{2} \left.s_{K_{1}}^{H}\right|_{m_{H^{\pm}} \rightarrow 0,\ Y_{t}^{2} \rightarrow X_{t}^{2},\ y_{t}^{2} \rightarrow x_{t}^{2},\ \eta \rightarrow -2 x_{t}^{2}}\ ,
  \end{split}\\
  \begin{split}
  s_{K_{1}}^{H} &= \left[-1 + \left(2 \eta - U_-\right) \left(1 - 6 U_{\times}\right) \left(1 - 4 U_{\times}\right) - 2 U_- x_t^2 y_t^2 \left(1 - 12 U_{\times}\right)\right] \mathrm{T}_{134}{\left(m_{\tilde{t}_i}^2,m_{\tilde{b}_1}^2,m_{H^{\pm }}^2\right)}\\
  &\quad + 2 \left\{-1 + \left[1 - 4 U_{\times}\right] \left[\eta \left(1 - 6 U_{\times}\right) - x_t^2 y_t^2 \left(1 - 16 U_{\times}\right)\right]\right\} \mathrm{T}_{134}{\left(m_{\tilde{t}_i}^2,m_{\tilde{t}_i}^2,m_{H^{\pm }}^2\right)}\\
  &\quad + \begin{aligned}[t]\Big\{
    & \left(\eta - \tfrac{U_-}{2}\right) \left[\tfrac{2 m_t^2}{\Delta_{\tilde{t}_i\tilde{t}_j}} \left(5 - 8 U_{\times}\right) + 2 U_{\times} \left(5 - 12 U_{\times}\right)\right] + \left[1 - 4 U_{\times}\right] \left[1 - U_{\times} - x_t^2 y_t^2\right] - \tfrac{5 m_t^2}{\Delta_{\tilde{t}_i\tilde{t}_j}}\\
    & + U_- \left[x_t^2 y_t^2 \left(1 - 12 U_{\times}\right) + y_t^2 \left(1 - 8 U_{\times}\right)\right] - 1 - y_t^2\Big\} \Delta_{\tilde{t}_i\tilde{t}_j} \mathrm{T}_{1134}{\left(m_{\tilde{t}_i}^2,m_{\tilde{t}_i}^2,m_{\tilde{b}_1}^2,m_{H^{\pm }}^2\right)}\end{aligned}\\
  &\quad + 4 m_{t}^{2} \begin{aligned}[t]\Big\{
    & \eta  \left[x_t^2 \left(5 - 12 U_{\times}\right) + 5 - 8 U_{\times}\right] + x_t^2 y_t^2 \left[-\left(5 x_t^2 \left(1 - 4 U_{\times}\right) + 5 - 16 U_{\times}\right)\right]\\
    & - x_t^2 \left[5 - 4 U_{\times}\right] - 5\Big\} \mathrm{T}_{1134}{\left(m_{\tilde{t}_i}^2,m_{\tilde{t}_i}^2,m_{\tilde{t}_i}^2,m_{H^{\pm }}^2\right)},\end{aligned}
  \end{split}\\\nextParentEquation
%
  \begin{split}
  s_{K_{2}} &= 8 \left\{\left(\mu ^2 - m_{\tilde{t}_i}^2\right) \left(\tfrac{U_- m_t^2 \left(x_t^2+\frac{1}{2}\right)}{\Delta _{\tilde{t}_i\tilde{t}_j}} + \tfrac{\left(1 - U_-\right)\left(1 - 4 U_{\times}\right)}{8}\right) - \tfrac{m_t^2 \left(x_t^2 + 1\right)\left(1 - U_-\right)}{2}\right\} \mathrm{T}_{1134}{\left(m_{\tilde{t}_i}^2,m_{\tilde{t}_i}^2,\mu ^2,0\right)}\hspace{-3em}\\
  &\quad + \left\{-U_- \left[\left(1 - 12 U_{\times}\right) \left(2 \tfrac{\mu^{2} - m_{\tilde{t}_i}^2}{\Delta_{\tilde{t}_i\tilde{t}_j}} + 1\right) - \tfrac{4 m_t^2}{\Delta_{\tilde{t}_i\tilde{t}_j}}\right] + 1 - 4 U_{\times}\right\} \mathrm{T}_{134}{\left(m_{\tilde{t}_i}^2,\mu^2,0\right)}\\
  &\quad + 4 m_t^2 \mathrm{T}_{1134}{\left(m_t^2,m_t^2,m_{\tilde{t}_i}^2,\mu ^2\right)} + c_{\beta}^2 s_{K_{2}}^{H} + s_{\beta}^{2} \left.s_{K_{2}}^{H}\right|_{m_{H^{\pm}} \rightarrow 0,\ Y_{t}^{2} \rightarrow X_{t}^{2},\ y_{t}^{2} \rightarrow x_{t}^{2},\ \eta \rightarrow -2 x_{t}^{2}}\ ,
  \end{split}\\
  \begin{split}
  s_{K_{2}}^{H} &= \left\{2 \left[1 - 4 U_{\times}\right] \left[1 + \tfrac{3 m_t^2}{\Delta_{\tilde{t}_i\tilde{t}_j}} \left(2 \eta - U_- m_t^2\right)\right] + 2 U_- y_t^2 \left[1 - 12 U_{\times}\right]\right\} \mathrm{T}_{134}{\left(m_{\tilde{t}_i}^2,m_{\tilde{b}_1}^2,m_{H^{\pm }}^2\right)}\\
  &\quad + 2 \left(1 - 4 U_{\times}\right) \left[1 + \tfrac{6 \eta  m_t^2}{\Delta_{\tilde{t}_i\tilde{t}_j}} + y_t^2 \left(1 - 16 U_{\times}\right)\right] \mathrm{T}_{134}{\left(m_{\tilde{t}_i}^2,m_{\tilde{t}_i}^2,m_{H^{\pm }}^2\right)}\\
  &\quad + \begin{aligned}[t]\Big\{
    & m_t^2 \left[4 x_t^2 + 7 - 4 U_{\times}\right] + \left[\tfrac{4 m_t^2}{\Delta_{\tilde{t}_i\tilde{t}_j}} - 3 \left(1 - 4 U_{\times}\right)\right] \left[2 \eta m_t^2 + U_- \left(Y_t^2 - m_t^2\right)\right]\\
    & + Y_t^2 \left[2 U_- + 1 - 4 U_{\times}\right]\Big\} \mathrm{T}_{1134}{\left(m_{\tilde{t}_i}^2,m_{\tilde{t}_i}^2,m_{\tilde{b}_1}^2,m_{H^{\pm }}^2\right)}\end{aligned}\\
  &\quad + 4 m_{t}^{2} \left[\tfrac{4\eta m_t^2}{\Delta_{\tilde{t}_i\tilde{t}_j}} + 2 x_t^2 + 4 + \left(1 - 4 U_{\times}\right) \left(1 - 3 \eta + 2 y_t^2 +5 x_t^2 y_t^2\right)\right] \mathrm{T}_{1134}{\left(m_{\tilde{t}_i}^2,m_{\tilde{t}_i}^2,m_{\tilde{t}_i}^2,m_{H^{\pm }}^2\right)},\hspace{-2em}
  \end{split}\\\nextParentEquation
  \begin{split}
  s_{K_{3}} &= \tfrac{c_{\beta}^2\mu^2}{s_{\beta}^{2}c_{\beta}^{2}}
  \begin{aligned}[t] \Big\{
    & 2 \left[\tfrac{m_t^2}{\Delta_{\tilde{t}_i\tilde{t}_j}} + U_{\times}\right] \left[\mathrm{T}_{1134}{\left(m_{\tilde{t}_i}^2,m_{\tilde{t}_i}^2,m_{\tilde{b}_1}^2,m_{H^{\pm }}^2\right)} + 2 \mathrm{T}_{1134}{\left(m_{\tilde{t}_i}^2,m_{\tilde{t}_i}^2,m_{\tilde{t}_i}^2,m_{H^{\pm }}^2\right)}\right]\\
    & + \tfrac{1 - 4 U_{\times}}{\Delta_{\tilde{t}_i\tilde{t}_j}}\left[\mathrm{T}_{134}{\left(m_{\tilde{t}_i}^2,m_{\tilde{b}_1}^2,m_{H^{\pm}}^2\right)} + \mathrm{T}_{134}{\left(m_{\tilde{t}_i}^2,m_{\tilde{t}_i}^2,m_{H^{\pm }}^2\right)}\right]\Big\},
  \end{aligned}\end{split}\\\nextParentEquation
  \begin{split}
  s_{L_{1}} &= \left\{\tfrac{2 U_- \left[Y_t^2 - 3 \eta^2 m_t^2\right] \left[\mu^2 - m_{\tilde{t}_i}^2\right]}{\Delta_{\tilde{t}_i\tilde{t}_j}^{2}} + \tfrac{\eta m_t^2}{\Delta_{\tilde{t}_i\tilde{t}_j}} \left[\left(1 - 3  U_-\right)\eta + 4 U_-\right] - \left[1 - U_-\right] \left[y_t^2 + 1\right]\right\} \mathrm{T}_{134}{\left(m_{\tilde{t}_i}^2,\mu ^2,0\right)}\\
  &\quad - \left\{\left[\mu^2 - m_{\tilde{t}_i}^2\right] \left[\tfrac{\eta m_t^2}{\Delta_{\tilde{t}_i\tilde{t}_j}} \left(\eta \left(1 - 3 U_-\right) + 4 U_-\right) - \left(1 - U_-\right) \left(y_t^2 + 1\right)\right] + \left[\eta - 2\right]^2 \left[1 - U_-\right] m_t^2\right\}\\
  &\qquad\times\mathrm{T}_{1134}{\left(m_{\tilde{t}_i}^2,m_{\tilde{t}_i}^2,\mu ^2,0\right)} + 2 \left[\mu^2 - m_{\tilde{t}_i}^2 + \left(2 \eta + 1\right) m_t^2\right] \mathrm{T}_{1134}{\left(m_t^2,m_t^2,m_{\tilde{t}_i}^2,\mu ^2\right)}\\
  &\quad+ c_{\beta}^2 \left.s_{L_{1}}^{H}\right|_{Y^{\prime 2}_{t} \rightarrow Y_{t}^{2},\ y^{\prime 2}_{t} \rightarrow y_{t}^{2},\ \eta^{\prime} \rightarrow \eta} + s_{\beta}^{2} \left.s_{L_{1}}^{H}\right|_{m_{H^{\pm}} \rightarrow 0,\ Y^{\prime 2}_{t} \rightarrow X_{t}^{2},\ y^{\prime 2}_{t} \rightarrow x_{t}^{2},\ \eta^{\prime} \rightarrow -2 x_{t}^{2}}\ ,
  \end{split}\\
  \begin{split}
  s_{L_{1}}^{H} &= \begin{aligned}[t]\bigg\{
    & \tfrac{4 \eta^2 m_t^2}{\Delta_{\tilde{t}_i\tilde{t}_j}} \left(\tfrac{3 m_t^2 \eta^{\prime}}{\Delta_{\tilde{t}_i\tilde{t}_j}} +1\right) - U_- \left[2 \tfrac{\left(m_t^2 - Y_t^{\prime 2}\right) \left(3 \eta^2 m_t^2 - Y_t^2\right)}{\Delta_{\tilde{t}_i\tilde{t}_j}^{2}} + \tfrac{4 \eta m_t^2}{\Delta_{\tilde{t}_i\tilde{t}_j}} + 1\right]\\
    &  - 2 \eta^{\prime} \left(\tfrac{2 m_t^2 y_t^2}{\Delta_{\tilde{t}_i\tilde{t}_j}} + 1\right) - 2 \eta \left(\tfrac{4 m_t^2 y_t^{\prime 2}}{\Delta_{\tilde{t}_i\tilde{t}_j}} - 1\right) - 4 y_t^{\prime 2} - 1\bigg\} \mathrm{T}_{134}{\left(m_{\tilde{t}_i}^2,m_{\tilde{b}_1}^2,m_{H^{\pm }}^2\right)}\end{aligned}\\
  &\quad+ 2 \begin{aligned}[t]\bigg\{
    & \tfrac{2 \eta^2 m_t^2}{\Delta_{\tilde{t}_i\tilde{t}_j}} \left(2 U_{\times} \left(4 x_t^2 + 3\right) - 2 y_t^{\prime 2} - 1\right) + \tfrac{2 m_t^2 \eta^{\prime} y_t^2}{\Delta_{\tilde{t}_i\tilde{t}_j}} + \left(y_t^2 + 1\right) y_t^{\prime 2} + 1\\
    & + \eta \left(\tfrac{4 m_t^2 y_t^{\prime 2}}{\Delta_{\tilde{t}_i\tilde{t}_j}} - 1\right) - x_t^2 \left(4 U_{\times} y_t^2 + 1\right)\bigg\} \mathrm{T}_{134}{\left(m_{\tilde{t}_i}^2,m_{\tilde{t}_i}^2,m_{H^{\pm }}^2\right)}\end{aligned}\\
  &\quad + \begin{aligned}[t]\bigg\{
    & m_t^2 \left\{\eta^2 \left[\left(1 - 3 U_-\right) y_t^{\prime 2} - 2\right] + 8 \left[\tfrac{\eta^{\prime}\left(y_t^2 + 1\right)}{4} + \tfrac{\eta}{2} - y_t^{\prime 2} + \tfrac{1 + U_-}{2} \left[\eta \left(y_t^{\prime 2} + \tfrac{1}{2}\right) - \tfrac{y_t^2}{4} - \tfrac{5}{4}\right]\right]\right\}\\
    & + \tfrac{\eta m_t^4 \left[\eta + \left(3 \eta - 4\right) \left(U_- - 2 \eta^{\prime}\right)\right]}{\Delta_{\tilde{t}_i\tilde{t}_j}} - Y_t^{\prime 2} \left(1 - U_-\right) \left(y_t^2 + 1\right)\bigg\} \mathrm{T}_{1134}{\left(m_{\tilde{t}_i}^2,m_{\tilde{t}_i}^2,m_{\tilde{b}_1}^2,m_{H^{\pm }}^2\right)}\end{aligned}\\
  &\quad + 4 m_t^2 \begin{aligned}[t]\bigg\{
    & \eta ^2 \left[\tfrac{m_t^2}{\Delta_{\tilde{t}_i\tilde{t}_j}} - \left(1 - 5 U_{\times }\right) y_t^{\prime 2} - 1\right] + \eta^{\prime} \left[\tfrac{m_t^2 \eta \left(4 - 3 \eta\right)}{\Delta_{\tilde{t}_i\tilde{t}_j}} + y_t^2 + 1\right] + 4 \eta \left[\left(1 - 2 U_{\times}\right) y_t^{\prime 2} + 1\right]\\
    & - y_t^{\prime 2} \left[x_t^2 \left(y_t^2 + 1\right) + 4\right] - y_t^2 - 5\bigg\} \mathrm{T}_{1134}{\left(m_{\tilde{t}_i}^2,m_{\tilde{t}_i}^2,m_{\tilde{t}_i}^2,m_{H^{\pm }}^2\right)},\end{aligned}
  \end{split}\\\nextParentEquation
  \begin{split}
  s_{L_{2}} &= \left\{\tfrac{\eta m_t^2}{\Delta_{\tilde{t}_i\tilde{t}_j}} \left[6 U_- \left(\tfrac{\mu^2 - m_{\tilde{t}_i}^2}{\Delta_{\tilde{t}_i\tilde{t}_j}} + \tfrac{1}{2}\right) - 1\right] - \tfrac{2 U_- m_t^2}{\Delta_{\tilde{t}_i\tilde{t}_j}}\right\} \mathrm{T}_{134}{\left(m_{\tilde{t}_i}^2,\mu ^2,0\right)} - 2 m_t^2 \mathrm{T}_{1134}{\left(m_t^2,m_t^2,m_{\tilde{t}_i}^2,\mu ^2\right)}\\
  &\quad + m_t^2 \left\{\left[\eta \left(1 - 3 U_-\right) + 2 U_-\right] \tfrac{\mu^2 - m_{\tilde{t}_i}^2}{\Delta_{\tilde{t}_i\tilde{t}_j}} + \left(\eta - 2\right) \left(1 - U_-\right)\right\} \mathrm{T}_{1134}{\left(m_{\tilde{t}_i}^2,m_{\tilde{t}_i}^2,\mu ^2,0\right)}\\
  &\quad+ c_{\beta}^2 \left.s_{L_{2}}^{H}\right|_{Y^{\prime 2}_{t} \rightarrow Y_{t}^{2},\ y^{\prime 2}_{t} \rightarrow y_{t}^{2},\ \eta^{\prime} \rightarrow \eta} + s_{\beta}^{2} \left.s_{L_{2}}^{H}\right|_{m_{H^{\pm}} \rightarrow 0,\ Y^{\prime 2}_{t} \rightarrow X_{t}^{2},\ y^{\prime 2}_{t} \rightarrow x_{t}^{2},\ \eta^{\prime} \rightarrow -2 x_{t}^{2}}\ ,
  \end{split}\\
  \begin{split}
  s_{L_{2}}^{H} &= m_t^2 \left\{\tfrac{\left(3 \eta - 2\right)\left[m_t^2 \left(2 \eta^{\prime} - U_-\right) + \left(1 + U_-\right) Y_t^{\prime 2}\right] - \eta\left(m_t^2 + 4 Y_t^{\prime 2}\right)}{\Delta_{\tilde{t}_i \tilde{t}_j}} - 2\eta - U_- - 3\right\} \mathrm{T}_{1134}{\left(m_{\tilde{t}_i}^2,m_{\tilde{t}_i}^2,m_{\tilde{b}_1}^2,m_{H^{\pm }}^2\right)}\\
  &\quad + 4 m_t^2 \left\{\tfrac{\left(3 \eta - 2\right) m_t^2 \eta^{\prime} - \eta\left[m_t^2 - \left(1 - 5 U_{\times}\right) Y_t^{\prime 2}\right]}{\Delta_{\tilde{t}_i \tilde{t}_j}} + \eta - 2 y_t^{\prime 2} \left(1 - 2 U_{\times}\right) - 2\right\} \mathrm{T}_{1134}{\left(m_{\tilde{t}_i}^2,m_{\tilde{t}_i}^2,m_{\tilde{t}_i}^2,m_{H^{\pm }}^2\right)}\\
  &\quad + \tfrac{2 m_t^2}{\Delta_{\tilde{t}_i \tilde{t}_j}}\left\{2 y_t^{\prime 2} + U_- - \eta \left[\tfrac{6 m_t^2 \eta^{\prime}}{\Delta_{\tilde{t}_i \tilde{t}_j}} - 3 U_- \tfrac{m_t^2 - Y_t^{\prime 2}}{\Delta _{\tilde{t}_i \tilde{t}_j}} + 2\right]\right\} \mathrm{T}_{134}{\left(m_{\tilde{t}_i}^2,m_{\tilde{b}_1}^2,m_{H^{\pm }}^2\right)}\\
  &\quad + \tfrac{4 m_t^2}{\Delta _{\tilde{t}_i\tilde{t}_j}} \left\{y_t^{\prime 2} - \eta \left[\tfrac{3 m_t^2 \eta^{\prime}}{\Delta_{\tilde{t}_i \tilde{t}_j}} + 2 \left(1 - 4 U_{\times}\right) y_t^{\prime 2} + 1\right]\right\} \mathrm{T}_{134}{\left(m_{\tilde{t}_i}^2,m_{\tilde{t}_i}^2,m_{H^{\pm }}^2\right)},
  \end{split}\\\nextParentEquation
  \begin{split}
  s_{L_{3}} &= 2 y_t^2 U_{\times} \left(6 U_- \left(\tfrac{\mu^2 - m_{\tilde{t}_i}^2}{\Delta_{\tilde{t}_i\tilde{t}_j}} + \tfrac{1}{2}\right) - 1\right) \mathrm{T}_{134}{\left(m_{\tilde{t}_i}^2,\mu ^2,0\right)}\\
  &\quad + 2 \left(\left(1 - 3 U_-\right) y_t^2 U_{\times} \left(\mu^2 - m_{\tilde{t}_i}^2\right) - \left(1 - U_-\right) \left(m_t^2 - Y_t^2 U_{\times}\right)\right) \mathrm{T}_{1134}{\left(m_{\tilde{t}_i}^2,m_{\tilde{t}_i}^2,\mu ^2,0\right)}\\
  &\quad+ c_{\beta}^2 \left.s_{L_{3}}^{H}\right|_{Y^{\prime 2}_{t} \rightarrow Y_{t}^{2},\ y^{\prime 2}_{t} \rightarrow y_{t}^{2},\ \eta^{\prime} \rightarrow \eta} + s_{\beta}^{2} \left.s_{L_{3}}^{H}\right|_{m_{H^{\pm}} \rightarrow 0,\ Y^{\prime 2}_{t} \rightarrow X_{t}^{2},\ y^{\prime 2}_{t} \rightarrow x_{t}^{2},\ \eta^{\prime} \rightarrow -2 x_{t}^{2}}\ ,
  \end{split}\\
  \begin{split}
  s_{L_{3}}^{H} &= - 4 y_t^2 U_{\times} \left\{\tfrac{6 m_t^2 \eta^{\prime}}{\Delta_{\tilde{t}_i\tilde{t}_j}} - 3 U_- \tfrac{m_t^2 - Y_t^{\prime 2}}{\Delta_{\tilde{t}_i \tilde{t}_j}} + 2\right\} \mathrm{T}_{134}{\left(m_{\tilde{t}_i}^2,m_{\tilde{b}_1}^2,m_{H^{\pm }}^2\right)}\\
  &\quad - 8 y_t^2 U_{\times} \left\{\tfrac{3 m_t^2 \eta^{\prime}}{\Delta_{\tilde{t}_i\tilde{t}_j}} + 2 \left(1 - 4 U_{\times}\right) y_t^{\prime 2} + 1\right\} \mathrm{T}_{134}{\left(m_{\tilde{t}_i}^2,m_{\tilde{t}_i}^2,m_{H^{\pm }}^2\right)}\\
  &\quad + 2 m_t^2 \left\{x_t^2 y_t^2 \left[\tfrac{6 m_t^2 \eta^{\prime} - \left(1 + 3 U_-\right) m_t^2 - \left(1 - 3 U_-\right) y_t^{\prime 2}}{\Delta_{\tilde{t}_i\tilde{t}_j}} + 2\right] - 2 y_t^{\prime} - U_- - 1\right\}\mathrm{T}_{1134}{\left(m_{\tilde{t}_i}^2,m_{\tilde{t}_i}^2,m_{\tilde{b}_1}^2,m_{H^{\pm }}^2\right)}\\
  &\quad + 8 m_t^2 \left\{x_t^2 y_t^2 \left[\tfrac{3 m_t^2 \eta^{\prime} - m_t^2 + \left(1 - 5 U_{\times}\right) Y_t^{\prime 2}}{\Delta _{\tilde{t}_i \tilde{t}_j}} + 1\right] - y_t^{\prime 2} - 1\right\} \mathrm{T}_{1134}{\left(m_{\tilde{t}_i}^2,m_{\tilde{t}_i}^2,m_{\tilde{t}_i}^2,m_{H^{\pm }}^2\right)},
  \end{split}\\\nextParentEquation
  \begin{split}
  s_{M} &= \tfrac{1 - U_-}{\Delta_{\tilde{t}_i\tilde{t}_j}} \mathrm{T}_{113}{\left(m_{\tilde{b}_1}^2,m_{\tilde{b}_1}^2,m_{\tilde{t}_i}^2\right)}
  + c_{\beta}^2 s_{M}^{H} + s_{\beta}^{2} \left.s_{M}^{H}\right|_{m_{H^{\pm}} \rightarrow 0,\ Y_{t}^{2} \rightarrow X_{t}^{2},\ y_{t}^{2} \rightarrow x_{t}^{2},\ \eta \rightarrow -2 x_{t}^{2}}\ ,
  \end{split}\\
  \begin{split}
  s_{M}^{H} &= \left(\tfrac{m_t^2}{\Delta_{\tilde{t}_i\tilde{t}_j}} \left(1 + U_- - 2 \eta\right) + \left(1 - U_-\right) y_t^2\right) \mathrm{T}_{1134}{\left(m_{\tilde{b}_1}^2,m_{\tilde{b}_1}^2,m_{\tilde{t}_i}^2,m_{H^{\pm }}^2\right)}.
  \end{split}\\\nextParentEquation
  \begin{split}
  s_{N} &= \tfrac{\Delta_{\tilde{t}_i\tilde{t}_j}}{2\Delta_{\tilde{t}_i\tilde{b}_1}} \left\{\tfrac{2\mu^2m_t^2}{c_{\beta}^2s_{\beta}^2\Delta_{\tilde{t}_i\tilde{t}_j}^2} + y_t^2 \left(U_- - 1 - \tfrac{2 m_t^2}{\Delta_{\tilde{t}_i\tilde{t}_j}}\right)\right\} \mathrm{T}_{134}{\left(m_t^2,m_{\tilde{t}_i}^2,\mu ^2\right)}\\
  &\quad+ \left(m_t^2 - m_{\tilde{t}_i}^2 + \mu^2\right) \mathrm{T}_{1134}{\left(m_t^2,m_t^2,m_{\tilde{t}_i}^2,\mu^2\right)},
  \end{split}\\\nextParentEquation
  \begin{split}
  s_{O} &= \tfrac{1}{\Delta_{\tilde{t}_i\tilde{b}_1}} \begin{aligned}[t] \Bigg\{
    & y_t^2 \left[-\tfrac{m_t^2}{\Delta_{\tilde{t}_i\tilde{t}_j}} \left(1 - U_- \left(\tfrac{m_t^2}{\Delta_{\tilde{t}_i\tilde{t}_j}} + 1\right)\right) + U_{\times} \left(\tfrac{4 m_t^2}{\Delta _{\tilde{t}_i \tilde{t}_j}} - U_- + 2\right) + \tfrac{U_- - 1}{2}\right]\\
    &+ \tfrac{\mu^2 m_t^2}{c_{\beta}^2s_{\beta}^2\Delta_{\tilde{t}_i\tilde{t}_j}^2} \left[1 - 2 U_{\times} - U_- \left(\tfrac{m_t^2}{\Delta_{\tilde{t}_i\tilde{t}_j}} + 1\right)\right]\Bigg\}\mathrm{T}_{13}{\left(m_{\tilde{t}_i}^2,\mu^2\right)}
  \end{aligned}\\
  &\quad+ \tfrac{4}{\Delta_{\tilde{t}_i\tilde{b}_1}} \begin{aligned}[t]\Bigg\{
    &\left(1 - 4 U_{\times}\right) \begin{aligned}[t] \bigg[
      & y_t^2 \left(\tfrac{m_t^2}{\Delta_{\tilde{t}_i\tilde{t}_j}} \left(\tfrac{X_t^2 - m_t^2}{\Delta_{\tilde{t}_i\tilde{t}_j}} + U_- - 1\right) + \tfrac{U_- - 1}{2}\right)\\
      &+ \tfrac{\mu^2 m_t^2}{c_{\beta}^2s_{\beta}^2\Delta_{\tilde{t}_i\tilde{t}_j}^2}\left(\tfrac{m_t^2}{\Delta_{\tilde{t}_i\tilde{t}_j}} + 1 - U_-\right)\bigg] - \tfrac{2 U_- U_{\times} \mu^2 m_t^2}{c_{\beta}^2s_{\beta}^2\Delta_{\tilde{t}_i\tilde{t}_j}^2}\Bigg\}\mathrm{T}_{13}{\left(m_{\tilde{t}_i}^2,m_{\tilde{t}_i}^2\right)}
  \end{aligned}\end{aligned}\\
  &\quad+\tfrac{4 m_t^4}{\Delta_{\tilde{t}_i\tilde{b}_1}\Delta_{\tilde{t}_j\tilde{b}_1}^2} \begin{aligned}[t] \Bigg\{
    & y_t^2 \left(1 - 4 U_{\times}\right) \left[\tfrac{m_t^2 - X_t^2}{\Delta_{\tilde{t}_i\tilde{t}_j}}\left(\tfrac{m_t^2 - X_t^2}{\Delta_{\tilde{t}_i\tilde{t}_j}} - 2 U_-\right) + 1 - 4 U_{\times}\right]\\
    &+ \tfrac{\mu^2}{c_{\beta}^2s_{\beta}^2\Delta_{\tilde{t}_i\tilde{t}_j}} \begin{aligned}[t] \bigg[
      & 2 \left(1 - x_t^2 \tfrac{3 m_t^2 - X_t^2}{\Delta_{\tilde{t}_i\tilde{t}_j}}\right) \left(\tfrac{m_t^2 \left(2 x_t^2 + U_-\right)}{\Delta_{\tilde{t}_i\tilde{t}_j}} - \tfrac{1}{2}\right)\\
      &+ \left(1 - 4 U_{\times}\right) \tfrac{X_t^4 - m_t^4}{\Delta_{\tilde{t}_i\tilde{t}_j}^2} + \tfrac{1}{8} \left(1 - 16 U_{\times}\right)\bigg]\Bigg\} \mathrm{T}_{13}{\left(m_{\tilde{t}_i}^2,m_{\tilde{t}_j}^2\right)}
  \end{aligned}\end{aligned}
  \end{split}\notag\\
  \begin{split}
  &\quad+\tfrac{m_t^4}{\Delta_{\tilde{t}_i\tilde{b}_1}\Delta_{\tilde{t}_j\tilde{b}_1}^2} \begin{aligned}[t] \Bigg\{
    & \tfrac{\mu^2}{c_{\beta}^2s_{\beta}^2\Delta_{\tilde{t}_i\tilde{t}_j}} \left[
      \tfrac{m_t^2}{\Delta_{\tilde{t}_i\tilde{t}_j}} \left(\left(2 x_t^2 + U_-\right) \left(\tfrac{m_t^2 - X_t^2}{\Delta_{\tilde{t}_i\tilde{t}_j}} - U_-\right) - 1\right) + \tfrac{7}{2} \left(1 - U_-\right)\right]\\
    &- y_t^2 \left[ U_- \left(\tfrac{X_t^2 - m_t^2}{\Delta_{\tilde{t}_i\tilde{t}_j}}\right)^2 + 2\left(1 - 4 U_{\times}\right) \left(\tfrac{X_t^2 - m_t^2}{\Delta_{\tilde{t}_i\tilde{t}_j}} + \tfrac{U_-}{2}\right)\right]\Bigg\} \mathrm{T}_{13}{\left(m_{\tilde{t}_i}^2,m_{\tilde{b}_1}^2\right)}
  \end{aligned}\\
  &\quad+\tfrac{\Delta_{\tilde{t}_i\tilde{t}_j}}{2 \Delta_{\tilde{t}_i\tilde{b}_1}} \begin{aligned}[t] \Bigg\{
    &\tfrac{m_{\tilde{t}_i}^2 - \mu^2}{\Delta_{\tilde{t}_i\tilde{t}_j}} \begin{aligned}[t] \bigg[
      & y_t^2 \left(1 - 4 U_{\times} - U_- \left(1 - \tfrac{2 m_t^2}{\Delta _{\tilde{t}_i \tilde{t}_j}} \left(\tfrac{X_t^2 - m_t^2}{\Delta_{\tilde{t}_i\tilde{t}_j}} + U_- - 1\right)\right)\right)\\
      &+ \tfrac{2 \mu^2 m_t^2}{c_{\beta}^2 s_{\beta}^2\Delta_{\tilde{t}_i\tilde{t}_j}^2} \left(U_- \left(\tfrac{m_t^2}{\Delta_{\tilde{t}_i\tilde{t}_j}} + 1\right) + 2 U_{\times} - 1\right)\bigg]
    \end{aligned}\\
    & + \left(y_t^2 + 1\right) \left[\left(U_- - 1\right) \left(\tfrac{m_t^2}{\Delta_{\tilde{t}_i\tilde{t}_j}} + 1\right)^2 + 4 U_{\times} \left(\tfrac{m_t^2}{\Delta_{\tilde{t}_i\tilde{t}_j}} + 1\right) - \left(U_- + 1\right) U_{\times}\right]\\
    & - \tfrac{\mu^2 m_t^2}{c_{\beta}^2s_{\beta}^2\Delta_{\tilde{t}_i\tilde{t}_j}^2} \left[\left(U_- - 1\right) \left(\tfrac{m_t^2}{\Delta_{\tilde{t}_i\tilde{t}_j}} + 1\right) + 2 U_{\times}\right]\Bigg\} \mathrm{T}_{134}{\left(m_{\tilde{t}_i}^2,\mu^2,0\right)}
  \end{aligned}\\
  &\quad+\tfrac{\mu^2 - m_{\tilde{t}_i}^2}{2} \left\{\tfrac{\mu^2 m_t^2 \left(U_- - 1\right)}{c_{\beta}^2s_{\beta}^2\Delta _{\tilde{t}_i \tilde{t}_j}^2} - \left(y_t^2 + 1\right) \left[\left(U_- - 1\right) \left(\tfrac{m_t^2}{\Delta_{\tilde{t}_i\tilde{t}_j}} + 1\right) + 2 U_{\times}\right]\right\} \mathrm{T}_{1134}{\left(m_{\tilde{t}_i}^2,m_{\tilde{t}_i}^2,\mu ^2,0\right)}\\
  &\quad+ c_{\beta}^2 \left.s_{O}^{H}\right|_{Y^{\prime 2}_{t} \rightarrow Y_{t}^{2},\ y^{\prime 2}_{t} \rightarrow y_{t}^{2},\ \mu^{\prime} \rightarrow \mu} + s_{\beta}^{2} \left.s_{O}^{H}\right|_{m_{H^{\pm}} \rightarrow 0,\ Y^{\prime 2}_{t} \rightarrow X_{t}^{2},\ y^{\prime 2}_{t} \rightarrow x_{t}^{2},\ \mu^{\prime} \rightarrow 0}\ ,
  \end{split}\\
  \begin{split}
  s_{O}^{H} &= \begin{aligned}[t] \Bigg\{
    & m_t^2 \left[1 - 2 U_{\times} - U_- \left(1 + \tfrac{m_t^2}{\Delta_{\tilde{t}_i\tilde{t}_j}}\right)\right] \left(y_t^2 - \tfrac{\mu^2}{c_{\beta}^2s_{\beta}^2 \Delta_{\tilde{t}_i \tilde{t}_j}}\right)\\
    &+ Y_t^2 \left[2 U_{\times} \left(-\tfrac{m_t^2}{\Delta_{\tilde{t}_i\tilde{t}_j}} + \tfrac{U_-}{2} - 1\right) + \frac{1}{2} \left(1 - U_-\right)\right]\Bigg\}\mathrm{T}_{13}{\left(m_{\tilde{t}_i}^2,m_{H^{\pm }}^2\right)}
    \end{aligned}\\
  &\quad- m_t^2 \begin{aligned}[t] \Bigg\{
    & \tfrac{2 m_t^2}{\Delta_{\tilde{t}_i\tilde{t}_j}} \begin{aligned}[t] \bigg[
      & \tfrac{\mu^2}{c_{\beta}^2s_{\beta}^2\Delta_{\tilde{t}_i\tilde{t}_j}} \left(\tfrac{\mu^{\prime 2}}{c_{\beta}^2s_{\beta}^2 \Delta_{\tilde{t}_i \tilde{t}_j}} - \left(x_t^2 + 1\right)^2\right) + \tfrac{\mu^{\prime 2}}{c_{\beta}^2s_{\beta}^2\Delta_{\tilde{t}_i\tilde{t}_j}} \left(x_t^4 + x_t^2 - 1 - \left(x_t^2 + 2\right) y_t^2\right)\\
      & + \left(x_t^2 + 1\right) \left(y_t^2 + 1\right) \left(y_t^{\prime 2} + 1\right)\bigg]\end{aligned}\\
    & +\left(U_- - 1\right) \left(y_t^2 + 1\right) \left[\tfrac{\mu^{\prime 2}}{c_{\beta}^2s_{\beta}^2\Delta_{\tilde{t}_i\tilde{t}_j}} - \left(x_t^2 + 1\right) \left(y_t^{\prime 2} + 1\right)\right]\Bigg\} \mathrm{T}_{1134}{\left(m_{\tilde{t}_i}^2,m_{\tilde{t}_i}^2,m_{\tilde{t}_i}^2,m_{H^{\pm }}^2\right)}
    \end{aligned}\\
  &\quad + \tfrac{\Delta_{\tilde{t}_i\tilde{t}_j}}{\Delta_{\tilde{t}_i\tilde{b}_1}} \begin{aligned}[t] \Bigg\{
    & \tfrac{U_- - 1}{2} \begin{aligned}[t] \bigg[
      & \tfrac{\mu^{\prime 2} - \mu^2}{c_{\beta}^2s_{\beta}^2\Delta_{\tilde{t}_i\tilde{t}_j}} + y_t^2 +y_t^{\prime 2} \left(\tfrac{2 m_t^2}{\Delta_{\tilde{t}_i\tilde{t}_j}}\left(1 - 2 y_t^2 \left(\tfrac{m_t^2}{\Delta_{\tilde{t}_i\tilde{t}_j}} + x_t^2 + 2 U_{\times}\right)\right) + y_t^2 + 1\right)\\
      &+ \tfrac{2 m_t^2}{\Delta _{\tilde{t}_i \tilde{t}_j}} \begin{aligned}[t] \bigg(
        &\tfrac{\mu^{\prime 2}}{c_{\beta}^2s_{\beta}^2\Delta_{\tilde{t}_i\tilde{t}_j}}\left(\left(-\tfrac{m_t^2}{\Delta_{\tilde{t}_i\tilde{t}_j}} + 1 - 2 U_{\times}\right) x_t^2 + \tfrac{m_t^2 \left(2 x_t^2 + 3\right) y_t^2}{\Delta_{\tilde{t}_i\tilde{t}_j}} + 1\right)\\
        &- \tfrac{\mu^2}{c_{\beta}^2s_{\beta}^2\Delta_{\tilde{t}_i\tilde{t}_j}}\left(\tfrac{m_t^2}{\Delta_{\tilde{t}_i\tilde{t}_j}}\left(\tfrac{\mu^{\prime 2}}{c_{\beta}^2s_{\beta}^2\Delta_{\tilde{t}_i\tilde{t}_j}} - 2 x_t^2 \left(x_t^2 + 1\right)\right) + x_t^2 + 2\right)\\
        &+ y_t^2 \left(1 - x_t^2 \left(\tfrac{2 m_t^2}{\Delta_{\tilde{t}_i\tilde{t}_j}} + 1\right)\right) + 1\bigg) + 1\bigg]
    \end{aligned}\end{aligned}\\
    &+ \tfrac{m_t^2}{\Delta_{\tilde{t}_i\tilde{t}_j}} \begin{aligned}[t] \bigg[
      &\tfrac{2 m_t^4}{\Delta _{\tilde{t}_i \tilde{t}_j}^2} \begin{aligned}[t] \bigg(
        &\tfrac{\mu^{\prime 2} \left(x_t^2\left(1 + 2 x_t^2\right) - 2 y_t^2 \left(1 + x_t^2\right)\right)}{c_{\beta}^2s_{\beta}^2\Delta_{\tilde{t}_i\tilde{t}_j}} + \tfrac{\mu^2}{c_{\beta}^2s_{\beta}^2\Delta_{\tilde{t}_i\tilde{t}_j}}\left(\tfrac{\mu^{\prime 2}}{c_{\beta}^2s_{\beta}^2\Delta_{\tilde{t}_i\tilde{t}_j}} - 2 x_t^2 \left(x_t^2 + 1\right)\right)\\
        &+ y_t^2 \left(2 x_t^2 y_t^{\prime 2} + y_t^{\prime 2} + x_t^2\right)\bigg)
        \end{aligned}\\
      &+ \tfrac{m_t^2}{\Delta _{\tilde{t}_i \tilde{t}_j}} \begin{aligned}[t] \bigg(
        &\tfrac{\mu^{\prime 2}\left(2 x_t^4 - 1\right)}{c_{\beta}^2s_{\beta}^2\Delta_{\tilde{t}_i\tilde{t}_j}} + \tfrac{\mu^2}{c_{\beta}^2s_{\beta}^2\Delta_{\tilde{t}_i\tilde{t}_j}}\left(\tfrac{\mu^{\prime 2}}{c_{\beta}^2s_{\beta}^2\Delta_{\tilde{t}_i\tilde{t}_j}} - 2 x_t^2 \left(x_t^2 + 1\right) + 1\right)\\
        &- y_t^{\prime 2} - y_t^2 \left(y_t^{\prime 2} + 2 x_t^2 \left(2 x_t^2 y_t^{\prime 2} + y_t^{\prime 2} + x_t^2\right) + 1\right) - 1\bigg)
        \end{aligned}\\
      &+ x_t^2 \left(\tfrac{\mu^{\prime 2} - \mu^2}{c_{\beta}^2s_{\beta}^2\Delta_{\tilde{t}_i\tilde{t}_j}} + y_t^{\prime 2} + y_t^{\prime 2} y_t^2 + y_t^2 + 1\right)\bigg]\Bigg\} \mathrm{T}_{134}{\left(m_{\tilde{t}_i}^2,m_{\tilde{t}_i}^2,m_{H^{\pm }}^2\right)}.
  \end{aligned}\end{aligned}\end{split}\notag\\
  \begin{split}
    &\quad + \tfrac{m_t^4\Delta_{\tilde{t}_i\tilde{t}_j}}{\Delta_{\tilde{t}_i\tilde{b}_1}\Delta_{\tilde{t}_j\tilde{b}_1}^2}\begin{aligned}[t] \Bigg\{
      & \tfrac{\mu^2\left(1-U_- y_t^{\prime 2}\right)}{c_{\beta}^2s_{\beta}^2\Delta_{\tilde{t}_i \tilde{t}_j}} - x_t^4\left[\left(U_- y_t^2 + 1\right) y_t^{\prime 2} + \tfrac{U_- + 1}{2} + \left(1 - 2 U_{\times}\right) \left(\tfrac{\mu^{\prime 2} - \mu^2}{c_{\beta}^2s_{\beta}^2\Delta_{\tilde{t}_i\tilde{t}_j}} + y_t^2\right)\right]\\
      &+ \tfrac{\mu^{\prime 2}\mu^2}{c_{\beta}^4s_{\beta}^4\Delta_{\tilde{t}_i\tilde{t}_j}^2}\left[\tfrac{U_- - 1}{2} + U_- U_{\times}\right] - \left(1 - 6 U_{\times}\right) \left[\tfrac{\mu^{\prime 2} \left(1 - 2 U_{\times}\right) + \mu^2}{c_{\beta}^2s_{\beta}^2\Delta_{\tilde{t}_i\tilde{t}_j}} + y_t^{\prime 2} + \tfrac{U_- + 1}{2}\right]\\
      &+ y_t^2 \left[\tfrac{\mu^{\prime 2}\left(1 - 8 U_- U_{\times}\right)}{c_{\beta}^2s_{\beta}^2\Delta_{\tilde{t}_i\tilde{t}_j}} - U_- \left(1 - 10 U_{\times}\right) y_t^{\prime 2} - 2 U_{\times}\right]\\
      &- x_t^2 \begin{aligned}[t] \bigg[
          & \tfrac{\mu^{\prime 2}\left(4 U_{\times} y_t^2 + U_- U_{\times} + 1\right)}{c_{\beta}^2s_{\beta}^2\Delta_{\tilde{t}_i\tilde{t}_j}} - \tfrac{2 \mu^2}{c_{\beta }^2s_{\beta}^2\Delta_{\tilde{t}_i\tilde{t}_j}} - 4 U_{\times} + 2 y_t^{\prime 2} \left(y_t^2\left(1 - 5 U_{\times}\right) + U_-\right)\\
          &+ \tfrac{U_- + 1}{2} \left(\tfrac{\mu^2}{c_{\beta}^2s_{\beta}^2\Delta_{\tilde{t}_i\tilde{t}_j}} + 2\right) + U_- \left(2 - 5 U_{\times}\right) \left(\tfrac{\mu^{\prime 2} - \mu^2}{c_{\beta}^2s_{\beta}^2\Delta_{\tilde{t}_i\tilde{t}_j}} + y_t^2\right)\bigg]
      \end{aligned}\\
      &+ \tfrac{m_t^2}{\Delta _{\tilde{t}_i \tilde{t}_j}} \begin{aligned}[t] \bigg[
          &-\tfrac{\mu^{\prime 2} 4 \left(1 - 6 U_{\times}\right) y_t^2}{c_{\beta}^2s_{\beta}^2\Delta_{\tilde{t}_i\tilde{t}_j}} - \tfrac{\mu ^2}{2 c_{\beta }^2 s_{\beta }^2 \Delta _{\tilde{t}_i \tilde{t}_j}}\left(\tfrac{\mu^{\prime 2} 12 U_{\times }}{c_{\beta}^2s_{\beta}^2\Delta_{\tilde{t}_i\tilde{t}_j}} + U_-\left(3 - 20 U_{\times}\right) + 1\right)\\
          &+ U_- \left(-\tfrac{4 U_{\times} \mu^{\prime 2}}{c_{\beta}^2s_{\beta}^2\Delta_{\tilde{t}_i\tilde{t}_j}} + \tfrac{2 \mu^{\prime 2}}{c_{\beta}^2s_{\beta}^2\Delta_{\tilde{t}_i\tilde{t}_j}} + 2 y_t^{\prime 2} - 10 U_{\times} y_t^2 + 3 y_t^2 + 1\right)\\
          &+ 4 \left(1 - 5 U_{\times}\right) y_t^{\prime 2} y_t^2 - 4 U_{\times} + 1\bigg]
      \end{aligned}\\
      &+ \tfrac{m_t^4}{\Delta _{\tilde{t}_i \tilde{t}_j}^2} \begin{aligned}[t] \bigg[
          &\tfrac{\mu^{\prime 2}\left(8 U_- y_t^2 + 2 U_{\times} - 1\right)}{c_{\beta}^2s_{\beta}^2\Delta_{\tilde{t}_i\tilde{t}_j}} + \tfrac{\mu^2}{c_{\beta }^2 s_{\beta }^2 \Delta _{\tilde{t}_i \tilde{t}_j}}\left(-\tfrac{3 U_- \mu^{\prime 2}}{c_{\beta}^2s_{\beta}^2\Delta_{\tilde{t}_i\tilde{t}_j}} + 3 - 10 U_{\times}\right)\\
          &- \left(3 - 10 U_{\times}\right) y_t^2 - \frac{U_-}{2} - y_t^{\prime 2} \left(5 U_- y_t^2 + 1\right) - \tfrac{1}{2}\bigg]
      \end{aligned}\\
      &+ \tfrac{m_t^6}{\Delta _{\tilde{t}_i \tilde{t}_j}^3} \left[\tfrac{2 \mu^{\prime 2}}{c_{\beta}^2s_{\beta}^2\Delta_{\tilde{t}_i\tilde{t}_j}}\left(\tfrac{\mu^2}{c_{\beta}^2s_{\beta}^2\Delta_{\tilde{t}_i\tilde{t}_j}} - 2 y_t^2\right) - \tfrac{U_- \mu^2}{c_{\beta}^2s_{\beta}^2\Delta_{\tilde{t}_i\tilde{t}_j}} + y_t^2 \left(2 y_t^{\prime 2} + U_-\right)\right]\\
      &+ \left(1 - 10 U_{\times}\right) \left(1 - 2 U_{\times}\right) \left(\tfrac{\mu^2}{c_{\beta}^2s_{\beta}^2\Delta_{\tilde{t}_i\tilde{t}_j}} - y_t^2\right)\Bigg\}
      \mathrm{T}_{134}{\left(m_{\tilde{t}_i}^2,m_{\tilde{b}_1}^2,m_{H^{\pm }}^2\right)}
    \end{aligned}\\
  &\quad+ \tfrac{m_t^4 \Delta _{\tilde{t}_i \tilde{t}_j}}{\Delta_{\tilde{t}_i\tilde{b}_1}\Delta _{\tilde{t}_j\tilde{b}_1}^2} \begin{aligned}[t] \Bigg\{
    & \tfrac{\mu^{\prime 2}}{c_{\beta}^2s_{\beta}^2\Delta_{\tilde{t}_i\tilde{t}_j}} \left(x_t^2 - y_t^2\right) \left[\tfrac{m_t^4 \left(12 x_t^4 + 1 - 4 U_{\times}\right)}{\Delta_{\tilde{t}_i\tilde{t}_j}^2} + 2 U_- U_{\times}\left(\tfrac{3 m_t^2}{\Delta_{\tilde{t}_i\tilde{t}_j}} - x_t^2\right) + 1 - 5 U_{\times}\right]\\
    &+ \tfrac{\mu^2}{2 c_{\beta}^2s_{\beta}^2\Delta_{\tilde{t}_i\tilde{t}_j}} \begin{aligned}[t] \bigg[
      & \tfrac{\mu^{\prime 2}}{c_{\beta}^2s_{\beta}^2\Delta_{\tilde{t}_i\tilde{t}_j}} + \tfrac{m_t^2 \left(4 U_- y_t^{\prime 2} + 8 U_{\times}^2 - 1\right)}{\Delta_{\tilde{t}_i\tilde{t}_j}} - 3 x_t^2\\
      &- 2 U_{\times} \left(-2 x_t^2 \left(U_- x_t^2 - 6 U_{\times} + 2\right) + U_- \left(1 + 6 U_{\times}\right)\right)\bigg]
    \end{aligned}\\
    &+ \left(1 - 4 U_{\times}\right) y_t^{\prime 2} y_t^2 \left[\frac{m_t^4}{\Delta_{\tilde{t}_i\tilde{t}_j}^2} + 2 U_- \left(x_t^2 - \tfrac{m_t^2}{\Delta_{\tilde{t}_i\tilde{t}_j}}\right) + x_t^4 + 1 - 6 U_{\times}\right]\Bigg\}\\
    &\times\mathrm{T}_{134}{\left(m_{\tilde{t}_i}^2,m_{\tilde{t}_j}^2,m_{H^{\pm }}^2\right)}
  \end{aligned}\\
  &\quad + \Delta_{\tilde{t}_i\tilde{t}_j} \begin{aligned}[t] \Bigg\{
    & \tfrac{m_t^4}{\Delta _{\tilde{t}_i \tilde{t}_j}^2} \begin{aligned}[t] \bigg[
    &\tfrac{\mu^2}{c_{\beta}^2s_{\beta}^2\Delta_{\tilde{t}_i\tilde{t}_j}} \left(2 x_t^2 + 1 - \tfrac{\mu^{\prime 2}}{c_{\beta}^2s_{\beta}^2\Delta_{\tilde{t}_i\tilde{t}_j}}\right) + \tfrac{\mu^{\prime 2}}{c_{\beta}^2s_{\beta}^2\Delta_{\tilde{t}_i\tilde{t}_j}} \left(2 y_t^2 - x_t^2 + 1\right)\\
    &- \left(y_t^2 + 1\right) \left(y_t^{\prime 2} + 2 x_t^2 + 1\right) - \tfrac{U_--1}{2} \left(y_t^2 + 1 - \tfrac{\mu^2}{c_{\beta}^2s_{\beta}^2\Delta_{\tilde{t}_i\tilde{t}_j}}\right)\bigg]
    \end{aligned}\\
    &+ \tfrac{U_--1}{2} \left[U_{\times} \left(\tfrac{\mu^{\prime 2} - \mu^2}{c_{\beta}^2s_{\beta}^2\Delta_{\tilde{t}_i\tilde{t}_j}} + y_t^2 + 1\right) - \tfrac{m_t^2}{\Delta_{\tilde{t}_i\tilde{t}_j}} \left(\tfrac{\mu^{\prime 2}\left(1 + 2 y_t^2\right)}{c_{\beta}^2s_{\beta}^2\Delta_{\tilde{t}_i\tilde{t}_j}} - 2 \left(y_t^2 + 1\right) y_t^{\prime 2}\right)\right]\\
    &+ \left(y_t^2 + 1\right) y_t^{\prime 2}\left(\tfrac{U_--1}{2} + U_{\times}\right)\Bigg\}\mathrm{T}_{1134}{\left(m_{\tilde{t}_i}^2,m_{\tilde{t}_i}^2,m_{\tilde{b}_1}^2,m_{H^{\pm }}^2\right)}
    \end{aligned}
  \end{split}
\end{align}
\end{subequations}
}%

\subsection{Genuine two-loop tadpoles\label{sec:tadpole2L}}

The explicit expressions for the genuine two-loop tadpoles of the Higgs bosons are given by
{\allowdisplaybreaks\footnotesize
\begin{subequations}
\begin{align}
  \Tadpole^{(2)}_{h} &= \frac{N_{c} s_{\beta} h_{t}^{3} m_{t}}{128 \sqrt{2} \pi^{4}} \bigg\{t_{A} + \sum_{\ontop{\,i\;=\;1}{j\;\neq\;i}}^{2}\left[-x_{t}^{2} t_{B} + \left(1 + x_{t}^{2}\right) t_{C} - x_{t}^{2} t_{D_{1}} - t_{D_{2}} + t_{D_{3}}\right]\bigg\},\\
  \Tadpole^{(2)}_{H} &= -\frac{N_{c} c_{\beta} h_{t}^{3} m_{t}}{128 \sqrt{2} \pi^{4}} \bigg\{t_{A} + \sum_{\ontop{\,i\;=\;1}{j\;\neq\;i}}^{2}\left[\tfrac{\eta}{2} t_{B} + \left(1 - \tfrac{\eta}{2}\right) t_{C} + \tfrac{\eta}{2} t_{D_{1}} - t_{D_{2}}\right]\bigg\},\\
  \Tadpole^{(2)}_{A} &= \sum_{\ontop{\,i\;=\;1}{j\;\neq\;i}}^{2} \frac{N_{c} h_{t}^{3} m_{t}}{128 \sqrt{2} \pi^{4} s_{\beta}}\frac{\Imag{X_{t}\mu^{*}}}{s_{\beta}c_{\beta}\Delta_{\tilde{t}_i\tilde{t}_j}} \bigg\{t_{B} - t_{C} + t_{D_{1}}\bigg\}.
\end{align}
\end{subequations}
%
%
%
%
\begin{subequations}
\begin{align}
  t_{A} &= s_{A} + \sum_{i=1}^{2}\left(m_t^2 - m_{\tilde{t}_i}^2 + \mu^2\right) \mathrm{T}_{1134}{\left(m_t^2,m_t^2,m_{\tilde{t}_i}^2,\mu^2\right)},\\\nextParentEquation
  t_{B} &= s_{D} + s_{G} + s_{H},\\\nextParentEquation
  \begin{split}
  t_{C} &= s_{I_{1}} + s_{I_{2}} + \tfrac{1}{2} \left(1 - U_-\right) \left(\mu^2 - m_{\tilde{t}_i}^2\right) \mathrm{T}_{1134}{\left(m_{\tilde{t}_i}^2,m_{\tilde{t}_i}^2,\mu ^2,0\right)}\\
  &\quad + c_{\beta}^2 t_{C}^{H} + s_{\beta}^{2} \left.t_{C}^{H}\right|_{m_{H^{\pm}} \rightarrow 0,\ Y_{t}^{2} \rightarrow X_{t}^{2},\ y_{t}^{2} \rightarrow x_{t}^{2},\ \eta \rightarrow -2 x_{t}^{2}}\ ,
  \end{split}\\
  \begin{split}
  t_{C}^{H} &= \left\{m_t^2 \left(\eta - \tfrac{1 + U_-}{2}\right) - \tfrac{1 - U_-}{2} Y_t^2\right\} \mathrm{T}_{1134}{\left(m_{\tilde{t}_i}^2,m_{\tilde{t}_i}^2,m_{\tilde{b}_1}^2,m_{H^{\pm }}^2\right)}\\
  &\quad+ 2 m_t^2 \left(\eta - x_t^2 y_t^2 - 1\right) \mathrm{T}_{1134}{\left(m_{\tilde{t}_i}^2,m_{\tilde{t}_i}^2,m_{\tilde{t}_i}^2,m_{H^{\pm }}^2\right)}\ ,
  \end{split}\\\nextParentEquation
  \begin{split}
  t_{D_{1}} &= \left\{\tfrac{1}{2} - U_- \left(\tfrac{\mu^2 - m_{\tilde{t}_i}^2}{\Delta_{\tilde{t}_i\tilde{t}_j}} + \tfrac{1}{2}\right)\right\} \mathrm{T}_{134}{\left(m_{\tilde{t}_i}^2,0,\mu^2\right)}\\
  &\quad + c_{\beta}^2 t_{D_{1}}^{H} + s_{\beta}^{2} \left.t_{D_{1}}^{H}\right|_{m_{H^{\pm}} \rightarrow 0,\ Y_{t}^{2} \rightarrow X_{t}^{2},\ y_{t}^{2} \rightarrow x_{t}^{2},\ \eta \rightarrow -2 x_{t}^{2}}\ ,
  \end{split}\\
  \begin{split}
  t_{D_{1}}^{H} &= \left\{\tfrac{2 \eta m_t^2 + U_- \left(Y_t^2 - m_t^2\right)}{\Delta_{\tilde{t}_i\tilde{t}_j}} + 1\right\} \mathrm{T}_{134}{\left(m_{\tilde{t}_i}^2,m_{\tilde{b}_1}^2,m_{H^{\pm }}^2\right)}\\
  &\quad + \left\{\tfrac{2 \eta  m_t^2 + Y_t^2 \left(1 - 4 U_{\times}\right)}{\Delta_{\tilde{t}_i\tilde{t}_j}} + 1\right\} \mathrm{T}_{134}{\left(m_{\tilde{t}_i}^2,m_{\tilde{t}_i}^2,m_{H^{\pm }}^2\right)},
  \end{split}\\\nextParentEquation
  t_{D_{2}} &= \tfrac{1 - U_-}{2} \mathrm{T}_{134}{\left(m_{\tilde{t}_i}^2,\mu ^2,0\right)} + c_{\beta}^2 t_{D_{2}}^{H} + s_{\beta}^{2} \left.t_{D_{2}}^{H}\right|_{m_{H^{\pm}} \rightarrow 0,\ Y_{t}^{2} \rightarrow X_{t}^{2},\ y_{t}^{2} \rightarrow x_{t}^{2},\ \eta \rightarrow -2 x_{t}^{2}}\ ,\\
  t_{D_{2}}^{H} &= \tfrac{2 y_t^2 + 1 + U_-}{2} \mathrm{T}_{134}{\left(m_{\tilde{t}_i}^2,m_{\tilde{b}_1}^2,m_{H^{\pm }}^2\right)} + \left(y_t^2 + 1\right) \mathrm{T}_{134}{\left(m_{\tilde{t}_i}^2,m_{\tilde{t}_i}^2,m_{H^{\pm }}^2\right)},\\\nextParentEquation
  t_{D_{3}} &= c_{\beta}^{2}\tfrac{\mu^2}{s_{\beta}^{2}c_{\beta}^{2}\Delta_{\tilde{t}_i\tilde{t}_j}} \left[\mathrm{T}_{134}{\left(m_{\tilde{t}_i}^2,m_{\tilde{b}_1}^2,m_{H^{\pm }}^2\right)} + \mathrm{T}_{134}{\left(m_{\tilde{t}_i}^2,m_{\tilde{t}_i}^2,m_{H^{\pm}}^2\right)}\right].
\end{align}
\end{subequations}
}%

\subsection{One-loop self-energies with counterterm insertions\label{sec:selfCT}}

The one-loop self-energies with counterterm insertion are part of the full two-loop self-energies. They are given in the following:
{\allowdisplaybreaks\footnotesize
\begin{subequations}
\begin{align}
  \begin{split}
  \Sigma^{(2)\,\text{ct}}_{hh} &= \frac{N_{c} s_{\beta}^{2} h_{t}^{2}}{16 \pi^{2}} \left\{s^{\text{ct}}_{A_{1}} + s^{\text{ct}}_{A_{2}}\right\}\\
  &\quad + \sum_{\ontop{\,i\;=\;1}{j\;\neq\;i}}^{2} \frac{N_{c} s_{\beta}^{2} h_{t}^{2}}{16 \pi^{2}}\begin{aligned}[t]\bigg\{
    & \left(1 + x_{t}^{2}\right) s^{\text{ct}}_{A_{3}} + \left(1 + x_{t}^{2}\right)^{2} s^{\text{ct}}_{A_{4}} + x_{t}^{2} \left(1 - 4 U_{\times}\right) s^{\text{ct}}_{A_{5}} - 2 x_{t}^{4} s^{\text{ct}}_{B_{1}}\\
    & + x_{t}^{2} s^{\text{ct}}_{B_{2}} + x_{t}^{2} \left(1 - 4 U_{\times}\right) s^{\text{ct}}_{D_{1}} + x_{t}^{2} \left(1 + x_{t}^{2}\right) s^{\text{ct}}_{D_{2}}\bigg\},\end{aligned}
  \end{split}\\
  \begin{split}
  \Sigma^{(2)\,\text{ct}}_{hH} &= -\frac{N_{c} s_{\beta} c_{\beta} h_{t}^{2}}{16 \pi^{2}} \left\{s^{\text{ct}}_{A_{1}} + s^{\text{ct}}_{A_{2}}\right\}\\
  &\quad - \sum_{\ontop{\,i\;=\;1}{j\;\neq\;i}}^{2} \frac{N_{c} s_{\beta} c_{\beta} h_{t}^{2}}{16 \pi^{2}}\begin{aligned}[t]\bigg\{
    & \left(1 - \tfrac{\eta}{4} + \tfrac{x_{t}^{2}}{2}\right) s^{\text{ct}}_{A_{3}} - \left(\tfrac{\eta}{2} - 1\right) \left(1 + x_{t}^{2}\right) s^{\text{ct}}_{A_{4}} - \tfrac{\eta}{2} \left(1 - 4 U_{\times}\right) s^{\text{ct}}_{A_{5}}\hspace{-3em}\\
    & + \eta x_{t}^{2} s^{\text{ct}}_{B_{1}} + \left(\tfrac{x_{t}^{2}}{2} - \tfrac{\eta}{4}\right) s^{\text{ct}}_{B_{2}} + \tfrac{\Imag{X_{t}\mu^{*}}}{s_{\beta}c_{\beta}\Delta_{\tilde{t}_i\tilde{t}_j}} \left(x_{t}^{2} s^{\text{ct}}_{B_{3}} - s^{\text{ct}}_{B_{4}}\right)\\
    & - \left(\tfrac{\eta}{2} + x_{t}^{2}\right)\left[\left(1 - 4 U_{\times}\right) s^{\text{ct}}_{C_{1}} + \left(1 + x_{t}^{2}\right) s^{\text{ct}}_{C_{2}}\right] + x_{t}^{2} s^{\text{ct}}_{D_{2}}\\
    & + \left(\tfrac{x_{t}^{2}}{2} - \tfrac{\eta}{4}\right)\left[\left(1 - 4 U_{\times}\right) s^{\text{ct}}_{D_{1}} + x_{t}^{2} s^{\text{ct}}_{D_{2}}\right] + \left(\tfrac{\eta^2}{4} - x_{t}^2 y_{t}^{2}\right) s^{\text{ct}}_{D_{3}}\bigg\},\hspace{-3em}\end{aligned}
  \end{split}\\
  \begin{split}
  \Sigma^{(2)\,\text{ct}}_{HH} &= \frac{N_{c} c_{\beta}^{2} h_{t}^{2}}{16 \pi^{2}} \left\{s^{\text{ct}}_{A_{1}} + s^{\text{ct}}_{A_{2}}\right\}\\
  &\quad + \sum_{\ontop{\,i\;=\;1}{j\;\neq\;i}}^{2} \frac{N_{c} c_{\beta}^{2} h_{t}^{2}}{16 \pi^{2}}\begin{aligned}[t]\bigg\{
    & -\left(\tfrac{\eta}{2} - 1\right) s^{\text{ct}}_{A_{3}} + \left(\tfrac{\eta}{2} - 1\right)^{2} s^{\text{ct}}_{A_{4}} - \tfrac{m_{t}^{2}\eta^{2} - Y_{t}^{2}}{\Delta_{\tilde{t}_i\tilde{t}_j}} s^{\text{ct}}_{A_{5}}\\
    & - \tfrac{\eta^{2}}{2} s^{\text{ct}}_{B_{1}} - \tfrac{\eta}{2} s^{\text{ct}}_{B_{2}} - \tfrac{\Imag{X_{t}\mu^{*}}}{s_{\beta}c_{\beta}\Delta_{\tilde{t}_i\tilde{t}_j}}\left(\eta s^{\text{ct}}_{B_{3}} + 2 s^{\text{ct}}_{B_{4}}\right)\\
    & + \left[\eta \left(1 - 4 U_{\times}\right) - 2 \tfrac{m_{t}^{2}\eta^{2} - Y_{t}^{2}}{\Delta_{\tilde{t}_i\tilde{t}_j}}\right] s^{\text{ct}}_{C_{1}} + \left(\tfrac{\eta}{2} - 1\right)\left(\eta + 2 x_{t}^{2}\right) s^{\text{ct}}_{C_{2}}\\
    & + \tfrac{\eta}{2}\left(1 - 4 U_{\times}\right) s^{\text{ct}}_{D_{1}} - \left(\tfrac{\eta}{2} - 1\right) x_{t}^{2} s^{\text{ct}}_{D_{2}} + \left(\tfrac{\eta^2}{2} - 2 x_{t}^2 y_{t}^{2}\right) s^{\text{ct}}_{D_{3}}\bigg\},\end{aligned}
  \end{split}\\
  \begin{split}
  \Sigma^{(2)\,\text{ct}}_{AA} &= \frac{N_{c} c_{\beta}^{2} h_{t}^{2}}{16 \pi^{2}} s^{\text{ct}}_{A_{1}}\\
  &\quad + \sum_{\ontop{\,i\;=\;1}{j\;\neq\;i}}^{2} \frac{N_{c} c_{\beta}^{2} h_{t}^{2}}{16 \pi^{2}}\begin{aligned}[t]\bigg\{
    & -\left(\tfrac{\eta^{2}}{4} - x_{t}^{2} y_{t}^{2}\right) s^{\text{ct}}_{A_{4}} + \tfrac{m_{t}^{2}\eta^{2} + \left(1 - 4 U_{\times}\right) Y_{t}^{2}}{\Delta_{\tilde{t}_i\tilde{t}_j}} s^{\text{ct}}_{A_{5}} + \left(\tfrac{\eta^2}{2} - 2 x_{t}^2 y_{t}^{2}\right) s^{\text{ct}}_{B_{1}}\\
    & + \tfrac{\Imag{X_{t}\mu^{*}}}{s_{\beta}c_{\beta}\Delta_{\tilde{t}_i\tilde{t}_j}} \eta s^{\text{ct}}_{B_{3}} + 2\left(\tfrac{\eta}{2} + \tfrac{m_{t}^{2}\eta^{2} + \left(1 - 4 U_{\times}\right) Y_{t}^{2}}{\Delta_{\tilde{t}_i\tilde{t}_j}}\right) s^{\text{ct}}_{C_{1}} - \left(\tfrac{\eta^2}{2} - 2 x_{t}^2 y_{t}^{2}\right) s^{\text{ct}}_{C_{2}}\\
    & - \tfrac{\eta}{2} s^{\text{ct}}_{D_{1}} + \left(\tfrac{\eta^2}{2} - 2 x_{t}^2 y_{t}^{2}\right)\left[\left(1 - 4 U_{\times}\right) s^{\text{ct}}_{D_{3}} + x_{t}^{2} s^{\text{ct}}_{D_{4}}\right]\bigg\},\end{aligned}
  \end{split}\\
  \begin{split}
  \Sigma^{(2)\,\text{ct}}_{hA} &= \sum_{\ontop{\,i\;=\;1}{j\;\neq\;i}}^{2} \frac{N_{c} s_{\beta} c_{\beta} h_{t}^{2}}{16 \pi^{2}} \frac{\Imag{X_{t}\mu^{*}}}{s_{\beta}c_{\beta}\Delta_{\tilde{t}_i\tilde{t}_j}}\begin{aligned}[t]\bigg\{
    & -\tfrac{1}{2} s^{\text{ct}}_{A_{3}} - \left(1 + x_{t}^{2}\right) s^{\text{ct}}_{A_{4}} - \left(1 - 4 U_{\times}\right) s^{\text{ct}}_{A_{5}}\\
    & + 2 x_{t}^{2} s^{\text{ct}}_{B_{1}} - \tfrac{1}{2} s^{\text{ct}}_{B_{2}} - \tfrac{s_{\beta}c_{\beta}\Delta_{\tilde{t}_i\tilde{t}_j}}{\Imag{X_{t}\mu^{*}}}\tfrac{\eta}{2}\left(x_{t}^{2} s^{\text{ct}}_{B_{3}} - s^{\text{ct}}_{B_{4}}\right)\\
    & - \left(1 - 4 U_{\times}\right)\left(s^{\text{ct}}_{C_{1}} + \tfrac{1}{2} s^{\text{ct}}_{D_{1}}\right) - \left(1 + x_{t}^{2}\right)\left(s^{\text{ct}}_{C_{2}} - x_{t}^{2} s^{\text{ct}}_{D_{4}}\right)\\
    & - \tfrac{x_{t}^{2}}{2} s^{\text{ct}}_{D_{2}} + \left[\tfrac{\eta}{2} + x_{t}^{2}\left(1 - 4 U_{\times}\right)\right] s^{\text{ct}}_{D_{3}}\bigg\},\end{aligned}
  \end{split}\\
  \begin{split}
  \Sigma^{(2)\,\text{ct}}_{HA} &= \sum_{\ontop{\,i\;=\;1}{j\;\neq\;i}}^{2} \frac{N_{c} c_{\beta}^{2} h_{t}^{2}}{16 \pi^{2}} \frac{\Imag{X_{t}\mu^{*}}}{s_{\beta}c_{\beta}\Delta_{\tilde{t}_i\tilde{t}_j}}\begin{aligned}[t]\bigg\{
    & \tfrac{1}{2} s^{\text{ct}}_{A_{3}} - \left(\tfrac{\eta}{2} - 1\right) s^{\text{ct}}_{A_{4}} + \tfrac{2m_{t}^{2}\eta}{\Delta_{\tilde{t}_i\tilde{t}_j}} s^{\text{ct}}_{A_{5}} + \tfrac{\eta}{2} s^{\text{ct}}_{B_{1}}\\
    & + \tfrac{1}{2} s^{\text{ct}}_{B_{2}} - \tfrac{s_{\beta}c_{\beta}\Delta_{\tilde{t}_i\tilde{t}_j}}{\Imag{X_{t}\mu^{*}}}\left[\left(\tfrac{\eta^2}{2} - x_{t}^2 y_{t}^{2}\right) s^{\text{ct}}_{B_{3}} + \tfrac{\eta}{2} s^{\text{ct}}_{B_{4}}\right]\\
    & + \tfrac{4 m_{t}^{2}}{\Delta_{\tilde{t}_i\tilde{t}_j}}\left(\eta + x_{t}^{2}\right) s^{\text{ct}}_{C_{1}} - \left(\eta + x_{t}^{2} - 1\right) s^{\text{ct}}_{C_{2}}\\
    & - \tfrac{2 m_{t}^{2}}{\Delta_{\tilde{t}_i\tilde{t}_j}} x_{t}^{2}\left(s^{\text{ct}}_{D_{1}} + \eta s^{\text{ct}}_{D_{3}}\right) + \tfrac{x_{t}^{2}}{2} s^{\text{ct}}_{D_{2}} + \left(\tfrac{\eta}{2} - 1\right) x_{t}^{2} s^{\text{ct}}_{D_{4}}\bigg\},\end{aligned}
  \end{split}\\
  \begin{split}
  \Sigma^{(2)\,\text{ct}}_{H^{\pm}H^{\pm}} &= \frac{N_{c} c_{\beta}^{2} h_{t}^{2}}{16 \pi^{2}} \bigg\{s^{\text{ct}}_{E} - s^{\text{ct}}_{F_1} - s^{\text{ct}}_{F_2}\bigg\}\\
  &\quad + \sum_{\ontop{\,i\;=\;1}{j\;\neq\;i}}^{2} \frac{N_{c} c_{\beta}^{2} h_{t}^{2}}{16 \pi^{2}} \begin{aligned}[t] \bigg\{
    & s^{\text{ct}}_{G} + s^{\text{ct}}_{H} + \tfrac{m_t^2 \left(\eta - \lvert\mathbf{U}_{\tilde{t}\,1i}\rvert^{2}\right) - Y_t^2 \lvert\mathbf{U}_{\tilde{t}\,1j}\rvert^{2}}{\Delta_{\tilde{t}_i\tilde{b}_1}} s^{\text{ct}}_{I} + s^{\text{ct}}_{J} + s^{\text{ct}}_{K} + s^{\text{ct}}_{L_1} + s^{\text{ct}}_{L_2}\\
    &+ s^{\text{ct}}_{M_1} + s^{\text{ct}}_{M_2}\bigg\}.\end{aligned}
  \end{split}
\end{align}
\end{subequations}
%
%
%
%
\begin{subequations}
\begin{align}
  \begin{split}
    s^{\text{ct}}_{A_{1}} &= 2 \left[\tfrac{\delta^{(1)}h_{t}}{h_{t}} + \tfrac{\delta^{(1)}Z_{\mathcal{H}_{2}}}{2}\right] \left[-2 \Anull{m_t^2} + \sum_{i\;=\;1}^{2}\Anull{m_{\tilde{t}_i}^2}\right]\\
    &\quad- 4 m_t^2 \tfrac{\delta^{(1)}m_{t}}{m_{t}} \Bnull{0,m_t^2,m_t^2} + \sum_{i\;=\;1}^{2}\delta^{(1)}m_{\tilde{t}_i\tilde{t}_i}^2 \Bnull{0,m_{\tilde{t}_i}^2,m_{\tilde{t}_i}^2},
  \end{split}\\
  s^{\text{ct}}_{A_{2}} &= - 8 m_t^2 \left[\tfrac{\delta^{(1)}m_{t}}{m_{t}} + \tfrac{\delta^{(1)}h_{t}}{h_{t}} + \tfrac{\delta^{(1)}Z_{\mathcal{H}_{2}}}{2}\right] \Bnull{0,m_t^2,m_t^2} - 16 m_t^4 \tfrac{\delta^{(1)}m_{t}}{m_{t}} \Cnull{0,0,0,m_t^2,m_t^2,m_t^2},\\
  s^{\text{ct}}_{A_{3}} &= 4 m_t^2 \tfrac{\delta^{(1)}m_{t}}{m_{t}} \Bnull{0,m_{\tilde{t}_i}^2,m_{\tilde{t}_i}^2},\\
  s^{\text{ct}}_{A_{4}} &= 4 m_t^2 \left\{\left[\tfrac{\delta^{(1)}h_{t}}{h_{t}} + \tfrac{\delta^{(1)}Z_{\mathcal{H}_{2}}}{2}\right] \Bnull{0,m_{\tilde{t}_i}^2,m_{\tilde{t}_i}^2} + \delta^{(1)}m_{\tilde{t}_i\tilde{t}_i}^2 \Cnull{0,0,0,m_{\tilde{t}_i}^2,m_{\tilde{t}_i}^2,m_{\tilde{t}_i}^2}\right\},\\
  s^{\text{ct}}_{A_{5}} &= 2 \left[\tfrac{\delta^{(1)}h_{t}}{h_{t}} + \tfrac{\delta^{(1)}Z_{\mathcal{H}_{2}}}{2}\right] \Anull{m_{\tilde{t}_i}^2} - \delta^{(1)}m_{\tilde{t}_i\tilde{t}_i}^2\left[\Bnull{0,m_{\tilde{t}_i}^2,m_{\tilde{t}_j}^2} - \Bnull{0,m_{\tilde{t}_i}^2,m_{\tilde{t}_i}^2}\right],\\\nextParentEquation
  s^{\text{ct}}_{B_{1}} &= \tfrac{4 \left(\lvert\mathbf{U}_{\tilde{t}\,11}\rvert^{2} - \lvert\mathbf{U}_{\tilde{t}\,12}\rvert^{2}\right) \lvert\mathbf{U}_{\tilde{t}\,12}\rvert^{2}}{X_{t}^{2}}\Real{\tfrac{\delta^{(1)}m_{\tilde{t}_1\tilde{t}_2}^2 \mathbf{U}_{\tilde{t}\,22}\mathbf{U}_{\tilde{t}\,11}^{*}}{m_{t}X_{t}}} \left[\Anull{m_{\tilde{t}_i}^2} - \tfrac{\Delta_{\tilde{t}_i\tilde{t}_j}}{2} \Bnull{0,m_{\tilde{t}_i}^2,m_{\tilde{t}_i}^2}\right],\\
  s^{\text{ct}}_{B_{2}} &= \tfrac{4 \left(\lvert\mathbf{U}_{\tilde{t}\,11}\rvert^{2} - \lvert\mathbf{U}_{\tilde{t}\,12}\rvert^{2}\right) \lvert\mathbf{U}_{\tilde{t}\,12}\rvert^{2}}{X_{t}^{2}}\Real{\tfrac{\delta^{(1)}m_{\tilde{t}_1\tilde{t}_2}^2 \mathbf{U}_{\tilde{t}\,22}\mathbf{U}_{\tilde{t}\,11}^{*}}{m_{t}X_{t}}} \Delta_{\tilde{t}_i\tilde{t}_j} \Bnull{0,m_{\tilde{t}_i}^2,m_{\tilde{t}_i}^2},\\
  s^{\text{ct}}_{B_{3}} &= \tfrac{4 \lvert\mathbf{U}_{\tilde{t}\,12}\rvert^{2}}{X_{t}^{2}}\Imag{\tfrac{\delta^{(1)}m_{\tilde{t}_1\tilde{t}_2}^2 \mathbf{U}_{\tilde{t}\,22}\mathbf{U}_{\tilde{t}\,11}^{*}}{m_{t}X_{t}}} \left[\Anull{m_{\tilde{t}_i}^2} - \tfrac{\Delta_{\tilde{t}_i\tilde{t}_j}}{2} \Bnull{0,m_{\tilde{t}_i}^2,m_{\tilde{t}_i}^2}\right],\\
  s^{\text{ct}}_{B_{4}} &= \tfrac{2 \lvert\mathbf{U}_{\tilde{t}\,12}\rvert^{2}}{X_{t}^{2}}\Imag{\tfrac{\delta^{(1)}m_{\tilde{t}_1\tilde{t}_2}^2 \mathbf{U}_{\tilde{t}\,22}\mathbf{U}_{\tilde{t}\,11}^{*}}{m_{t}X_{t}}} \Delta_{\tilde{t}_i\tilde{t}_j} \Bnull{0,m_{\tilde{t}_i}^2,m_{\tilde{t}_i}^2},\\\nextParentEquation
  s^{\text{ct}}_{C_{1}} &= \left[\tfrac{\delta^{(1)}\mu}{\mu} - \tfrac{\delta^{(1)}t_{\beta}}{t_{\beta}}\right] \Anull{m_{\tilde{t}_i}^2},\\
  s^{\text{ct}}_{C_{2}} &= 2 m_t^2 \left[\tfrac{\delta^{(1)}\mu}{\mu} - \tfrac{\delta^{(1)}t_{\beta}}{t_{\beta}}\right] \Bnull{0,m_{\tilde{t}_i}^2,m_{\tilde{t}_i}^2},\\\nextParentEquation
  s^{\text{ct}}_{D_{1}} &= \tfrac{\delta^{(1)}x_{t}^{2}}{x_{t}^{2}} \Anull{m_{\tilde{t}_i}^2},\\
  s^{\text{ct}}_{D_{2}} &= 2 m_t^2 \tfrac{\delta^{(1)}x_{t}^{2}}{x_{t}^{2}} \Bnull{0,m_{\tilde{t}_i}^2,m_{\tilde{t}_i}^2},\\
  s^{\text{ct}}_{D_{3}} &= \tfrac{s_{\beta}c_{\beta}\Delta_{\tilde{t}_i\tilde{t}_j} \delta^{(1)}\phi_X}{\Imag{X_{t}\mu^{*}}} \Anull{m_{\tilde{t}_i}^2},\\
  s^{\text{ct}}_{D_{4}} &= 2 m_{t}^{2} \tfrac{s_{\beta}c_{\beta}\Delta_{\tilde{t}_i\tilde{t}_j} \delta^{(1)}\phi_X}{\Imag{X_{t}\mu^{*}}} \Bnull{0,m_{\tilde{t}_i}^2,m_{\tilde{t}_i}^2},\\\nextParentEquation
  \begin{split}
    s^{\text{ct}}_{E} &= 2 \left[\tfrac{\delta^{(1)}h_{t}}{h_{t}} + \tfrac{\delta^{(1)}Z_{\mathcal{H}_{2}}}{2}\right] \bigg[\Anull{m_{\tilde{b}_1}^2} - 2 \Anull{m_t^2} + \sum_{\ontop{\,i\;=\;1}{j\;\neq\;i}}^{2}\lvert\mathbf{U}_{\tilde{t}\,1j}\rvert^{2} \Anull{m_{\tilde{t}_i}^2}\bigg]\\
    &\quad - 4 m_t^2 \tfrac{\delta^{(1)}m_{t}}{m_{t}} \Bnull{0,m_t^2,m_t^2} + \delta^{(1)}m_{\tilde{b}_1\tilde{b}_1}^2 \left(1 - \tfrac{\eta m_t^2 \Delta_{\tilde{t}_1\tilde{t}_2}}{\Delta_{\tilde{t}_1\tilde{b}_1}\Delta_{\tilde{t}_2\tilde{b}_1}}\right) \Bnull{0,m_{\tilde{b}_1}^2,m_{\tilde{b}_1}^2},
  \end{split}\\\nextParentEquation
  s^{\text{ct}}_{F_1} &= \tfrac{\Delta_{\tilde{t}_1\tilde{t}_2}^2 \left[2 x_t^2 \left(m_t^2 - Y_t^2\right) + \Delta _{\tilde{t}_1 \tilde{t}_2} \eta \left(\lvert\mathbf{U}_{\tilde{t}\,11}\rvert^{2} - \lvert\mathbf{U}_{\tilde{t}\,12}\rvert^{2}\right)\right]}{\Delta_{\tilde{t}_1\tilde{b}_1}\Delta_{\tilde{t}_2\tilde{b}_1}} \tfrac{\lvert\mathbf{U}_{\tilde{t}\,12}\rvert^{2}}{X_t^2} \Real{\tfrac{\delta^{(1)}m_{\tilde{t}_1\tilde{t}_2}^2 \mathbf{U}_{\tilde{t}\,22}\mathbf{U}_{\tilde{t}\,11}^{*}}{m_{t}X_{t}}} \Anull{m_{\tilde{b}_1}^2},\\
  s^{\text{ct}}_{F_2} &= \tfrac{\Delta_{\tilde{t}_1\tilde{t}_2}^2}{\Delta_{\tilde{t}_1\tilde{b}_1}\Delta_{\tilde{t}_2\tilde{b}_1}} \tfrac{2 \Imag{X_{t}\mu^{*}}}{c_{\beta}s_{\beta}} \tfrac{\lvert\mathbf{U}_{\tilde{t}\,12}\rvert^{2}}{X_t^2} \Imag{\tfrac{\delta^{(1)}m_{\tilde{t}_1\tilde{t}_2}^2 \mathbf{U}_{\tilde{t}\,22}\mathbf{U}_{\tilde{t}\,11}^{*}}{m_{t}X_{t}}} \Anull{m_{\tilde{b}_1}^2},\\\nextParentEquation
  s^{\text{ct}}_{G} &= -\tfrac{\lvert\mathbf{U}_{\tilde{t}\,1i}\rvert^{2}\left(m_t^2 \Delta_{\tilde{t}_j\tilde{b}_1} + Y_t^2 \Delta_{\tilde{t}_i\tilde{b}_1}\right)}{\Delta_{\tilde{t}_i\tilde{b}_1}\Delta _{\tilde{t}_j\tilde{b}_1}}\delta^{(1)}m_{\tilde{b}_1\tilde{b}_1}^2 \Bnull{0,m_{\tilde{b}_1}^2,m_{\tilde{b}_1}^2},\\\nextParentEquation
  s^{\text{ct}}_{H} &= \left[\lvert\mathbf{U}_{\tilde{t}\,1j}\rvert^{2} \left(\tfrac{Y_t^2}{\Delta_{\tilde{t}_i\tilde{b}_1}} + 1\right) + \tfrac{m_t^2 \left(\lvert\mathbf{U}_{\tilde{t}\,1i}\rvert^{2} - \eta\right)}{\Delta_{\tilde{t}_i\tilde{b}_1}}\right] \delta^{(1)}m_{\tilde{t}_i\tilde{t}_i}^2 \Bnull{0,m_{\tilde{t}_i}^2,m_{\tilde{t}_i}^2},\\\nextParentEquation
  s^{\text{ct}}_{I} &= 2 \left[\tfrac{\delta^{(1)}h_{t}}{h_{t}} + \tfrac{\delta^{(1)}Z_{\mathcal{H}_{2}}}{2} + \tfrac{\delta^{(1)}m_{\tilde{b}_1\tilde{b}_1}^2 - \delta^{(1)}m_{\tilde{t}_i\tilde{t}_i}^2}{2 \Delta_{\tilde{t}_i\tilde{b}_1}}\right] \left[\Anull{m_{\tilde{b}_1}^2} - \Anull{m_{\tilde{t}_i}^2}\right],\\\nextParentEquation
  s^{\text{ct}}_{J} &= \tfrac{\eta - 2 \lvert\mathbf{U}_{\tilde{t}\,1i}\rvert^{2}}{\Delta_{\tilde{t}_i\tilde{b}_1}} m_t^2 \tfrac{\delta^{(1)}m_{t}}{m_{t}} \left[\Anull{m_{\tilde{b}_1}^2} - \Anull{m_{\tilde{t}_i}^2}\right],\\\nextParentEquation
  s^{\text{ct}}_{K} &= \tfrac{m_t^2 \left(\eta + 2 x_t^2\right) - \Delta_{\tilde{t}_i\tilde{t}_j} \lvert\mathbf{U}_{\tilde{t}\,1j}\rvert^{2} \left(\eta + 2 y_t^2\right)}{\Delta_{\tilde{t}_i\tilde{b}_1}}\left[\tfrac{\delta^{(1)}\mu}{\mu} - \tfrac{\delta^{(1)}t_{\beta}}{t_{\beta}}\right] \left[\Anull{m_{\tilde{b}_1}^2} - \Anull{m_{\tilde{t}_i}^2}\right],\\\nextParentEquation
  s^{\text{ct}}_{L_1} &= -\tfrac{2 m_t^2 x_t^2 - \Delta_{\tilde{t}_i\tilde{t}_j} \eta \lvert\mathbf{U}_{\tilde{t}\,1j}\rvert^{2}}{2 \Delta_{\tilde{t}_i\tilde{b}_1}} \tfrac{\delta^{(1)}x_{t}^{2}}{x_{t}^{2}} \left[\Anull{m_{\tilde{b}_1}^2} - \Anull{m_{\tilde{t}_i}^2}\right],\\
  s^{\text{ct}}_{L_2} &= \tfrac{2 \Imag{X_{t}\mu^{*}}}{c_{\beta}s_{\beta}} \tfrac{\lvert\mathbf{U}_{\tilde{t}\,1j}\rvert^{2}}{\Delta_{\tilde{t}_i\tilde{b}_1}} \delta^{(1)}\phi_X \left[\Anull{m_{\tilde{b}_1}^2} - \Anull{m_{\tilde{t}_i}^2}\right],\\\nextParentEquation
  s^{\text{ct}}_{M_1} &= \left\{2 \lvert\mathbf{U}_{\tilde{t}\,1j}\rvert^{2} - \tfrac{\Delta_{\tilde{t}_i\tilde{t}_j} \left[2 x_t^2 \left(m_t^2 - Y_t^2\right) + \Delta_{\tilde{t}_i\tilde{t}_j} \eta U_-\right]}{\Delta_{\tilde{t}_i\tilde{b}_1}} \tfrac{\lvert\mathbf{U}_{\tilde{t}\,1j}\rvert^{2}}{X_t^2}\right\} \Real{\tfrac{\delta^{(1)}m_{\tilde{t}_i\tilde{t}_j}^2 \mathbf{U}_{\tilde{t}\,2j}\mathbf{U}_{\tilde{t}\,1i}^{*}}{m_{t}X_{t}}} \Anull{m_{\tilde{t}_i}^2},\\
  s^{\text{ct}}_{M_2} &= -\tfrac{\Delta_{\tilde{t}_i\tilde{t}_j}}{\Delta_{\tilde{t}_i\tilde{b}_1}} \tfrac{2 \Imag{X_{t}\mu^{*}}}{c_{\beta}s_{\beta}} \tfrac{\lvert\mathbf{U}_{\tilde{t}\,1j}\rvert^{2}}{X_t^2} \Imag{\tfrac{\delta^{(1)}m_{\tilde{t}_i\tilde{t}_j}^2 \mathbf{U}_{\tilde{t}\,2j}\mathbf{U}_{\tilde{t}\,1i}^{*}}{m_{t}X_{t}}} \Anull{m_{\tilde{t}_i}^2}.
\end{align}
\end{subequations}
}%

\subsection{One-loop tadpoles with counterterm insertions\label{sec:tadpoleCT}}

The one-loop tadpoles with counterterm insertion are part of the two-loop tadpoles of the Higgs bosons. They are given by
{\allowdisplaybreaks\footnotesize
\begin{subequations}
\begin{align}
  \Tadpole^{(2)\,\text{ct}}_{h} &= \frac{N_{c} s_{\beta} h_{t} m_{t}}{8 \sqrt{2} \pi^{4}} \bigg\{t^{\text{ct}}_{A_{1}} + \sum_{\ontop{\,i\;=\;1}{j\;\neq\;i}}^{2}\left[\left(1 + x_{t}^{2}\right) t^{\text{ct}}_{A_{2}} + x_{t}^{2} t^{\text{ct}}_{B_{1}} + \tfrac{x_{t}^{2}}{2} t^{\text{ct}}_{D_{1}}\right]\bigg\},\\
  \Tadpole^{(2)\,\text{ct}}_{H} &= \frac{N_{c} c_{\beta} h_{t} m_{t}}{8 \sqrt{2} \pi^{4}} \bigg\{-t^{\text{ct}}_{A_{1}} + \sum_{\ontop{\,i\;=\;1}{j\;\neq\;i}}^{2}\left[\left(\tfrac{\eta}{2} - 1\right) t^{\text{ct}}_{A_{2}} + \tfrac{\eta}{2} t^{\text{ct}}_{B_{1}} + \tfrac{\Imag{X_{t}\mu^{*}}}{s_{\beta}c_{\beta}\Delta_{\tilde{t}_i\tilde{t}_j}} t^{\text{ct}}_{B_{2}} + \left(\tfrac{\eta}{2} + x_{t}^{2}\right) t^{\text{ct}}_{C} - \tfrac{x_{t}^{2}}{2} t^{\text{ct}}_{D_{1}}\right]\bigg\},\\
  \Tadpole^{(2)\,\text{ct}}_{A} &= -\sum_{\ontop{\,i\;=\;1}{j\;\neq\;i}}^{2} \frac{N_{c} c_{\beta} h_{t} m_{t}}{8 \sqrt{2} \pi^{4}}\frac{\Imag{X_{t}\mu^{*}}}{s_{\beta}c_{\beta}\Delta_{\tilde{t}_i\tilde{t}_j}} \bigg\{t^{\text{ct}}_{A_{2}} + t^{\text{ct}}_{B_{1}} - \tfrac{s_{\beta}c_{\beta}\Delta_{\tilde{t}_i\tilde{t}_j}}{\Imag{X_{t}\mu^{*}}}\tfrac{\eta}{2} t^{\text{ct}}_{B_{2}} + t^{\text{ct}}_{C} - x_{t}^{2} t^{\text{ct}}_{D_{2}}\bigg\}.
\end{align}
\end{subequations}
%
%
%
%
\begin{subequations}
\begin{align}
  t^{\text{ct}}_{A_{1}} &= -2 \left[\tfrac{\delta^{(1)}h_{t}}{h_{t}} + \tfrac{\delta^{(1)}Z_{\mathcal{H}_{2}}}{2}\right] \Anull{m_t^2} - \tfrac{\delta^{(1)}m_{t}}{m_{t}}\left(2 \Anull{m_t^2} + 4 m_t^2 \Bnull{0,m_t^2,m_t^2} - \sum_{i=1}^{2}\Anull{m_{\tilde{t}_i}^2}\right),\\
  t^{\text{ct}}_{A_{2}} &= \left[\tfrac{\delta^{(1)}h_{t}}{h_{t}} + \tfrac{\delta^{(1)}Z_{\mathcal{H}_{2}}}{2}\right] \Anull{m_{\tilde{t}_i}^2} + \delta^{(1)}m_{\tilde{t}_i\tilde{t}_i}^2 \Bnull{0,m_{\tilde{t}_i}^2,m_{\tilde{t}_i}^2},\\\nextParentEquation
  t^{\text{ct}}_{B_{1}} &= \tfrac{\left(\lvert\mathbf{U}_{\tilde{t}\,11}\rvert^{2} - \lvert\mathbf{U}_{\tilde{t}\,12}\rvert^{2}\right) \lvert\mathbf{U}_{\tilde{t}\,12}\rvert^{2}\Delta_{\tilde{t}_i\tilde{t}_j}}{m_{t}^{2}x_{t}^{2}}\Real{\tfrac{\delta^{(1)}m_{\tilde{t}_1\tilde{t}_2}^2 \mathbf{U}_{\tilde{t}\,22}\mathbf{U}_{\tilde{t}\,11}^{*}}{m_{t}X_{t}}}\Anull{m_{\tilde{t}_i}^2},\\
  t^{\text{ct}}_{B_{2}} &= \tfrac{\lvert\mathbf{U}_{\tilde{t}\,12}\rvert^{2}\Delta_{\tilde{t}_i\tilde{t}_j}}{m_{t}^{2}x_{t}^{2}}\Imag{\tfrac{\delta^{(1)}m_{\tilde{t}_1\tilde{t}_2}^2 \mathbf{U}_{\tilde{t}\,22}\mathbf{U}_{\tilde{t}\,11}^{*}}{m_{t}X_{t}}}\Anull{m_{\tilde{t}_i}^2},\\\nextParentEquation
  t^{\text{ct}}_{C} &= s^{\text{ct}}_{C_{1}},\\\nextParentEquation
  t^{\text{ct}}_{D_{1}} &= s^{\text{ct}}_{D_{1}},\\
  t^{\text{ct}}_{D_{2}} &= s^{\text{ct}}_{D_{3}}.
\end{align}
\end{subequations}
}%

\subsection{Renormalization constants for subrenormalization\label{sec:renconst}}

The required renormalization constants are explicitly expressed in the following:
{\allowdisplaybreaks\footnotesize
\begin{subequations}
\begin{align}
  \delta^{(1)}T_h &= \frac{3s_{\beta}h_{t}}{8\sqrt{2}\pi^{2}}\biggl\{-\sum_{i\;=\;1}^{2}\left(m_{t} + \Real{X_{t}\mathbf{U}_{\tilde{t}\,i1}\mathbf{U}_{\tilde{t}\,i2}^{*}}\right)\Anull{m_{\tilde{t}_{i}}^{2}} + 2 m_{t} \Anull{m_{t}^{2}}\biggl\},\\
  \delta^{(1)}T_H &= \frac{3c_{\beta}h_{t}}{8\sqrt{2}\pi^{2}}\biggl\{\sum_{i\;=\;1}^{2}\left(m_{t} + \Real{Y_{t}\mathbf{U}_{\tilde{t}\,i1}\mathbf{U}_{\tilde{t}\,i2}^{*}}\right)\Anull{m_{\tilde{t}_{i}}^{2}} - 2 m_{t} \Anull{m_{t}^{2}}\biggl\},\\
  \delta^{(1)}T_A &= -\frac{3c_{\beta}h_{t}}{8\sqrt{2}\pi^{2}}\sum_{i\;=\;1}^{2}\Imag{Y_{t}\mathbf{U}_{\tilde{t}\,i1}\mathbf{U}_{\tilde{t}\,i2}^{*}}\Anull{m_{\tilde{t}_{i}}^{2}},\\\nextParentEquation
  \begin{split}
  \delta^{(1)}m_{H^{\pm}}^2 &= \begin{aligned}[t] \frac{3c_{\beta}^{2}h_{t}^{2}}{16\pi^{2}}\biggl\{
    & \sum_{i\;=\;1}^{2}\left[\lvert\mathbf{U}_{\tilde{t}\,i2}\rvert^{2}\Anull{m_{\tilde{t}_{i}}^{2}} + \lvert m_{t}\mathbf{U}_{\tilde{t}\,i1}^{*} + Y_{t}\mathbf{U}_{\tilde{t}\,i2}^{*}\rvert^{2}\Real{\Bnull{0,m_{\tilde{t}_{i}}^{2},m_{\tilde{b}_{1}}^{2}}}\right]\\
    & -2m_{t}^{2}\Real{\Bnull{0,0,m_{t}^{2}}} + \Anull{m_{\tilde{b}_{1}}^{2}} \biggl\},\end{aligned}
  \end{split}\\\nextParentEquation
  \delta^{(1)}Z_{\mathcal{H}_{1}} &= 0,\\
  \delta^{(1)}Z_{\mathcal{H}_{2}} &= -\frac{3h_{t}^{2}}{16\pi^{2}\epsilon},\\\nextParentEquation
  \begin{split}
  \frac{\delta^{(1)}M_W^2}{M_W^2} &=
  \begin{aligned}[t]
    \frac{3s_{\beta}^{2}}{16\pi^{2}}\frac{h_{t}^{2}}{m_{t}^{2}}\biggl\{
    & 2\left(2 \Real{\Bnnull{0,0,m_{t}^{2}}} - m_{t}^{2} \Real{\Bnull{0,0,m_{t}^{2}}}\right) + \Anull{m_{\tilde{b}_{1}}^{2}}\\
    & +\sum_{i\;=\;1}^{2}\lvert\mathbf{U}_{\tilde{t}\,i1}\rvert^{2}\left(\Anull{m_{\tilde{t}_{i}}^{2}} - 4\Real{\Bnnull{0,m_{\tilde{b}_{1}}^{2},m_{\tilde{t}_{i}}^{2}}}\right)\biggr\},
  \end{aligned}
  \end{split}\\
  \begin{split}
  \frac{\delta^{(1)}M_Z^2}{M_Z^2} &=
  \begin{aligned}[t]
    \frac{3s_{\beta}^{2}}{16\pi^{2}}\frac{h_{t}^{2}}{m_{t}^{2}}\biggl\{
    & -m_{t}^{2} \Real{\Bnull{0,m_{t}^{2},m_{t}^{2}}} + \sum_{i\;=\;1}^{2}\lvert\mathbf{U}_{\tilde{t}\,i1}\rvert^{2}\lvert\mathbf{U}_{\tilde{t}\,i2}\rvert^{2}\Anull{m_{\tilde{t}_{i}}^{2}}\\
    & - 4 \Real{\Bnnull{0,m_{\tilde{t}_{1}}^{2},m_{\tilde{t}_{2}}^{2}}}\lvert\mathbf{U}_{\tilde{t}\,11}\rvert^2\lvert\mathbf{U}_{\tilde{t}\,12}\rvert^2\biggr\},
  \end{aligned}
  \end{split}\\\nextParentEquation
  \begin{split}
  \frac{\delta^{(1)}m_{t}}{m_{t}} &=
    \frac{h_{t}^{2}}{32\pi^{2}} \begin{aligned}[t] \biggl\{
    & - \Real{\Beins{m_{t}^{2},\mu^2,m_{\tilde{t}_{1}}^{2}} + \Beins{m_{t}^{2},\mu^2,m_{\tilde{t}_{2}}^{2}} + \Beins{m_{t}^{2},\mu^2,m_{\tilde{b}_{1}}^{2}}}\\
    & + c_{\beta}^{2}\frac{\delta^{(1)}m_{t}^H}{m_{t}} + s_{\beta}^{2} \left.\frac{\delta^{(1)}m_{t}^H}{m_{t}}\right|_{m_{H^{\pm}} \rightarrow 0}\biggl\},
  \end{aligned}\end{split}\\
  \frac{\delta^{(1)}m_{t}^H}{m_{t}} &= \Real{\Bnull{m_{t}^{2},m_{H^{\pm}}^{2},m_{t}^{2}} + \Beins{m_{t}^{2},m_{H^{\pm}}^{2},m_{t}^{2}} - \Beins{m_{t}^{2},0,m_{H^{\pm}}^{2}} + \Beins{m_{t}^{2},m_{t}^{2},m_{H^{\pm}}^{2}}},\\\nextParentEquation
  \begin{split}
  \delta^{(1)}m_{\tilde{t}_{i}\tilde{t}_{i}}^{2} &= \frac{h_{t}^{2}}{16 \pi^2} \begin{aligned}[t]\bigg\{
    & - 2 m_{\tilde{t}_i}^2 \Real{\Beins{m_{\tilde{t}_i}^2,m_t^2,\mu^2}} + 8 U_{\times} \Anull{m_{\tilde{t}_i}^2} + \left(1 - 8 U_{\times}\right) \Anull{m_{\tilde{t}_j}^2}\\
    & - 2 \left(\lvert\mathbf{U}_{\tilde{t}\,1j}\rvert^{2} + 1\right) \Anull{\mu^2} + \lvert\mathbf{U}_{\tilde{t}\,1j}\rvert^{2} \Anull{m_{\tilde{b}_1}^2}\\
    & - 2 m_t^2 \Real{\Bnull{m_{\tilde{t}_i}^2,m_t^2,\mu^2}} - 2 m_{\tilde{t}_i}^2 \lvert\mathbf{U}_{\tilde{t}\,1j}\rvert^{2} \Real{\Beins{m_{\tilde{t}_i}^2,0,\mu^2}}\\
    & + c_{\beta}^2 \delta^{(1)}m_{\tilde{t}_{i}\tilde{t}_{i}}^{2H} + s_{\beta}^2 \left.\delta^{(1)}m_{\tilde{t}_{i}\tilde{t}_{i}}^{2H}\right|_{m_{H^{\pm}} \rightarrow 0,\ Y_{t}^{2} \rightarrow X_{t}^{2},\ y_{t}^{2} \rightarrow x_{t}^{2},\ \eta \rightarrow -2 x_{t}^{2}}\bigg\}, \qquad j \neq i,
    \end{aligned}\end{split}\\
  \begin{split}
  \delta^{(1)}m_{\tilde{t}_{i}\tilde{t}_{i}}^{2H} &= Y_t^2 \left(1 - 2 U_{\times}\right) \Real{\Bnull{m_{\tilde{t}_i}^2,m_{H^{\pm}}^2,m_{\tilde{t}_j}^2}} - 2 m_t^2 \left(\eta - x_t^2 y_t^2 - 1\right) \Real{\Bnull{m_{\tilde{t}_i}^2,m_{H^{\pm }}^2,m_{\tilde{t}_i}^2}}\\
  &\quad + \left(m_t^2 \lvert\mathbf{U}_{\tilde{t}\,1i}\rvert^{2} + Y_t^2 \lvert\mathbf{U}_{\tilde{t}\,1j}\rvert^{2} - \eta m_t^2\right) \Real{\Bnull{m_{\tilde{t}_i}^2,m_{H^{\pm}}^2,m_{\tilde{b}_1}^2}} + \left(\lvert\mathbf{U}_{\tilde{t}\,1j}\rvert^{2} + 1\right) \Anull{m_{H^{\pm}}^2},
  \end{split}\\\nextParentEquation
  \begin{split}
  \delta^{(1)}m_{\tilde{t}_{1}\tilde{t}_{2}}^{2} &= \frac{h_{t}^{2} \mathbf{U}_{\tilde{t}\,11}\mathbf{U}_{\tilde{t}\,22}^{*}\lvert\mathbf{U}_{\tilde{t}\,12}\rvert^{2}\Delta_{\tilde{t}_1\tilde{t}_2}}{16 \pi^2 m_{t} X_{t}^{*}} \begin{aligned}[t]\bigg\{
    & \sum_{i\;=\;1}^{2}\left[4 U_- \Anull{m_{\tilde{t}_i}^2} - m_{\tilde{t}_i}^2 \Real{\Beins{m_{\tilde{t}_i}^2,0,\mu^2}}\right]\\
    & - 2 \Anull{\mu^2} + \Anull{m_{\tilde{b}_1}^2}\\
    & - \tfrac{\I c_{\beta}^{2} \Imag{X_{t}\mu^{*}}\Delta_{\tilde{t}_1\tilde{t}_2}}{2 c_{\beta} s_{\beta} X_{t}^{2}} \sum_{i,\,j\;=\;1}^{2}\begin{aligned}[t] \bigg\{& \Real{\Bnull{m_{\tilde{t}_i}^2,m_{H^{\pm}}^{2},m_{\tilde{t}_j}^2}}\\ & + \Real{\Bnull{m_{\tilde{t}_i}^2,m_{H^{\pm}}^{2},m_{\tilde{b}_1}^2}}\bigg\}\end{aligned}\\
    & + c_{\beta}^2 \delta^{(1)}m_{\tilde{t}_{1}\tilde{t}_{2}}^{2H} + s_{\beta}^2 \left.\delta^{(1)}m_{\tilde{t}_{1}\tilde{t}_{2}}^{2H}\right|_{m_{H^{\pm}} \rightarrow 0,\ y_{t}^{2} \rightarrow x_{t}^{2},\ \eta \rightarrow -2 x_{t}^{2}}\bigg\},
  \end{aligned}\end{split}\\
  \begin{split}
  \delta^{(1)}m_{\tilde{t}_{1}\tilde{t}_{2}}^{2H} &= \Anull{m_{H^{\pm}}^{2}} - \tfrac{\Delta_{\tilde{t}_1\tilde{t}_2}^2}{2 X_{t}^{2}} \sum_{\ontop{\,i\;=\;1}{j\;\neq\;i}}^{2} \left[\tfrac{\eta  U_-}{2} - x_t^2 y_t^2 + U_{\times}\right]\Real{\Bnull{m_{\tilde{t}_i}^2,m_{H^{\pm}}^{2},m_{\tilde{b}_1}^2}}\\
  &\quad - \tfrac{\Delta_{\tilde{t}_1\tilde{t}_2}^2}{2 X_{t}^{2}}\sum_{\ontop{\,i\;=\;1}{j\;\neq\;i}}^{2} U_-\left[\tfrac{\eta}{2} - x_t^2 y_t^2\right] \Real{\Bnull{m_{\tilde{t}_i}^2,m_{H^{\pm}}^{2},m_{\tilde{t}_i}^2} + \Bnull{m_{\tilde{t}_j}^2,m_{H^{\pm}}^{2},m_{\tilde{t}_i}^2}},
  \end{split}\\\nextParentEquation
  \delta^{(1)}m_{\tilde{t}_{2}\tilde{t}_{1}}^{2} &= \left(\delta^{(1)}m_{\tilde{t}_{1}\tilde{t}_{2}}^{2}\right)^*,\\\nextParentEquation
  \begin{split}
  \frac{\delta^{(1)}\mu}{\mu} &= -\frac{3 h_{t}^{2}}{32\pi^{2}} \left\{\Real{B_{1}{\left(\mu^2,m_{t}^{2},m_{\tilde{b}_{1}}^{2}\right)}} + \sum_{i\;=\;1}^{2}\lvert\mathbf{U}_{\tilde{t}\,i2}\rvert^{2}\Real{B_{1}{\left(\mu^2,0,m_{\tilde{t}_{i}}^{2}\right)}}\right\}.
  \end{split}
\end{align}
\end{subequations}
}%

\end{appendix}

\pdfbookmark[1]{References}{refs}

\end{document}